\documentclass[a4paper,9pt,twocolumn,twosides]{extarticle}

\usepackage{multirow}
\usepackage{bbding}
\usepackage{latexsym,amssymb,amsmath}
\usepackage{amsthm}
\usepackage{extarrows}
\usepackage{xspace}
\usepackage{colortbl}
\usepackage{comment}
\usepackage{url}
\usepackage{enumerate}
\usepackage[inline]{enumitem}
\usepackage{subfigure}
\usepackage{graphicx}
\graphicspath{{images/}}
\usepackage{calc}
\usepackage{mathtools}
\usepackage{microtype}
\usepackage{tabularx}
\usepackage{booktabs}
\usepackage{empheq}
\usepackage{hyperref}
\usepackage{threeparttable}
\usepackage{rotating}
\usepackage{algorithm}
\usepackage{varwidth}
\usepackage[noend]{algpseudocode}
\usepackage{hhline}
\usepackage{tikz}
\usetikzlibrary{shapes}
\usetikzlibrary{shapes.multipart}
\usetikzlibrary{positioning}
\usetikzlibrary{decorations}
\usetikzlibrary{decorations.pathmorphing} 
\usetikzlibrary{fit}		
\usetikzlibrary{backgrounds}	
\usetikzlibrary{matrix,arrows}
\usetikzlibrary{calc}
\usetikzlibrary{through}
\usetikzlibrary{decorations.pathreplacing}
\usetikzlibrary{arrows,snakes,automata,backgrounds,petri,positioning,decorations.pathmorphing,fit}
\usepackage[margin={1.5cm,1.5cm}]{geometry}

\usepackage{standalone}

\begin{document}

\newtheorem{claim}{Claim}

\newcommand{\symAlph}[2]{\textsf{symAlph}_{#1,#2}}
\newcommand{\abstraction}{\textsf{Abstr}}
\newcommand{\concret}{\textsf{Concr}}

\newcommand{\runset}[1]{\textsf{Runs}(#1)}
\newcommand{\brunset}[1]{\textsf{Runs}_\uabound(#1)}
\newcommand{\substDB}[2]{\textsf{Substitute}(#2,#1)}

\newcommand{\action}{\alpha}
\newcommand{\actiona}{\alpha}
\newcommand{\actionb}{\beta}
\newcommand{\actionc}{\gamma}
\newcommand{\actiond}{\delta}

\newcommand{\successor}{\textsf{succ}}
\newcommand{\phiruns}{\varphi^\textsf{Runs}_{\s}}
\newcommand{\MSO}{\texttt{MSO}}

\newcommand{\gadom}[1]{\textsc{Gadom}(#1)}
\newcommand{\existsg}{\exists^g}
\newcommand{\existsl}{\exists^\ell}
\newcommand{\forallg}{\forall^g}
\newcommand{\foralll}{\forall^\ell}
\newcommand{\posdom}[1]{\textsc{PosDom}(#1)}
\newcommand{\gvar}{\nu}
\newcommand{\Active}{\textsf{Active}}
\newcommand{\dbset}[2]{\textsf{DB-Inst-Set}(#1,#2)}
\newcommand{\getguard}[1]{#1{\cdot}\textsf{guard}}
\newcommand{\getdel}[1]{#1{\cdot}\del}
\newcommand{\getadd}[1]{#1{\cdot}\add}
\newcommand{\getfree}[1]{#1{\cdot}\textsf{free}}
\newcommand{\getnew}[1]{#1{\cdot}\textsf{new}}

\newcommand{\msodb}{\texttt{MSO-FO}\xspace}

\newcommand{\getbound}[1]{#1{\cdot}\textsf{bound}}
\newcommand{\boundvars}[1]{\getbound{#1}}

\newcolumntype{C}{>{\centering\arraybackslash}X}
\newcolumntype{R}{>{\raggedleft\arraybackslash}X}
\newcolumntype{L}{>{\raggedright\arraybackslash}X}

\newcommand{\confGraph}{\mathcal{C}}
\newcommand{\msonw}{\ensuremath{\texttt{MSO}_\texttt{NW}}}
\newcommand{\stepex}{\varphi}
\newcommand{\qedboxfull}{\vrule height 5pt width 5pt depth 0pt}
\newcommand{\qedfull}{\hfill{\qedboxfull}}

\newtheorem{exampleAux}{Example}[section]
\newenvironment{example}{\begin{exampleAux}\small\upshape}{\qedfull\end{exampleAux}}
\newtheorem{theorem}{Theorem}[section]
\newtheorem{lemma}[theorem]{Lemma}
\newtheorem{proposition}[theorem]{Proposition}
\newtheorem{corollary}[theorem]{Corollary}
\newtheorem{definition}{Definition}[section]
\newtheorem{remark}{Remark}[section]
\newtheorem{fact}{Fact}

\newcommand{\im}[1]{\textsc{im}(#1)}

\newcommand{\per}{\mbox{\bf .}}                  

\newcommand{\cld}{,\ldots,}                      
\newcommand{\ld}[1]{#1 \ldots #1}                 
\newcommand{\cd}[1]{#1 \cdots #1}                 
\newcommand{\lds}[1]{\, #1 \; \ldots \; #1 \,}    
\newcommand{\cds}[1]{\, #1 \; \cdots \; #1 \,}    

\newcommand{\conj}{\mathit{conj}}

\newcommand{\var}{\mathit{Var}}
\newcommand{\fresh}{\mathit{Fresh}}
\newcommand{\fr}[1]{\underline{#1}}

\newcommand{\set}[1]{\{#1\}}
\newcommand{\tup}[1]{\langle #1\rangle}
\newcommand{\pair}[1]{\left({#1}\right)}

\newcommand{\domain}{\Delta}
\newcommand{\schema}{\mathcal{R}}
\newcommand{\const}{\domain_0}

\newcommand{\I}{I}
\newcommand{\history}{H}

\newcommand{\idb}{\I_0}
\newcommand{\act}{\textsc{acts}}

\newcommand{\adom}[1]{\textsc{adom}(#1)}
\newcommand{\ADOM}[1]{\adom{#1}}
\newcommand{\sys}{\text{DMS}\xspace}
\newcommand{\syss}{{\sys}s\xspace}
\newcommand{\esys}{\textsc{dms}(\exlang)\xspace}
\newcommand{\esyss}{{\esys}s\xspace}
\newcommand{\dcds}{\text{DCDS}\xspace}

\newcommand{\guard}{\fodb{Q}}
\newcommand{\add}{\mathit{Add}}
\newcommand{\del}{\mathit{Del}}

\newcommand{\s}{\mathcal{S}}
\newcommand{\cgraph}{\mathcal{C}}

\newcommand{\ts}[1][]{\ensuremath{\Upsilon_{#1}}}

\newcommand{\trans}[1][]{\xrightarrow{#1}}
\newcommand{\transcl}[1][]{\makebox{\ensuremath{\xrightarrow{#1}\!\!\!^{\ast}}}}
\newcommand{\goto}[2]{#1 \trans #2}
\newcommand{\gotos}[3]{#1 \trans[#3] #2}

\newcommand{\reach}[2]{#1 \transcl #2}
\newcommand{\reachs}[3]{#1 \transcl[#3] #2}

\newcommand{\reachproblem}{{\sys}\text{-}\textsc{reach}\xspace}
\newcommand{\rreachproblem}{{\sys}\text{-}\textsc{rep}\text{-}\textsc{reach}\xspace}
\newcommand{\evsatproblem}{{\sys}\text{-}\textsc{evSat}\xspace}
\newcommand{\edmsreachproblem}{{\esys}\text{-}\textsc{reach}\xspace}

\newcommand{\ctrans}{\hookrightarrow}
\newcommand{\ctranscl}{\ctrans \!\!\!^{\ast}}
\newcommand{\cgoto}[2]{#1 \ctrans #2}
\newcommand{\creach}[2]{#1 \ctranscl #2}
\newcommand{\chaltproblem}{{\cmac}\text{-}\textsc{Halt}\xspace}
\newcommand{\tchaltproblem}{{\tcmac}\text{-}\textsc{Halt}\xspace}
\newcommand{\creachproblem}{{\cmac}\text{-}\textsc{Reach}\xspace}
\newcommand{\tcreachproblem}{{\tcmac}\text{-}\textsc{Reach}\xspace}

\newcommand{\live}{\textsc{live}}

\newcommand{\tcmac}{\textsc{2cm}\xspace}
\newcommand{\cmac}{\textsc{cm}\xspace}
\newcommand{\cm}{\mathcal{M}}
\newcommand{\creg}{n}
\newcommand{\cstate}{Q}
\newcommand{\state}{q}
\newcommand{\cprog}{\Pi}
\newcommand{\inc}{\mathtt{inc}}
\newcommand{\dec}{\mathtt{dec}}
\newcommand{\ifz}{\mathtt{ifz}}

\newcommand{\ans}[2]{\mathit{ans}(#1,#2)}

\newcommand{\true}{\mathsf{true}}
\newcommand{\false}{\mathsf{false}}

\newcommand{\rel}[1]{\mathit{#1}}
\newcommand{\cval}[1]{\mathsf{#1}}

\newcommand{\rell}{\rel{R}}

\newcommand{\compl}[1]{\textsc{#1}}
\newcommand{\ptime}{\compl{pTime}}
\newcommand{\exptime}{\compl{expTime}}
\newcommand{\pspace}{\compl{pSpace}}
\newcommand{\expspace}{\compl{expSpace}\xspace}
\newcommand{\twoexpspace}{\compl{2expSpace}\xspace}

\newcommand{\undecidable}{{undecidable}}
\newcommand{\decidable}{\textbf{decidable}}

\newcommand{\constr}{\Gamma}

\newcommand{\instance}{inst\xspace}
\newcommand{\mapof}[2]{{\it {#1}}\left({#2}\right)}
\newcommand{\tuple}[1]{\left\langle {#1}\right\rangle}
\def\by#1{\mathop{{\hbox{\setbox0=\hbox{$\scriptstyle{#1\quad}$}{$%
\mathrel{\mathop{\setbox1=\hbox to \wd0{\rightarrowfill}\ht1=3pt\dp1=-2pt\box1}\limits^{#1}}%
$}}}}}
\newcommand{\setcomp}[2]{\left\{{#1}|\;{#2}\right\}}
\newcommand{\fin}{\texttt{fin}\xspace}
\newcommand{\init}{\texttt{init}\xspace}
\newcommand{\target}{\texttt{target}\xspace}
\newcommand{\tr}{\texttt{true}\xspace}
\newcommand{\fa}{\texttt{false}\xspace}
\newcommand{\ste}{\texttt{state}\xspace}
\newcommand{\ctr}{\texttt{idle}\xspace}
\newcommand{\pre}{\texttt{pre}\xspace}
\newcommand{\ready}{\texttt{ready}\xspace}
\newcommand{\trs}{\texttt{trans}\xspace}
\newcommand{\present}[1]{\texttt{val}\left({#1}\right)}
\newcommand{\tval}[1]{\present{#1}}
\newcommand{\sigof}[2]{{\texttt{sig}}\left({#1} \texttt{ in } {#2}\right)}
\newcommand{\allsigsof}[1]{{\texttt{Sigs}}\left({#1}\right)}
\newcommand{\bitv}{\texttt{bitv}\xspace}
\newcommand{\typeof}[2]{\texttt{type}\left({#1} \texttt{ in } {#2}\right)}
\newcommand{\typeofsig}[2]{\texttt{type} \left( \texttt{sig} \left({#1} \texttt{ in } {#2}\right) \right)}
\newcommand{\sigoftype}[2]{\texttt{sig}\left( \texttt{type} \left({#1} \texttt{ in } {#2}\right) \right)}
\newcommand{\relval}{\texttt{rval}\xspace}
\newcommand{\boolval}{\texttt{bval}\xspace}
\newcommand{\relvalof}[2]{\relval\left({#1} \texttt{ in } {#2}\right)}
\newcommand{\boolvalof}[2]{\boolval\left({#1} \texttt{ in } {#2}\right)}
\newcommand{\typefrom}{\texttt{typefrom}\xspace}
\newcommand{\typefromof}[1]{\typefrom\left({#1}\right)}
\newcommand{\sigfrom}{\texttt{sigfrom}\xspace}
\newcommand{\sigfromof}[1]{\sigfrom\left({#1}\right)}
\newcommand{\signull}{\texttt{flag\_sig}}
\newcommand{\typenull}{\texttt{flag\_type}}
\newcommand{\signullof}[1]{\signull\left({#1}\right)}
\newcommand{\typenullof}[1]{\typenull\left({#1}\right)}
\newcommand{\suboftype}{\ominus}
\newcommand{\addoftype}{\oplus}
\newcommand{\maximof}[1]{\texttt{max}\left({#1}\right)}
\newcommand{\atom}{\texttt{atom}\xspace}
\newcommand{\explicit}[1]{{explicit\xspace}\left({#1}\right)}
\newcommand{\compactfact}[1]{\texttt{compact-fact}\xspace\left({#1}\right)}
\newcommand{\expandfact}[1]{\texttt{expand-fact}\xspace\left({#1}\right)}
\newcommand{\compactrel}[1]{\texttt{compact-rel}\xspace\left({#1}\right)}
\newcommand{\compactdb}[1]{\texttt{compact-db-inst}\xspace\left({#1}\right)}
\newcommand{\expanddb}[1]{\texttt{expand-db-inst}\xspace\left({#1}\right)}
\newcommand{\compactact}[1]{\texttt{compact-act}\xspace\left({#1}\right)}
\newcommand{\pos}{\textsf{pos\xspace}}
\newcommand{\fix}{\textsf{fix\xspace}}
\newcommand{\fixof}[1]{{\fix}\left({#1}\right)}
\newcommand{\cons}{\textsf{cons\xspace}}
\newcommand{\consof}[1]{{\cons}\left({#1}\right)}
\newcommand{\sub}{\textsf{sub\xspace}}
\newcommand{\subof}[1]{{\sub}\left({#1}\right)}
\newcommand{\bound}{\textsf{bound\xspace}}
\newcommand{\boundof}[1]{{\bound}\left({#1}\right)}
\newcommand{\confc}{\textsf{conf\xspace}}
\newcommand{\isoconf}{\textsf{iso-conf\xspace}}
\newcommand{\isoconfof}[1]{{\isoconf}\left({#1}\right)}
\newcommand{\isoop}{\textsf{iso-op\xspace}}
\newcommand{\isoopof}[1]{{\isoop}\left({#1}\right)}

\newcommand{\lang}{\mathcal{L}}
\newcommand{\clang}[1]{\mathtt{#1}}
\newcommand{\folang}{\clang{FOL}(\schema)}
\newcommand{\exlang}{\clang{\exists FOL}}
\newcommand{\ucqlang}{\clang{UCQ}}
\newcommand{\cqlang}{\clang{CQ}}
\newcommand{\cqlangn}{\clang{CQ_{\lnot}}}

\newcommand{\Hit}{\textsc{Hit}\xspace}


\newcommand{\uabound}{\mathbf{b}}
\newcommand{\maxseqno}{\texttt{MAX\_seq\_no}}
\newcommand{\seqno}[1]{\texttt{seq\_no}(#1)}
\newcommand{\seqnof}{\texttt{seq\_no}}
\newcommand{\seqnonew}[1]{\texttt{seq\_no}'(#1)}
\newcommand{\seqnonewf}{\texttt{seq\_no}'}
\newcommand{\recent}[1]{\textsf{Recent}_\uabound(#1)}
\newcommand{\Nat}{\mathbb{N}}
\newcommand{\Zed}{\mathbb{Z}}
\newcommand{\kgoto}[3]{#2 \trans_{#1} #3}
\newcommand{\kgotos}[4]{#2 \trans[#4]_{#1} #3}
\newcommand{\kreach}[2]{#1 \transcl_\uabound #2}
\newcommand{\kreachproblem}{$\uabound$-{\sys}\text{-}\textsc{Reach}\xspace}
\newcommand{\kmcproblem}{$\uabound$-{\sys}\text{-}\textsc{MC}\xspace}
\newcommand{\ktrans}{\trans_{\uabound}}
\newcommand{\kprimereachproblem}{$\uabound'$-{\sys}\text{-}\textsc{Reach}\xspace}
\newcommand{\kform}{\varphi^k_{\s,q}}
\newcommand{\kformval}{\varphi^k_{\s}}
\newcommand{\formRecent}{\varphi^\text{Recent}}
\newcommand{\nestrel}{\triangleright}
\newcommand{\nwtl}{\ensuremath{\textsc{nwtl}^+}}

\newcommand{\pop}[1]{\uparrow_{#1}}
\newcommand{\push}[1]{\downarrow_{#1}}
\newcommand{\word}{\text{block}}
\newcommand{\block}{\textsf{block}}
\newcommand{\step}{\textsf{step}}
\newcommand{\sameblock}{\textsf{Block}^{=}}
\newcommand{\Eq}{\textsf{Eq}}
\newcommand{\Eqij}[2]{\textsf{Eq}^{#1,#2}}

\newcommand{\Anw}{\Sigma}
\newcommand{\Apush}{\Anw_\downarrow}
\newcommand{\Apop}{\Anw_\uparrow}
\newcommand{\Aint}{\Anw_\text{int}}
\newcommand{\nnew}{\mathbf{n}}
\newcommand{\Live}{\textsf{live}}
\newcommand{\translate}[1]{\lfloor{#1}\rfloor_{\action, s, x}}

\newcommand{\nbxina}[1]{\text{n-x}_{#1}}
\newcommand{\nbyina}[1]{\text{n-y}_{#1}}
\newcommand{\maxy}{\mathbf{\eta}}
\newcommand{\f}[1]{\uabound+#1}

\newcommand{\relm}[3]{\textsf{Rel-#1}(#2)@#3^\circleddash}
\newcommand{\reln}[3]{\textsf{Rel-#1}(#2)@#3^\oplus}

\newcommand{\arity}{a}
\newcommand{\arityb}{b}

\newcommand{\fodb}[1]{#1}

\newcommand{\forun}[1]{#1}
\newcommand{\fovars}{\textsf{Vars}_\textsf{FO}}
\newcommand{\sovars}{\textsf{Vars}_\textsf{SO}}
\newcommand{\datavars}{\textsf{Vars}_\textsf{data}}
\newcommand{\vars}{\textsf{Vars}}
\newcommand{\freevars}[1]{\textsf{Free-Vars}(#1)}
\newcommand{\freshvars}[1]{\textsf{Fresh-Vars}(#1)}

\newcommand{\fakt}{\textsf{fa}}

\newcommand{\enrolled}{\textsc{Enrolled}}
\newcommand{\graduated}{\textsc{Graduated}}

\newcommand{\brecent}{\textsf{Recent}_\uabound}
\newcommand{\brecentof}[2]{\textsf{Recent}_\uabound(#1,#2)}

\newcommand{\generator}{\textsf{Gen}}
\newcommand{\agenerator}{\textsf{AGen}}

\newcommand{\seqnoset}[1]{\mathsf{SNset}(#1)}
\newcommand{\rec}{\mathsf{rec}}
\newcommand{\recset}[1]{\mathsf{recIdxSet(#1)}}
\newcommand{\symSubs}{\mathsf{SymSubs}}
\newcommand{\translation}[1]{\lfloor #1 \rfloor}
\newcommand{\maxarity}{\mathfrak{a}}
\newcommand{\maxvars}{\mathfrak{n}}


\newcommand{\start}[1]{
\begin{tikzpicture}[baseline=(char.base)]
\node(char)[draw,fill=white,
  shape=rounded rectangle
  ]
  {\textsc{beg} \ensuremath{#1}};
\end{tikzpicture}
}
\newcommand{\finish}[1]{
\begin{tikzpicture}[baseline=(char.base)]
\node(char)[draw,fill=white,
  shape=rounded rectangle
  ]
  {\textsc{end} \ensuremath{#1}};
\end{tikzpicture}
}
\newcommand{\valuation}[2]{
\begin{tikzpicture}[baseline=(char.base)]
\node(char)[draw,fill=white,
  shape=rounded rectangle
  ]
  {#1 : \ensuremath{ #2} };
\end{tikzpicture}
}
\newcommand{\frameletter}[1]{
\begin{tikzpicture}[baseline=(char.base)]
\node(char)[draw,fill=white, inner sep = 1pt, 
  shape=rounded rectangle, minimum height=.35cm
  ]
  {\ensuremath{ #1} };
\end{tikzpicture}
}

\newcommand{\dtrans}[1]{\Upsilon_{#1}}
\newcommand{\exact}[1]{\mathsf{#1}}
\newcommand{\inputvar}[1]{\mathbf{#1}}

\sloppy

\title{\textbf{Recency-Bounded Verification of
 Dynamic Database-Driven Systems}\\
 (Extended Version)}

\author{
	Parosh Aziz Abdulla\\
	Uppsala University\\
	\texttt{parosh@it.uu.se}
\and 
	 C.~Aiswarya\\
	Uppsala University\\
	\texttt{aiswarya@cmi.ac.in}
\and 
	Mohamed Faouzi Atig\\
	Uppsala University\\
	\texttt{mohamed\_faouzi.atig@it.uu.se}
\and 
	Marco Montali\\
	Free Univ.~of Bozen/Bolzano\\
	\texttt{montali@inf.unibz.it}
\and 
	Othmane Rezine\\
	Uppsala University\\
	\texttt{othmane.rezine@it.uu.se}
}

\date{}

\maketitle

\begin{abstract}
We propose a formalism to model database-driven systems, called database manipulating systems (\sys). The actions of a \sys modify the current instance of a relational database by adding new elements into the database, deleting tuples from the relations and adding tuples to the relations. The elements which are modified by an action are chosen by (full) first-order queries.  \sys is a highly expressive model and can be thought of as a succinct representation of an infinite state relational transition system, in line with similar models proposed in the literature. We propose monadic second order logic (\msodb) to reason about sequences of database instances appearing along a run. Unsurprisingly, the linear-time model checking problem of \sys against \msodb is undecidable. Towards decidability, we propose under-approximate model checking of \sys, where the under-approximation parameter is the ``bound on recency''. In a $k$-recency-bounded run, only the most recent $k$ elements in the current active domain may be modified by an action. More runs can be verified by increasing the bound on recency. Our main result shows that recency-bounded model checking of \sys against \msodb is decidable, by a reduction to the satisfiability problem of MSO over nested words.
\end{abstract}

\medskip


\section{Introduction}

In the last 15 years, research in business process management (BPM) and workflow technology has progressively shifted its emphasis from a purely control-flow, activity-centric perspective to a more holistic approach that considers also how data are manipulated and evolved by the process \cite{Reic12}. In particular, two lines of research emerged at the intersection of database theory, BPM and formal methods: one focused on modeling languages and technologies for specifying and enacting data-aware business processes \cite{MeSW11}, and the other tailored to their analysis and verification \cite{CaDM13}. 

The first line of research gave birth to a plethora of new languages and execution platforms, culminating in the so-called \emph{object-centric} \cite{KuWR11} and \emph{artifact-centric} paradigms \cite{Nigam03:artifacts}, respectively exemplified by frameworks like PHILharmonicFlows \cite{KunR11} and IBM GSM (Guard-Stage-Milestone) \cite{DaHV11}. Notably, GSM became the core of the recently published CMMN OMG standard on (adaptive) case management\footnote{\url{http://www.omg.org/spec/CMMN/}}. In this paper, we will use \emph{dynamic database-driven systems} as an umbrella term for all such platforms.

The second line of research focused on
understanding the boundaries of decidability and complexity for the verification of dynamic database-driven systems. Two main trends can be identified along this line. 
The first trend was initiated in the late 1990s with the introduction of relational transducers \cite{AVFY00}, and continued with new results over progressively richer variants of the initial model, such as systems equipped with arithmetic \cite{DHPV09,Damaggio2011:Artifact}, systems decomposed into interacting web services \cite{DeSV07}, and systems operating over XML databases \cite{BoST13}. 
The modelling formalisms introduced in this direction operate over a read-only, input database that is fixed during the system evolution, and use quantifier-free FO formulae to query such a database. The obtained answers can be stored into a read-write state database, whose size is fixed a-priori. Verification problems include control-state reachability \cite{BoST13}, or model checking \cite{DHPV09,Damaggio2011:Artifact} against formulae expressed in FO variants of temporal logics with a  limited form of FO quantification across state.
Furthermore, verification is \emph{input-parametric}, that is, studied independently from the configuration of data in the initial input database. 

In contrast, the second trend studies dynamic systems where the initial state is known. 

Hence their execution semantics  can be captured by means of a single \emph{relational transition system} (RTS), that is, a (possibly) infinite-state transition system whose states are labeled with database instances \cite{Vard05}. 
Further, the actions allow for bulk read-write operations over the database, possibly injecting fresh values taken from an infinite domain. The injection of such values accounts for the input of new information from the external environment (e.g., through user interaction or communication with external systems/services), or the insertion of globally unique identifiers (GUIDS).  

Verification of dynamic database-driven systems is challenging due to the infinite state-space generated. Several works \cite{BLP:KR:12,BCDD13,SMTD13,ICSOC10,ICSOC11,Bagheri2011:Artifacts} succeeded in obtaining decidability  by imposing restrictions that yielded finite-state abstractions of the entire system. In \cite{LomM14}, decidability is obtained for unbounded-state dynamic database-driven systems in the restrictive case where the  database schema contains a single  unary relation.

In this paper we propose an under-approximation based on recency of the elements, which allows unbounded state-space. With this restriction we show decidability for the model checking problem against monadic second-order logic over sequences of database instances.

More specifically, we introduce \emph{database-manipulating systems} (DMSs) to model dynamic database-driven systems. 
Salient features of DMS include guarding every action using (unrestricted) first-order queries on the current database, addition and deletion of tuples in the database, and addition of new elements in the database (which results in a growing active domain).

On top of this model, we study  linear-time model checking, using \emph{monadic second-order logic over runs} (\msodb) to reason about sequences of database instances appearing along the DMS runs. \msodb employs FO queries as its atomic formulae, and supports  FO data-quantifications across distinct time points. This powerful  logic can express popular verification problems such as  reachability, repeated reachability, fairness, liveness, safety, FO-LTL, etc. For example, the property that ``every enrolled student eventually graduates'' can be formalized in \msodb as:
\[ \forall x \forall u . \mathit{Enrolled}(u) @ x \Rightarrow \exists y . y > x \land \mathit{Graduated}(u) @ y\]
where $x$ and $y$ are \emph{position} variables, used to predicate about the different time points encountered along a run, while $u$ is a data variable, which matches with values stored in the databases present at these time points. This property corresponds to the FO-LTL formula $\forall u. \mathbf{G} \mathit{Enrolled}(u) \Rightarrow \mathbf{F} \mathit{Graduated}(u)$. More sophisticated properties can be encoded by leveraging the expressive power of \msodb, such as that between the enrolment of a student to a course and the moment in which the student passes that course, there is an even number of times in which the student fails that course. 

As a first result we show that, unsurprisingly, already propositional reachability turns out to be undecidable to check, even for extremely limited DMSs. Instead of attacking this negative result by limiting the expressive power of the DMS specification formalism, we consider \emph{under-approximate verification}, restricting our attention only to those runs that satisfy a given criterion. In particular, we consider as the under-approximation parameter the \emph{bound on recency}. In a $\uabound$-recency-bounded run, only the most recent $\uabound$ elements in the current active domain may be modified (i.e., updated or deleted) by an action, but the behavior of the action may be influenced by the entire content of the database. More runs are verified by increasing the bound on recency.
In particular, model checking of safety properties converges to exact model checking in the limit.

Our main result shows that \emph{recency-bounded model checking of \sys against \msodb is decidable.} Towards a proof, we encode runs of a recency-bounded DMS as an (infinite) nested word \cite{AlurM09}. We then show that the correctness of the encoding can be expressed in MSO over nested words, consequently isolating those runs that correspond to the actual possible behaviors induced by the DMS. At the same time, we describe how to translate the \msodb property of interest into a corresponding MSO formula over nested words. In this way, we are able to reduce recency-bounded model checking of \sys against \msodb to the satisfiability problem of MSO over nested words, which is known to be decidable \cite{AlurM09}.

\section{Preliminaries}
\label{section:Preliminaries}
We start by introducing the preliminaries necessary for the
development of our framework and results.

\smallskip
\textbf{Databases.}
We fix a (data) domain $\domain$, which is a countably infinite set of
data values, acting as standard names.
A relational schema $\schema$ is a finite  set
$\set{R_1/\arity_1,\ldots,R_n/\arity_n}$ of relation names $R_i$, each coming with its own arity $\arity_i$.
A database instance $\I$ over schema $\schema$ and domain $\domain$  is the union set $\cup_{i:1\le i \le n} R_i^\I$,
where $R_i^\I \subseteq \set{R_i} \times 
\domain^{\arity_i}$
represents the content of relation $R_i$ in the database instance $\I$.
If $\I$  contains a tuple (or a \textit{fact}) $\tup{R_i, e_1,\ldots,e_{a_i}}$, we write
 $R_i(e_1,\ldots,e_{a_i})\in\I$.
A nullary relation $p/0$ (also known as proposition) can be either instantiated as the singleton set $\set {p()}$
or the empty set $\emptyset$.
In the former case, we say the proposition is \emph{true}, and 
write $p \in \I$.
In the latter case $p \notin \I$ and  we say $p$ is \emph{false}.

We denote the set of all database instances over $\schema$ and $\domain$ by $\dbset{\schema}{\domain}$.
The \emph{active domain} of $\I$, denoted $\adom{\I}$,
is the subset of $\domain$ such that $e \in \adom{\I}$ if and only if $e$ occurs in some fact in $\I$
(i.e. there exist $\tup{R_i, e_1,\ldots,e_{a_i}}\in\I$ such that $e = e_j$ for some $j:1\le j \le a_i$).
Given two database instances $\I_1, \I_2 \in \dbset{\schema}{\domain}$, we define $I_1 + I_2$ to be the database instance $\I \in \dbset{\schema}{\domain}$ obtained by taking the relation-wise union.
Similarly we define $\I_1 - \I_2$ where we take the relation-wise set difference.
Simply put, $\I_1 + \I_2 = \I_1 \cup \I_2$ and $\I_1 - \I_2 = \I_1 \setminus \I_2$.

\smallskip
\noindent
\textbf{Queries.}
We use queries to access databases and extract data values of
interest. Queries are expressed in \emph{FOL with equality} over the schema $\schema$ ($\folang$
for short). Let $\datavars = \{ u, v, u_1, \dotsc \}$ be the set of FO
data-variables ranging over the data values in $\domain$. 
 A \emph{$\folang$ query}  is given by the following syntax:
\[
 \fodb{Q}::= \fodb{\true} \mid \fodb{ R(u_1,\ldots,u_\arity)}  \mid \fodb{\lnot Q} \mid\fodb{Q_1\land Q_2} \mid
              \fodb{\exists u. Q} \mid \fodb{u_1 = u_2}
\]
where $R/\arity \in \schema$, and $u, u_i$ are variables from $ \datavars$. 
We use standard abbreviations like $\fodb{Q_1 \lor Q_2} = \fodb{\neg (\neg Q_1 \land
\neg Q_2)}$,  $\fodb{\forall u. Q} = \fodb{\neg \exists u. \neg Q}$,  etc.
We also denote with  $\freevars{\fodb{Q}}$ the set of free variables
appearing in a query $Q$.

For a set $V \subseteq \datavars$, a \textit{substitution}  $\sigma$
of $V$ is a function that maps every variable in $V$ to a  value
in $\domain$ (i.e., $\sigma :  V \rightarrow \domain$). Given a
substitution $\sigma: V \rightarrow \domain$ and set $V' \subseteq V$,
we define the restriction of $\sigma$ on $V'$ as the substitution
$\sigma': V' \to \domain$ such that $\sigma'(u) = \sigma(u)$ for every $u \in V'$. We denote the restriction of $\sigma$ to $V'$ by $\sigma |_ {V'}$. 

Given a database instance $\I $ over $\schema$ and $\domain$,  a $\folang$
query $\fodb{Q}$ over $\schema$, and a substitution $\sigma:
\freevars{\fodb{Q}} \to \domain$, we write $\I, \sigma \models
\fodb{Q}$ if the query $\fodb{Q}$ under the substitution $\sigma$
\emph{holds} in  database $\I$.
The semantics are as expected, and can be found for completeness in Appendix~\ref{app:query-semantics}. 
The set   of 
  \emph{answers} of $\fodb{Q}$ over $\I$, denoted
  $\ans{\fodb{Q}}{\I}$,  is the set of all substitutions $\sigma:
  \freevars{\fodb{Q}} \to \domain $  such that $\I, \sigma \models
  \fodb{Q}$. When $\freevars{Q} = \emptyset$ (i.e., $Q$ is a boolean
  query), we set $\ans{\fodb{Q}}{\I}$ to be the empty substitution
  $\set \epsilon$ whenever $\I, \set \epsilon \models \fodb{Q}$ (or $\I \models \fodb Q$ for short), and
  we assign  $\ans{\fodb{Q}}{\I}$ to the empty set $\emptyset$
  whenever $\I, \set \epsilon \not\models \fodb Q$ (or $\I \not\models
  \fodb{Q}$ for short).

\begin{example}\label{ex:formula-active}
We describe a query $\fodb{\Active(u)}$ with a single free variable
$u$, to check whether $u$ is present in some tuple of some relation, no matter what the other elements of the tuple are:
\[
\Active(u) \equiv 
\!\!\!\!\!
\bigvee_{{R/\arity} \in \schema} 
\!\!\!\!\!
 \exists u_1,
\dotsc , u_{\arity}  
\!\!\!\!\!
\bigvee_{1 \le j \le \arity} 
\!\!\!\!\!
 R(u_1, \dotsc,
u_{j-1}, u, u_{j+1}, \dotsc, u_{\arity})
\]
$\fodb{\Active(u)}$ characterises $\adom{\I}$.
In fact, $\ans{\fodb{\Active(u)}}{\I}$ is $\{ \tuple{u \mapsto e} ~\mid ~ e \in \adom{\I} \}$. 
\end{example}

\smallskip
\noindent
\textbf{Substitutions in database instances.}
Let $V\subseteq \datavars$ be a set of variables. Consider a substitution $\sigma: V \to \domain$ that assigns each variable to an element from $\domain$. Let $\I \in \dbset{\schema}{V}$ be a database instance over schema $\schema$ and the variables $V$. We define $\substDB{\sigma}{\I} \in \dbset{\schema}{\domain}$ to be the database instance obtained from $\I$ by substituting every occurrence of  variable $u$ by $\sigma(u)$, for each variable $u \in V$. 

\section{Framework}
\label{sec:dms-definition}

We introduce our model for dynamic database-driven systems. 
A \emph{Database-Manipulating System ($\sys$)}
over domain $\domain$ and schema $\schema$ is a pair
$\s = \tup{\idb,\act}$, where:
\begin{itemize}
\item $\idb \in \dbset{\schema}{\domain}$ is the \emph{initial database instance} over $\schema$
  and $\domain$, with
  $\adom{\idb} = \emptyset$. $\idb$ gives truth-values to the nullary relations (also known as propositions), and has empty non-nullary relations
  \item $\act$ is a set of \emph{(guarded) actions}. An action  $\action$ is a tuple $\action =  \tup{\vec{u}, \vec{v}, \guard,\del,\add}$, where
  \begin{itemize}
  \item $\vec{u}$ and $\vec{v}$ are disjoint finite subsets of
    $\datavars$, respectively denoting \emph{action parameters} and
     \emph{fresh-input variables}.
  \item $\guard$ is a $\folang$ query, called the \emph{guard} of $\action$. 
  \item $ \vec{u} =\freevars{\fodb{Q}}$. 
  \item  $\del \in \dbset{\schema}{\vec{u}}$ is a database instance over the variables $\vec{u}$ and the schema $\schema$. 
 \item  $\add \in \dbset{\schema}{\vec{u} \uplus \vec{v}}$ is a
   database instance over the variables $\vec{u}\uplus \vec{v}$ and
   $\schema$, with $\vec{v} \subseteq \adom{\add}$.  The set $\vec{v}$
   contains the so-called \emph{fresh-input variables} of $\action$. 
  \end{itemize}
\end{itemize}

Given an action $\action =  \tup{\vec{u}, \vec{v}, \guard,\del,\add}$,
we refer to:  $\vec{u}$ by $\getfree{\action}$, 
 $\vec{v}$ by $\getnew{\action}$, 
 $\guard$ by $\getguard{\action}$, 
 $\del$ by $\getdel{\action}$, and 
 $\add$ by $\getadd{\action}$. 

Intuitively, a $\sys$ operates as follows. At any instant, it
maintains a database instance from $\dbset{\schema}{\domain}$ and a
history-set $\history \subseteq \domain$ of elements encountered along
its execution.  It starts with the initial database instance $\idb$,
and the empty history-set ($\history = \emptyset$). At an instant, the
$\sys$ can update the current database instance and the history-set by
applying an action. An action is applied in three steps. In the first
step  the current database is queried using $\guard$ to retrieve some
elements of interest from its active domain. In the second step, some
tuples involving the retrieved elements are removed from the current
database, as dictated by the variable-database instance
$\del$.
Finally, new tuples may be added to the relations of the current
database instance, as dictated by $\add$. The newly inserted tuples may
contain fresh values that were not present in the history-set, and
that are injected through the fresh-input variables. We give the formal execution semantics below. 

\smallskip
\noindent
\textbf{Execution semantics.}
The execution semantics of a \sys $\s =
\tup{\idb,\act}$ over $\schema$ and $\domain$ is defined in terms of a (possibly
infinite) \emph{configuration graph} $\confGraph_\s$, which has the
form of a relational transition system \cite{Vard05,BCDD13} equipped with
additional information about the data values encountered so far. Each configuration is a pair
$\tup{\I,\history}$, where $\I \in \dbset{\schema}{\domain}$ is a database instance over $\schema$
and $\domain$, and $\history \subseteq \domain$ is a
\emph{history-set}, i.e., the set of values encountered
in the history of the current execution of the system. 

Let $\tup{\I,\history}$ be a configuration and $\action =  \tup{\vec{u}, \vec{v}, \guard,\del,\add}$ be an action. Consider a substitution $\sigma$ from $\vec{u} \uplus \vec{v}$ to
  $\domain$. We say that $\sigma$ is an \emph{instantiating substitution} for $\action$ at $\tup{\I, \history}$ if it satisfies the following:
  \begin{itemize}
   \item for every variable $u_i \in \vec{u}$, $\sigma(u_i) \in
  \adom{\I}$ (action parameters are substituted with values from the
  current active domain);
\item for every variable $v_i \in \vec{v}$, $\sigma(v_i) \not \in \history$
  (fresh-input variables are substituted with history-fresh values);
\item $\sigma|_{\vec{v}}$ is injective (fresh-input variables are assigned to pairwise
  distinct values);
\item $\I , \sigma|_{\vec{u}} \models \guard$ (the action guard is satisfied).

  \end{itemize}

For a pair of configurations
$\tup{\I,\history}$ and $\tup{\I',\history'}$, an action $\action = \tup{\vec{u}, \vec{v}, \guard,\del,\add} \in \act$,  and a substitution $\sigma$ from $\vec{u} \uplus \vec{v}$ to
  $\domain$, we have an edge
  $\gotos{\tup{\I,\history}}{\tup{\I',\history'}}{\action:\sigma}$ in
  $\confGraph_\s$, if the following conditions hold:
 \begin{itemize}
\item $\sigma$ is an instantiating substitution for $\action$ at $\tup{\I,\history}$;
 \item $ \I ' = (\I - \substDB{\sigma}{\del}) + \substDB{\sigma}{ \add}$;
 \item    $\history' = \history \cup \set{\sigma(v_i) \mid v_i \in \vec{v}}$.
 \end{itemize}

An \emph{extended run} $\hat \rho$ of  $\s$ is an infinite sequence 
$$\tup{\I_0, \history_0} \trans[\action_0 : \sigma_0] \tup{\I_1, \history_1} \trans[\action_1 : \sigma_1] \tup{\I_2, \history_2} \trans[\action_2 : \sigma_2] \tup{\I_3, \history_3}\dots$$
where $\I_0$ is the initial database instance of $\s$, and $\history_0 = \emptyset$. Note that, by definition, ${\history}_i = {\cup}_{0 \le k \le i} \adom{{\I}_k}$. 
The \emph{run} $\rho$ generated by the extended run $\hat \rho$ is the
sequence $\I_0, \I_1, \I_2 \dots$ of database instances appearing
along $\hat \rho$.  The \emph{set of all runs} of a \sys $\s$ is
denoted by $\runset{\s}$.

\begin{figure*}[ht]
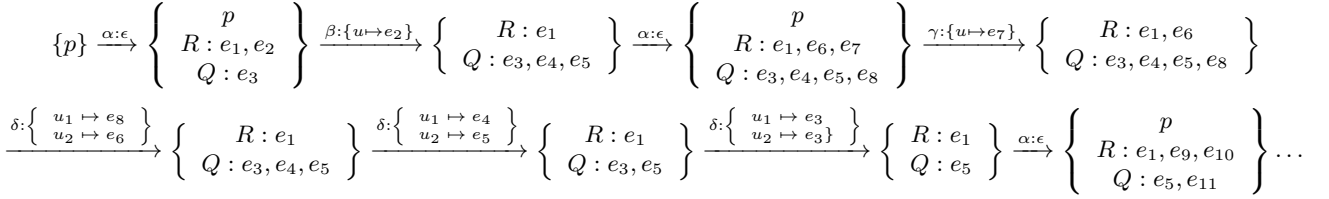

$$\{ p\}
\xrightarrow{\actiona:\epsilon}
\left\{\begin{array}{c}
p \\
R: e_1, e_2\\
 Q: e_3
\end{array}  \right\}
\xrightarrow{\actionb: \{u \mapsto e_2\}}
\left\{\begin{array}{c}
R: e_1\\
 Q: e_3, e_4, e_5
\end{array}  \right\}
\xrightarrow{\actiona:\epsilon}
\left\{\begin{array}{c}
p\\
R: e_1,e_6, e_7\\
 Q: e_3, e_4, e_5, e_8
\end{array}  \right\}
\xrightarrow{\actionc:\{ u \mapsto e_7\} }
\left\{\begin{array}{c}
R: e_1,e_6\\
 Q: e_3, e_4, e_5, e_8
\end{array}  \right\}
$$
$$
\xrightarrow{\actiond:\left\{ {\scriptsize\begin{array}{l}u_1 \mapsto e_8\\ u_2 \mapsto e_6\end{array}}\right\}}
\left\{\begin{array}{c}
R: e_1\\
 Q: e_3, e_4, e_5\\
\end{array}  \right\}
\xrightarrow{\actiond: \left\{ {\scriptsize\begin{array}{l} u_1 \mapsto e_4\\ u_2 \mapsto e_5\end{array}} \right\}}
\left\{\begin{array}{c}
R: e_1\\
 Q: e_3, e_5
\end{array}  \right\}
\xrightarrow{\actiond:\left\{ {\scriptsize \begin{array}{l}u_1 \mapsto e_3\\ u_2 \mapsto e_3\}\end{array}} \right\}}
\left\{\begin{array}{c}
R: e_1\\
Q: e_5\\
\end{array}  \right\}
\xrightarrow{\actiona:\epsilon}
\left\{\begin{array}{c}
p\\
R: e_1,e_9,e_{10}\\
Q: e_5,e_{11}
\end{array}  \right\}
\dots
$$
\caption{A run of Example~\ref{ex:dms-running-example1}. Recall that the schema is $\{p/0, R/1, Q/1\}$. For ease of readability, we omit the tuple notation $\tuple{e}$ and simply write $e$.}
\label{fig:run-of-example-1}
\end{figure*}

\begin{example}  \label{ex:dms-running-example1}
Consider a schema $\schema = \{ p/0, R/1, Q/1\}$, and a domain $\domain = \{ e_1, e_2, \ldots \}$.
Consider a \sys over $\schema$ and $\domain$, $\s = \tuple{\idb = \set{p}, \act = \{\actiona, \actionb, \actionc, \actiond \} }$ where
\begin{align*}
\actiona =& \tup{ \emptyset,  \set{v_1,v_2,v_3} , \true, \emptyset,  \{R(v_1), R(v_2),Q(v_3), p \} }\\
\actionb =& \tup{ \set{u},  \set{v_1,v_2} , p \wedge R(u), \{p, R(u)\},  \{Q(v_1), Q(v_2)\} }\\
\actionc =& \tup{ \set{u},  \emptyset,  p \wedge \neg Q(u), \{p, R(u)\}, \emptyset}\\
\actiond =& \langle \set{u_1,u_2},  \emptyset,  \neg p {\wedge}  Q(u_1)\,  \wedge (R(u_2) \vee Q(u_2) ),\\
& \{ Q(u_1), R(u_2)\},\emptyset \rangle
\end{align*}
A  run of the above system is depicted in
Figure~\ref{fig:run-of-example-1}. Notice that once an element is
deleted from the current database instance, it is never re-introduced, due to the history-fresh policy.
\end{example}

 \sys{s} are very expressive. The following example,
following the artifact-centric paradigm \cite{Nigam03:artifacts,DBLP:journals/debu/CohnH09,Hull2008:Artifact},
gives a glimpse about their modeling power.

\begin{example} \label{ex:agency}
Example in
Appendix~\ref{sec:example}
provides the full formalization of a \sys dealing with an agency that
advertises restaurant offers and manages the corresponding
bookings. Specifically, the process supports B2C interactions where
agents select and publish restaurant offers, while customers issue booking
requests. The process is centred around the two key business artifacts of \emph{offer} and
\emph{booking}. Intuitively, each agent can publish a dinner offer
related to some restaurant; if another, more interesting offer is
received by the agent, she puts the previous one on hold, so that it
will be picked up again later on by the same or another agent (when it will be among the most
interesting ones). Each offer can result in a corresponding booking by
a customer, or removed by the agent if nobody is interested in
it. Offers are customizable, hence each booking goes through
a preliminary phase in which the customer indicates who she wants to
bring with her to the dinner, then the agent proposes a customized
prize for the offer, and finally the customer decides whether to
accept it or not. This example is unbounded in many dimensions. On the one hand, unboundedly many offers can be advertised over time. On the other hand, unboundedly many bookings for the same offer can be created (and then canceled), and each such booking could lead to introduce unboundedly many hosts during the drafting stage of the booking.
\end{example}

We show in the following that several restrictions of the \sys model can be relaxed
without affecting its expressive power, nor compromising our technical
results. Such relaxations are essential towards capturing related
models in the literature
\cite{Bagheri2011:Artifacts,BLP:KR:12,BCDD13}, as well as concrete
specification languages like IBM GSM \cite{SMTD13}.

\noindent
\textbf{Adding constants to a \sys.}
We can extend 
 \sys and $\msodb$  to take into account a finite subset of
 distinguished \textit{constant values}
$\Delta_0\subseteq\Delta$ that can be used
to specify the content of the initial database instance $\I_0$, and
that may be explicitly mentioned in the definition of actions.
Given a \sys equipped with constant values $\Delta_0$,
we show in
Appendix~\ref{constant:less}
how to construct a
constant-free \sys over the data domain $\Delta' =
\Delta\setminus\Delta_0$, 
so that the configuration graphs of the two \syss are isomorphic.
The size of the constant-free \sys schema is exponential in 
the maximum arity of the relations.

\noindent
\textbf{Allowing Arbitrary Input Values.}
The semantics of a \sys requires the input values introduced via fresh
variables 
to not have occurred in the history of the run of the \sys.
We prove in Appendix~\ref{gen:freshness:proof} that this restriction can be lifted,
allowing for the input variables to be mapped to any possible value from the data domain.

\noindent
\textbf{Non-distinct input values.}
The semantics of the \sys requires that the fresh variables are
injectively mapped to distinct values.
We show in Appendix~\ref{gen:injectivity:proof} that this constraint is not restrictive.

\noindent
\textbf{Retrieving all answers of a query for bulk action in one step.}
We have used a \textit{retrieve-one-answer-per-step} semantics rather
than a \textit{retrieve-all-answers-per-step} semantics, which would
support the modeling of bulk operations over the database, in the
style of \cite{BCDD13}. Intuitively, in a \sys a bulk operation consists in an
action that is applied \emph{for all} the answers of its guard.

Such a bulk operation can be simulated by the iterative,
non-interruptible application of different standard actions, using
special accessory relation to control their execution. In summary,
this is done in three phases. In the first phase, the external
parameters of the bulk operation are inserted into a dedicated
\emph{input relation}, so as to maintain them fixed throughout the other two
phases. At the same time, a lock proposition is set, guaranteeing that
no other action will interrupt the execution of the next two
phases. In the second phase, an ``answer accumulation'' action is
repeatedly executed, incrementally filling an accessory \emph{answer relation} with
the answers obtained from the guard of the bulk operation. This is
needed because such answers must be computed \emph{before} applying
the bulk update. The second
phase terminates when all such answers have been transferred into the
accessory relation. In the third phase, the actual bulk
update is applied in two passes, by iteratively considering each tuple in the answer
relation, first applying all deletions, and then all additions.
When the third phase terminates, the lock is unset, enabling
the possibility of applying other actions.
Full details of this construction are given in Appendix~\ref{app:bulk}.


\section{MSO logic for \sys: \Large{\msodb}}

We propose a powerful logical formalism to reason about the linear runs of a \sys. The formalism, called \msodb, combines full monadic second-order logic to reason about the linear-time properties of runs, with atomic formulae consisting of $\folang$ queries, which are used to reason about the content of the encountered database instances.

We use $\fovars =\{x, y,  x_1, \dotsc \}$ to denote first-order position variables, $\sovars = \{ X, X_1, \ldots \}$  to denote second-order position variables and $\datavars = \{ u, v, u_1, \ldots \}$ to denote first-order data variables.
We let $\vars = \fovars \uplus \sovars \uplus \datavars$. 

\smallskip
\noindent
\textbf{Syntax. } Formulae $\forun{\phi}$ of $\msodb$ over schema $\schema$
are  given by the following syntax:
\begin{eqnarray*}
\forun{\phi}
 ::= 
\fodb{Q}@x 
\,|\,  
\forun{x\!<\!y}
 \,|\,
 \forun{ x\!\in\!X} 
 \,|\,
  \neg \forun{\phi}
 \,|\,
  \forun{\phi} \wedge \forun{\phi} 
 \,|\,
 \exists x. \forun{\phi}  
 \,|\,
   \exists X. \forun{\phi}
 \,|\,
 \existsg u.  \forun{\phi} 
\end{eqnarray*}
where   $x, y$ are first-order position variables, $X$ is a
second-order position variable,  $u$ is a first-order data variable,
and $\fodb{Q}$ is a $\folang$ query. 
We write $\forun{\forallg u. \phi } $ to denote $  \forun{ \neg \existsg u. \neg \phi} $.
Further we  make
use of standard abbreviations: 
$\forun{\forall x. \phi } \equiv  \forun{ \neg \exists x. \neg \phi} $, $\forun{\forall X. \phi } \equiv  \forun{ \neg \exists X. \neg \phi} $, etc.

The set of free variables of a formula $\forun{\phi}$  is denoted $\freevars{\forun{\phi}}$. For a set $V \subseteq \vars$, a substitution  $\sigma$ of $V$ is a mapping that maps every first-order position variable to a natural number (i.e., $\sigma |_ {\fovars} : \fovars \cap V \rightarrow \Nat$), every second-order position variable to  a subset of natural numbers (i.e., $\sigma |_ {\sovars} : \sovars \cap V \rightarrow 2^\Nat$) and every data variable  to an element from the domain $\domain$ (i.e., $\sigma |_ {\datavars} : \datavars \cap V \rightarrow \domain$).

\smallskip
\noindent
\textbf{Semantics. }
 A run $\rho$ is an infinite sequence of database instances over $\schema$ and $\domain$: $\rho = \I_0 , \I_1, \I_2, \I_3 \dots$\\
The \emph{global active domain} of the run $\rho$, denoted $\gadom{\rho}$ is the union of all active domains along the run. $\gadom{\rho} = \bigcup_{i \ge 0} \adom{\I_i}$.
An \msodb formula $\forun{\phi}$ is evaluated over an infinite run $\rho = \I_0 , \I_1, \I_2, \I_3 \dots$ under a substitution $\sigma$ of   $\freevars{\forun{\phi}}$.
If the formula holds in the run $\rho$ under the substitution $\sigma$, we write $\rho, \sigma \models \forun{\phi}$.
The semantics is as expected for the standard cases (see Appendix~\ref{app:mso-semantics}). For the particular cases, we have:
\noindent\begin{description}
\item $\rho, \sigma \models \forun{\fodb{Q}@x}$ if 
$\I_i, \sigma' \models \fodb Q$, 
$i = \sigma(x)$
and $\sigma' = \sigma|_{\freevars{\fodb{Q}}}$
\item $\rho, \sigma \models {\existsg u. \forun{\phi}  }$  if there exists $e \in \gadom{\rho}$, such that 
 $\rho, \sigma' \models \forun{ \phi}$,
 where $\gadom{\rho} = \bigcup_{i \ge 0} \adom{\I_i}$ and
 $\sigma'(u) = e\text{, and }$
 $\sigma'(\xi) = \sigma(\xi)\text{ if }\xi \neq u$.  
\end{description}

When the formula  $\forun{\phi}$ is a sentence (i.e, $\freevars{ \forun{\phi}} = \emptyset$), it can be interpreted on a run $\rho$ under the empty substitution, denoted $\rho \models \forun{\phi}$.

\begin{example}\label{ex:runset-formula} Consider the set $\runset{\s}$ of all runs of a \sys $\s = \tup{ \idb, \act}$. This set is \msodb definable by a formula $\phiruns$.  The formula uses set variable $X_\action$ to denote the set of positions where an $\action$ action was taken. It can be easily expressed in \msodb that the sets $(X_\action)_{\action \in \act}$ form a partition of $\Nat$. Further, we need to express the local consistency. For this, we need to say the following:
$\forall x \bigwedge_{\action \in \act} \left(x \in X_\action \Rightarrow  \varphi_\action(x)\right)$ where  $\varphi_\action(x)$ expresses the local consistency by action $\action$. 
If $\action =  \tup{\vec{u}, \vec{v}, \guard,\del,\add}$, then
$\varphi_\action(x)$ can be expressed as follows, where 
variables $\xi_i \in \vec{u} \uplus \vec{v}$, 
\begin{eqnarray*}
\existsg \vec{u}, \vec{v}.  \bigwedge_{u \in \vec u } \Active(u)@x \wedge \bigwedge_{v \in \vec v } \left(\forall y. \, y \le x \Rightarrow \neg \Active(v)@y\right) \wedge \fodb{Q}@x \\ \wedge 
\exists y. \successor(x,y) \wedge \!\!\! \bigwedge_{R/\arity \in \schema} \!\!\!
\left(\!\!\!\begin{tabular}{l}$\wedge_{\tup{\xi_1 \ldots \xi_\arity} \in R^\add}   R(\xi_1 \ldots \xi_\arity)@(y) \wedge$ \\
$\wedge_{\tup{\xi_1 \ldots \xi_\arity} \in R^\del \setminus R^\add} \neg    R(\xi_1 \ldots \xi_\arity)@(y) $
\end{tabular}\!\!\!\!\right)
\end{eqnarray*}
In the above, $\successor(x,y)$ states that $y$ is the successor position of $x$, which can be easily expressed in MSO.
\end{example}

\begin{example}Many standard verification problems on \sys can be expressed in \msodb since we can characterise the runs of a \sys (cf. Example~\ref{ex:runset-formula}). Of particular interest is the simplest verification problem: propositional reachability. Given a \sys $s$ over $\schema$ and $\domain$ and a proposition $p/0 \in \schema$, is it possible that an execution of $s$ ever reaches a database instance $\I$ with  $p^\I = \set{p}$ ?
This can be reduced to the satisfiability checking of $\exists \rho. \rho \models \left(\phiruns \wedge \exists x. p @x\right)$.
\end{example}

\noindent\textbf{Model checking. }
We now present the model checking problem of a \sys against MSO(\sys):
\noindent
{
\centering
\begin{tabular}{|ll|}
\hline
\textsc{Problem:} &\textsc{\texttt{MSO/DMS}-MC}\\
\textsc{Input:} & A \sys $\s$,  a \msodb   formula $\forun{\phi}$.\\
\textsc{Question:} & Does $\rho \models \forun{\phi}$, for every  $\rho \in \runset{\s}$?\\
\hline
\end{tabular}
}
\medskip

The next example shows how \textsc{\texttt{MSO/DMS}-MC}  can be
phrased in such a way that database
constraints are incorporated in the analysis of the \sys of interest.

\begin{example}
The presence of database constraints in the dynamic system under study
is a key feature, which has been extensively studied in the literature
\cite{DeSV07,DHPV09,Damaggio2011:Artifact,BoST13,BCDD13}. In our
setting, arbitrary FO constraints can be seamlessly added, adopting
the semantics, as in \cite{BCDD13}, that the
application of an action is blocked whenever the resulting database
instance violates one of the constraints. Given a \sys
$\s$, an $\msodb$ formula $\phi $ and a constraint specification on
the database instances as a $\folang$ sentence $\phi_c$, we can reduce
the model checking problem of the constrained \sys against $\phi$ to
an unconstrained model checking problem over $\s$, using as formula:
$(\forall x. \phi_c @ x) \Rightarrow \phi$. 
\end{example}

\begin{theorem}
\textsc{\texttt{MSO/DMS}-MC} is undecidable. 
\end{theorem}

We prove the above theorem by showing the undecidability of propositional reachability.
The negation of the propositional reachability itself can be reduced to the model checking problem,
by giving the input $\s$ and $\forall x. \neg p @x$ for the latter.
The proofs are conducted through a reduction from the reachability problem of a two counter Minsky machine
and can be found in Appendix~\ref{reach:undecidable:proofs}.
In particular, we show that propositional reachability is undecidable as soon as the $\sys$ has one of the following:
i) a binary predicate in $\schema$ even though the guards are only union of conjunctive queries ($\clang{UCQ}$),
ii) two unary predicates in $\schema$ and the guards allow $\clang{FOL}$.

\section{Recency-boundedness}

As mentioned in the previous section, 
even propositional reachability
is undecidable unless  the relational schema of the database is severely restricted. This motivates the study of under-approximate analysis of the \sys. 
We propose an under-approximation that is parametrised (by an integer $\uabound$) and is exhaustive. That is,  more behaviours are captured (in other words, more runs can be analysed) with higher values of $\uabound$, and in the limit it captures all finite behaviours of the \sys. The under-approximate analysis works over arbitrary (unrestricted) schema. 
Our under approximation is called recency boundedness. 

\textbf{$\uabound$-restricted actions.} In a recency bounded $\sys$  the actions are restricted  to act only on the $\uabound$ most  recent elements in the database. The guards can query the entire database, but the data values that can be retrieved as the result of a query will be only from the $\uabound$ recent elements of the database instance. Thus the deletions cannot involve  less recent elements. The newly added data values cannot participate in a relation with less recent elements either. This restriction still allows the transitions to reason about all elements in the current database instance. 
For example, the properties that all elements must satisfy (regardless of their recency), may be stated as a clause in the  guard of an action. However, all elements cannot be \emph{acted} on i.e.\ they cannot be deleted, nor new facts involving them can be added. 

The most recent $\uabound$ elements are taken relatively to the current database instance. Thus it is possible that an old element which is not in the $\uabound$-recency window eventually enters the $\uabound$-recency window. This happens if more recent elements were deleted from the current database instance, exposing the concerned element. 

\textbf{Sequence numbers.} In order to reason about recency, we assume that every element $e$ gets a sequence number $\seqno{e}$ when it is added to the database. An element which is added later/more recently gets a higher sequence number. 
If there are multiple fresh elements that are added in one action, these elements are given different and unique sequence numbers in the order in which they appear. Thus, these fresh elements are ordered amongst themselves, and their sequence number is higher than any other sequence number present in the current active domain. Since we have a countably infinite supply of sequence numbers,  we do not reuse sequence numbers. That means, even if an element is deleted from the database, its sequence number will not be used by a later element.  The sequence numbers may be also thought of as a way of (abstractly) time-stamping elements as  they enter the active domain. 

\textbf{$\brecent $.} Given a database instance $\I$ and a sequence-numbering $\seqnof: \adom{\I} \to \Nat$, we define the \emph{$\uabound$-recent active domain} of $\I$ wrt.\ $\seqnof$, denoted $\brecentof{\I}{\seqnof}$, to be  the \emph{maximal} set $D \subseteq \ADOM{\I}$ with $| D | \le \uabound$, such that for every (recent) element  $e' \in D$ and every (non-recent) element $ e \in \ADOM{\I} \setminus D$, we have $\seqno{e} < \seqno{e'}$. That is, the set $\brecentof{\I}{\seqnof}$ contains the $\uabound$ most-recent elements from $\ADOM{\I}$ according to the sequence numbering $\seqnof$. Notice that, thanks to maximality,  $| \brecentof{\I}{\seqnof} | < \uabound$ if, only if $| \ADOM{\I} | < \uabound$.

We are now ready to formally define the $\uabound$-bounded execution semantics for {\sys}s. 

\textbf{The $\uabound$-bounded configuration graph}  $\confGraph_\s^\uabound$ of a \sys $\s$ is given as follows. 
A configuration is a tuple $\langle \I, \history, \seqnof  \rangle$ where $\seqnof :  \history  \rightarrow \Nat$ is an injective function assigning sequence numbers to the data values  in the history-set.
For an action $\action =  \tup{\vec{u}, \vec{v}, \guard,\del,\add}\in \act$ and a substitution $\sigma$ from $\vec{u} \uplus \vec{v}$ to
$\domain$, we  write $\kgotos{\uabound}{\langle \I, \history, \seqnof  \rangle}{\langle \I', \history', \seqnof'  \rangle}{\action:\sigma }$ if  
\begin{enumerate}
\item $\gotos{\langle \I, \history  \rangle}{\langle \I', \history'  \rangle}{\action:\sigma}$ in $\confGraph_\s$.
\item $\sigma(u) \in \brecentof{\I}{\seqnof}$ for each $u \in \vec{u}$.
(That is, the values retrieved by the query must be among the $b$-most recent elements of the current database instance \I.)
\item $\seqnof'$ is an injective map from $\history'$ to $\Nat$. It agrees with $\seqnof$ on all data values in $\history$ (note that $\history \subseteq \history'$). For each fresh-input variable $v \in \vec{v}$, $\seqnonew{\sigma (v)} > \seqnof(e)$ for all $e \in \history$. (That is, the fresh elements that are added to the database get higher sequence numbers than the elements in $\history$ since they are more recent.)
\item If $\vec{v} = \tup{v_1, \ldots, v_\maxy}$ then for every $1 \le i < j \le \maxy$, we have $\seqnonew{v_i} < \seqnonew{v_j}$. (The sequence number of the fresh elements are ordered according to their appearance in $\vec v$.)
\end{enumerate}

Notice that Item 2 is a condition on the substitutions, rather than on transitions.
Thus $\confGraph_\s^\uabound$ has fewer edges than $\confGraph_\s$.
In Item 4 what is important is that each fresh element gets a pairwise different sequence number which is higher than that of the entire history. But then, in our decidability proof, we need to guess the order between fresh elements at every step. Fixing an order beforehand simplifies the encoding later on.

A \emph{$\uabound$-bounded extended run} $\hat \rho$ of  $\s$ is an infinite sequence 
$\tup{\I_0, \history_0, \seqnof_0} \trans[\action_0 : \sigma_0]_{\uabound} \tup{\I_1, \history_1,\seqnof_1} \trans[\action_1 : \sigma_1]_{\uabound} \tup{\I_2, \history_2, \seqnof_2} \trans[\action_2 : \sigma_2]_{\uabound} \tup{\I_3, \history_3, \seqnof_3}\dots$\hfill
where $\I_0$ is the initial database instance of $\s$, $\history_0 = \emptyset$ and $\seqnof_0$ is the empty (trivial) sequence-numbering. 
The \emph{$\uabound$-bounded run} $\rho$ generated by the $\uabound$-bounded extended run $\hat \rho$ is the sequence $\I_0, \I_1, \I_2 \dots$ of database instances appearing along $\hat \rho$.  The \emph{set of all  $\uabound$-bounded runs} of a \sys $\s$ is denoted $\brunset{\s}$.

\begin{example}
The run depicted in Figure~\ref{fig:run-of-example-1} is a 2-recency-bounded run.
\end{example}
\begin{example}\label{ex:lifo}
Consider the restaurant booking agency example sketched in
Section~\ref{sec:dms-definition} and detailed in
Appendix~\ref{sec:example}
.
Since the agency has a fixed
number of agents/customers, this number indirectly witnesses also how many
booking offers can be simultaneously managed by the
company. Suppose now that the company works with the following
strategy: an agent temporarily freezes the management of an offer
because a more interesting (in terms of potential revenue and/or
expiration time) offer is received. 
Furthermore, let us assume that once a booking is closed, it is
stored in the database for historical/audit reasons, but never modified in
the future courses of execution.

The DMS capturing this example can
consequently query the entire (unbounded) logged history of bookings so as, e.g., to check whether a
customer finalized at least a given number of bookings in the
past. This query can be used to characterize when a
customer is gold and, in turn, to tune the actual DMS behavior
depending on this. Furthermore, the DMS can manipulate unboundedly
many offers over time, following the ``last-in first-out'' strategy
that an offer is picked up or resumed only if the management of all
higher-priority offers has been completed, and no higher-priority
offer is received. 

If we now put a bound on the maximum number of hosts that can
be added by a customer to a booking, we can derive a number $k_{mb}$ that indicates
how many values need to be simultaneously manipulated in the worst
case so as to handle the current, highest-priority offers. This, in
turn, tells us that recency-bounded model checking of this unbounded DMS
coincides with exact model checking when the bound is $\geq k_{mb}$.
\end{example}

\textbf{Recency-bounded model checking. }
The 
problem is parametrised by a bound on  recency. 
\noindent
\begin{tabular}{|l p{6cm}|}
\hline
\textsc{Problem:} &\textsc{Recency-bounded-\texttt{MSO/DMS}-MC}\\
\textsc{Input:} &A \sys $\s$,  a \msodb  formula $\forun{\phi}$, a natural number $\uabound$,\\
\textsc{Question:} &Does $\rho \models \forun{\phi}$, for every  $\rho \in \brunset{\s}$?\\
\hline
\end{tabular}
\begin{theorem}
\textsc{Recency-bounded-\texttt{MSO/DMS}-MC} is decidable.
\end{theorem}
The proof of the above theorem is developed in the next section.

\section{Decidability of Recency-Bounded Model Checking}

We prove the decidability of recency bounded model checking problem by means of a symbolic encoding of runs. The symbolic encoding takes the form of finitely labelled nested words \cite{AlurM09}. We show that the set of all valid encodings of recency bounded runs is expressible in monadic second-order logic over nested words. We also show that the {\msodb} specification over runs can be translated syntactically to monadic second-order logic over nested words. Thus we reduce the $\uabound$-recency-bounded model checking problem to satisfiability problem of monadic second-order logic over nested words, which is decidable~\cite{AlurM09}. 

The encoding of $\uabound$-bounded runs using nested words and expressing their validity in MSO over nested words is given in Section~\ref{sec:encoding} and Section~\ref{sec:validity} respectively. The translation of \msodb specifications into MSO over nested word encodings is given in Section~\ref{sec:mso-translation}. 
First we will explain the symbolic abstraction used for the encoding in Section~\ref{sec:symbolic-abstraction} and recall nested words in Section~\ref{sec:nested-words}.

\subsection{Symbolic abstraction}\label{sec:symbolic-abstraction}
Consider a {$\uabound$-bounded run:
$\rho = \I_0, \I_1, \I_2, \I_3 \dotsc$

Each database instance $\I_i$ that appears on this run $\rho$ is potentially unbounded. For the sake of decidability we want our symbolic representation to be a word over finite alphabet. 
Towards this we will first consider a $\uabound$-bounded extended run $\hat{\rho} = \tup{\I_0, \history_0, \seqnof_0} \trans[\action_0 : \sigma_0]_{\uabound} \tup{\I_1, \history_1,\seqnof_1} \trans[\action_1 : \sigma_1]_{\uabound} \tup{\I_2, \history_2, \seqnof_2} \trans[\action_2 : \sigma_2]_{\uabound} \tup{\I_3, \history_3, \seqnof_3}\dots$\hfill  generating $\rho$, and the sequence of $\langle\text{action \textbf{:} substitution}\rangle$ pairs appearing along $\hat{\rho}$. 

A sequence $\generator = \langle \action_0: \sigma_0 \rangle \langle \action_1: \sigma_1 \rangle \langle \action_2: \sigma_2 \rangle \dots $ of $\langle\text{action \textbf{:} substitution}\rangle$ pairs generates a  unique $\hat{\rho}$ (if it exists) by following the semantics. However  $\generator$ is also not finitely labelled. The substitutions $\sigma_i$ maps variables to domain $\domain$, leaving the set of all such substitutions an infinite set. 

Hence we go for the recency-indexing abstraction of  a substitution. The recency-indexing abstraction, instead of mapping a variable to an element $e$, maps it to its relative recency in the current database. 
The recency-indexing  abstraction $s$ of a substitution $\sigma$ is determined by the current sequence-numbering. We explain this below. 

Consider a $\uabound$-bounded extended run $\hat \rho =$
$$  \left(\kgotos{\uabound}{\tup{\I_j,\history_j, \seqnof_j}}{\tup{\I_{j+1},\history_{j+1}, \seqnof_{j+1}}}{\action_j:\sigma_j} \right)_{j \ge 0} $$
For each substitution $\sigma_j: \vec{u}_j \uplus \vec{v}_j \to \domain$ appearing in $\hat{\rho}$, where $\vec{u}_j = \getfree{\action_j}$ and $\vec{v}_j = \getnew{\action_j}$,  we have $\sigma_j (u) \in \brecent(\I_j, \seqnof_j)$ for all $u \in \vec{u_j}$ thanks to the $\uabound$-boundedness. 

The recency-indexing abstraction of $\sigma_j$ at $\I_j$ wrt. the sequence numbering $\seqnof_j$ is a mapping $s_j : \vec{u}_j \uplus \vec{v}_j \to \set{-n, -n+1, \dots, 0, 1, \ldots \uabound-1}$ where $n = | \vec{v_j}|$ such that
\begin{itemize}
\item[r1.] if $\vec{v}_j = \tup{v_1, \ldots, v_n}$ then $s_j(v_i) = -i$
\item[r2.] for $u \in \vec{u}_j$, $s_j(u) \in \set{0, 1, \ldots, \uabound-1}$
\item[r3.] for $u \in \vec{u}_j$, $s_j(u)$ represents the recency of $\sigma_j(u)$ at $\I$ wrt. the sequence-numbering $\seqnof_j$. More precisely, $s_j(u) = i$ if $i = |\{ e \in \adom{\I_j} \mid \seqnof_j(e) > \seqnof_j(\sigma_j(u))\}|$. For example, if $\sigma_j(u)$ is the most-recent element, then $s_j(u) =0$.
\end{itemize}

Notice that given a recency bound $\uabound$ and a $\sys$ $\s = \tup{\idb,\act}$, the set of all symbolic substitutions $s$ is finite. Let us denote this set by $\symSubs(\s, \uabound)$. Let 
$\symSubs(\action, \uabound) = \{ s: \vec{u} \uplus \vec v \to \set{-n, -n+1, \ldots, 0, 1, \ldots, \uabound -1} \mid \vec{u} = \getfree{\action}, \vec v = \getnew{\action}, n = | \vec{v} | \text{ and $s$ satisfies conditions r1 and r2 above} \} $. 
We have $\symSubs(\s, \uabound) = \uplus_{\action \in \act} \symSubs(\action, \uabound)$.
Let the symbolic alphabet $\symAlph{\s}{\uabound}$ be the finite set $\{ \langle \action, s \rangle \mid \action \in \act \text{ and } s \in \symSubs(\action, \uabound)\}$.
\medskip

To every $\uabound$-bounded extended run $\hat \rho$, we can identify a corresponding word $w_{\hat \rho} \in (\symAlph{\s}{\uabound})^\omega$ by taking the recency-indexing abstraction of the substitutions. Let's denote this correspondence by a mapping $\abstraction$ from $\uabound$-bounded extended runs to $ (\symAlph{\s}{\uabound})^\omega$. That is  $\abstraction(\hat \rho) = w_{\hat \rho}$. We extend in the natural way the definition of the abstraction function to finite prefixes of $\uabound$-bounded extended runs.

The mapping $\abstraction$ is not injective.  However, if two $\uabound$-bounded extended runs $\hat \rho$ and $\hat \rho'$ have the same abstract generating sequence $w = \abstraction(\hat \rho) = \abstraction(\hat \rho')$, then $\hat \rho$ and $\hat \rho'$ are equivalent modulo permutations of the data domain $\domain$
(i.e. there exists a bijection $\lambda: \gadom{\hat \rho} \rightarrow \gadom{\hat \rho'}$ such that
$\lambda$ is an isomorphism from $\I_i$ onto $\I'_i$ for every $i\ge0$, see Appendix~\ref{claim:permutation:proof} for the detailed proof).
This notion of invariance under renaming is very well known in computer science and is discussed in \cite{STTT2016:Montali}.
If we assume a total ordering on the domain $\domain$,
 then we can define a canonical $\hat \rho$ as the representative of all such equivalent ones. Let the data domain be $\{e_1, e_2, \ldots \}$ with the ordering $e_i < e_j$ if $i < j$. A $\uabound$-bounded extended run $\hat \rho$  is canonical if it satisfies the following  invariants along the run:
\begin{itemize}
\item For every $i$, for every $j$, if  $e_j \in \history_i$ then $\seqnof_i(e_j) = j$.
\item For every $i$, if $v_k \in \getnew{\action_i}$ is the $k^{th}$ fresh input variable, then $\sigma(v_k) = e_{n+k}$ where $n = |\history_i|$. 
\end{itemize}

The second invariant implies the  following:
\begin{itemize}
\item For every $i$, $\history_i$ is of the form $\{e_1, \dots ,e_n \}$ for some $n \in \Nat$. That is, there are no gaps in the history.
\end{itemize}

The mapping $\abstraction$ is not surjective either.
We will define a partial concretizing function $\concret$ from infinite words $w \in (\symAlph{\s}{\uabound})^\omega$ to $\uabound$-bounded extended runs such that if $w$ is a valid abstraction then $\concret(w)$ is the canonical extended run $\hat \rho$ with $\abstraction(\hat \rho) =w$.
In order to do so, let us first denote the $k$-long prefix of $w$ (respectively $\hat \rho$) by $w_k$ (respectively $\hat{\rho}_k$).
Similarly to the abstraction function, we also extend the concretisation function to finite prefixes of infinite words from $(\symAlph{\s}{\uabound})^\omega$.
It is easy to see that $w$ is a valid abstract run if and only if, for every $k\ge 0$, $w_k$ is the prefix of a valid abstract run.
In that case, $\hat \rho = \concret(w)$ amounts to the limit of $\concret(w_k) = \hat \rho_k$ when $k \rightarrow +\infty$.
Also, for $k\ge0$, if $w_k$ is the prefix of a valid run, then $\hat \rho_k = \concret(w_k)$ is of the form
$\left(\kgotos{\uabound}{\tup{\I_j,\history_j, \seqnof_j}}{\tup{\I_{j+1},\history_{j+1}, \seqnof_{j+1}}}{\action_j:\sigma_j} \right)_{ 0 \le j  <k } $.
We define in what follows $\concret(w_k)$ by induction on its length $k$.

For the empty word $w_0 = \epsilon \in (\symAlph{\s}{\uabound})^\ast$ we define $\concret(w_0) =\tup {\idb, \history_0, \seqnof_0}$
where $\history_0 = \emptyset $ and $\seqnof_0 = \epsilon$, the empty mapping.
Suppose $w_{k+1}= w_{k}\langle \action_k, s \rangle$ where  $\action_k = \tup { \vec u, \vec v, \guard, \del, \add }$.
$\concret(w_{k+1})$ is not defined if $\concret(w_{k})$ is not defined.
Suppose $\concret(w_{k})$ is defined and is of the form
$\left(\kgotos{\uabound}{\tup{\I_j,\history_j, \seqnof_j}}{\tup{\I_{j+1},\history_{j+1}, \seqnof_{j+1}}}{\action_j:\sigma_j} \right)_{ 0 \le j  <k }$.
$\concret(k+1)$ is defined if, and  only if, the following condition holds:

\medskip
\textbf{Condition} \textsf{Cnd} \textbf{:}
There exists a substitution $\sigma: \vec u \to \adom{\I_{k}}$ such that
\begin{itemize}
\item $\I_{k}, \sigma \models \guard$ and
\item $s$ restricted to $\vec u $ is the recency-indexing abstraction of $\sigma$ at $\I_{k}$ wrt.\ $\seqnof_{k}$.
\end{itemize}

\medskip
Assuming condition \textsf{Cnd} holds, $w_{k+1}$ is also the prefix of a valid abstract run
and $\concret(k+1)$ is defined as follows.
Let $n$ be the size of $\history_k$, i.e. $n = |\history_k|$.
We define the substitution $\sigma_k: \vec{u}\uplus\vec{v} \rightarrow \Delta$ as follows:
$\sigma_k|_{\vec{u}} = \sigma$ and $\sigma_k(v_i) = e_{n+i}$ for every $i: 1\le i \le |\vec{v}|$.
Notice that 1) $\sigma_k(u)=\sigma(u)\in\adom{\I_k}$ for every $u\in\vec{u}$,
2) $\sigma_k(v)\notin\history_k$ for every $v\in\vec{v}$,
3) $\sigma_k|_{\vec{v}}$ is injective,
and 4) since $\I_k, \sigma \models \guard$,
and $\sigma_k|_{\vec{u}} = \sigma$, we have that $\I_k, \sigma_k|_{\vec{u}} \models \guard$.
Thus, $\sigma_k$ is an instantiating substitution for $\action_k$ at $\tup{\I_{k},\history_{k}}$.
Moreover, if we define the set $\history_{k+1} = \history_k \cup \set{e_{n+1},\ldots,e_{n+|\vec{v}|}}$
and the database instance $\I_{k+1} = (\I_k - \substDB{\sigma_k}{\del}) + \substDB{\sigma_k}{\add}$,
then we have that
$\gotos{\tup{\I_{k},\history_{k}}}{\tup{\I_{k+1},\history_{k+1}}}{\action_k,\sigma_k}$.

Furthermore, since the restriction of $s$ to $\vec u$ is the recency-indexing abstraction of $\sigma$ at $\I_k$ wrt.\ $\seqnof_k$,
we deduce that $\sigma_k(u) = \sigma(u)\in\brecentof{\I_k}{\seqnof_k}$.
Thus, the transition $\gotos{\tup{\I_{k},\history_{k}}}{\tup{\I_{k+1},\history_{k+1}}}{\action_k,\sigma_k}$
is also allowed by the $\uabound$-recency semantics and,
assuming that we define $\seqnof_{k+1}$ by
$\seqnof_{k+1}|_{\history_k} = \seqnof_{k}$ and by $\seqnof_{k+1}(e_{n+i}) = n+i$ for every $i:1\le i \le |\vec{v}|$,
we have that $\kgotos{\uabound}{\tup{\I_{k},\history_{k},\seqnof_{k}}}{\tup{\I_{k+1},\history_{k+1},\seqnof_{k+1}}}{\action_k,\sigma_k}$
and $\concret(k+1) = \left(\kgotos{\uabound}{\tup{\I_j,\history_j, \seqnof_j}}{\tup{\I_{j+1},\history_{j+1}, \seqnof_{j+1}}}{\action_j:\sigma_j} \right)_{ 0 \le j  < k+1 }$.  

\medskip

Now, for an infinite word $w \in (\symAlph{\s}{\uabound})^\omega$,  $\concret(w)$ is defined to be the limit of $\hat \rho_k = \concret(w_k)$ for $k\ge 0$.
If   defined, 
 $\concret(w)$ is a canonical run. }Further, $\abstraction(\concret(w)) = w$. Furthermore, for every $w$ such that $w = \abstraction(\hat \rho)$  for $\uabound$-bounded  run $\hat \rho$, $\concret(w)$ is defined, and $\concret(w)$ and $\hat \rho$ are equivalent modulo permutations of the data domain. In particular,  if $\hat \rho$ is a $\uabound$-bounded canonical  run, then $\hat \rho = \concret(\abstraction(\hat \rho))$.

\begin{example}
The abstract generation sequence corresponding to the run in Figure~\ref{fig:run-of-example-1} is:\\
$
\langle
	\actiona: \set{v_1 \mapsto -1, v_2 \mapsto -2, v_3 \mapsto - 3}
\rangle ~
\langle
	\actionb: \set{u\mapsto 1, v_1 \mapsto -1, v_2 \mapsto -2 }
\rangle
\langle
	\actiona: \set{v_1 \mapsto -1, v_2 \mapsto -2, v_3 \mapsto - 3}
\rangle ~
\langle
	\actionc: \set{u\mapsto 1}
\rangle ~
\langle
	\actiond: \set{u_1\mapsto 0, u_2\mapsto 1}
\rangle ~
\langle
	\actiond: \set{u_1\mapsto 1, u_2\mapsto 0}
\rangle ~
\langle
	\actiond: \set{u_1\mapsto 1, u_2\mapsto 1}
\rangle ~
\langle
	\actiona: \set{v_1 \mapsto -1, v_2 \mapsto -2, v_3 \mapsto - 3}
\rangle
\dots
$
\end{example}

In order to check the consistency of an abstract generating sequence,
we need to check that condition \textsf{Cnd} holds at every step of the sequence.
To achieve this within a formalism having ``decidable theories'', we add more structure to the abstract generating sequence
by embedding it into a nested word, which we recall in the next section.

\subsection{Nested words}\label{sec:nested-words}

A \emph{visible alphabet} $\Anw$ is a finite alphabet partitioned into push letters $\Apush$, pop letters $\Apop$ and internal letters $\Aint$. That is, $\Anw = \Apush \uplus \Apop \uplus \Aint$.  Given a word $w = a_1a_2\ldots$ over the visible alphabet $\Anw$, we say $i$ is a \emph{$\Apush$ position} if $a_i \in \Apush$. Similarly we define $\Apop$ positions and $\Aint$ positions. 

A nested word is a pair $(w, \nestrel)$ where $w$ is a word over a $\Anw$ and ${\nestrel} \subset {\set{1,\ldots, |w|}^2}$ is the \emph{maximal} binary nesting relation  relating $\Apush$ positions to $\Apop$ positions such that:
\begin{itemize}
\item if $i \nestrel j$ then $i < j$. The nesting relation preserves the linear order.
\item if $i \nestrel j $ and $i' \nestrel j'$ are two distinct pairs (either $i \neq i'$ or $j \neq j'$) then $|\set{ i, i', j, j' }| = 4$. Two different nesting edges are vertex-disjoint.
\item for every $i \nestrel j$ and $i ' \nestrel j'$ we do not have $i < i' < j < j'$. The nesting edges must not cross. 
\item if $i \nestrel j$ and $i < i' < j$ for some $\Apush$  position $i'$, then there exists $j'$ such that $i' \nestrel j'$. Similarly if $i \nestrel j$ and $i < j' < j$ for some $\Apop$  position $j'$, then there exists $i'$ such that $i' \nestrel j'$.
\end{itemize}

\begin{example} \label{ex:nested-word}
A nested word over the visible alphabet given by $\Apush = \{\push a , \push b\}$ ,  $\Apop = \{\pop a , \pop b\}$  and $\Aint = \{\bullet \}$ is given below:

\smallskip
\noindent
 \begin{tikzpicture}[->,>=stealth',shorten >=1pt,auto,node distance=.8cm,
  thick,main node/.style={circle, inner sep = 0pt, draw=blue!50,fill=blue!5,minimum size=12pt,font=\sffamily\small}]

  \node[main node, label =90:1] (1) {$\push a$};
  \node[main node, label =90:2] (2) [right of =1] {$\push a$};
  \node[main node, label =90:3] (3) [right of =2]  {$\pop a$};
  \node[main node, label =90:4] (4) [right of =3] {$\push b$};
    \node[main node, label =90:5] (5) [right of =4]  {$\push a$};
  \node[main node, label =90:6] (6) [right of =5] {$\pop b$};
  \node[main node, label =90:7] (7) [right of =6] {$\bullet$};
  \node[main node, label =90:8] (8) [right of =7]  {$\pop b$};
  \node[main node, label =90:9] (9) [right of =8] {$\push b$};
    \node[main node, label =90:10] (10) [right of =9]  {$\push a$};
  \node[main node, label =90:11] (11) [right of =10] {$\pop a$};

\draw[->,rounded corners=0.01cm,] (2) -- ++(0,-0.5) -- ++(.8, 0) --(3);
\draw[->,rounded corners=0.01cm,shorten >=1pt, ] (5) -- ++(0,-0.5) -- ++(.8, 0) --(6);
\draw[->,rounded corners=0.01cm,shorten >=1pt, ] (10) -- ++(0,-0.5) -- ++(.8, 0) --(11);
\draw[->,rounded corners=0.01cm,shorten >=2pt, ] (4) -- ++(0,-0.7) -- ++(3.2, 0) --(8);
\draw[->,rounded corners=0.01cm,shorten >=2pt, ] (1) -- ++(0,-.9) -- ++(8.2, 0);
\draw[->,rounded corners=0.01cm,shorten >=2pt, ] (9) -- ++(0,-0.7) -- ++(1.8, 0);

\end{tikzpicture}

\end{example}

Note that, given a word $w$ over a visible alphabet $\Anw$, the nesting relation $\nestrel$ is uniquely defined.

\textbf{Monadic Second-Order Logic (\msonw) over nested words} extends MSO over words with an additional binary predicate $\nestrel$ that links a matching push-pop pair.  This is in fact the same logic of \cite{LaST94} where the guessed second-order matching variable is built-in in the structure. We assume an unbounded supply of position variables $\{x, y,  \ldots\}$ and set variables $\{X, Y, \ldots \}$. The syntax of 
$\msonw$  
 is given by:
$$\varphi := a(x) \mid x < y \mid x \nestrel y \mid \neg \varphi \mid \varphi \vee \varphi \mid \exists x. \varphi \mid \exists X. \varphi$$
Here $a$ ranges over the visible  alphabet $\Anw$. The position  variables $x, y$ range over positions of the nested word. The set variable $X$ ranges over sets of positions of the nested word.  The semantics is as expected.

\begin{example} Let $x$ and $y$ be two free first-order variables. Suppose we want to state that the first $\push a $ labelled position after $x$ and the first $\pop b $ labelled position after $y$ are related by a nesting edge. This property can be stated by a formula with two free variables:
\begin{eqnarray*}
\stepex_{a,b}(x,y)  \equiv  
\exists x_1 \exists y_1 \, {\push a} (x_1) \wedge {\pop b} (y_1) \wedge  x < x_1 \wedge  y < y_1  \wedge x_1 \nestrel y_1  \\
 \wedge \forall z\,  (x < z < x_1 \Rightarrow \neg {\push a} (z)) \wedge ( y < z < y_1  \Rightarrow \neg {\pop b} (z))
\end{eqnarray*}
On  Example~\ref{ex:nested-word}, all pairs of positions $(i,j)$ with $2 \le i \le 4$ and $1 \le j \le 5$ satisfy the above formula. 
\end{example}

\noindent \begin{fact}[\cite{AlurM09}]
Satisfiability of {\msonw} is decidable. 
\end{fact}

\subsection{Encoding a  run as a nested word}\label{sec:encoding}

Let us fix a \sys $\s = \tup{\idb,\act}$, over a set of values $\Delta$ and a schema $\schema$, and a recency bound $\uabound$ for the rest of this section. We will first provide the visible alphabet, and then describe the encoding.
 
 \medskip
 
 \textbf{Visible alphabet of the encoding.} The visible alphabet $\Anw = \Aint \uplus \Apop \uplus \Apush$ where

 \begin{itemize}
  \item 
$\Aint = \{ \frameletter{\action:s} \mid  \langle \action, s \rangle  \in  \symAlph{\s}{ \uabound}  \}  \cup \{ \frameletter{\idb} \} $
 
   \item $\Apop = \{ \pop 0, \dotsc ,\pop {\uabound-1}\}$
  \item $\Apush = \{ \push{ - \maxy}, \dotsc, \push 0, \dotsc, \push {\uabound-1}   \}$ where $\maxy = \max_{\action \in \act} {|\getnew{\action}|}$
 \end{itemize}

 The internal letters represent the symbolic abstraction described in Section~\ref{sec:symbolic-abstraction}. Further,  we provide a letter $\frameletter{\idb}$ to represent the initial database $\idb$.

The pop letters and push letters as well as the nesting relation will be used to trace the elements (or datavalues) in an encoding. We explain this more in detail when describing the encoding.

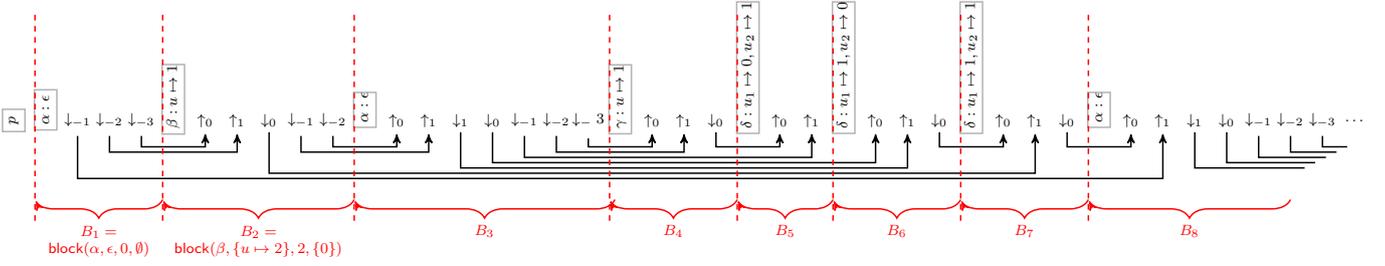
\begin{figure*}
\scalebox{.7}{
 \begin{tikzpicture}[->,>=stealth',shorten >=1pt,auto,node distance=.6cm,
  thick,main node/.style={circle, inner sep = 0pt, minimum size=12pt,font=\sffamily\small}]
\node[main node] (i0) {};
  \node[main node] (1) [right of = i0]{};
  \node[main node] (2) [right of =1] {$\push {-1}$};
  \node[main node] (3) [right of =2]  {$\push {-2}$};
  \node[main node] (4) [right of =3] {$\push {-3}$};
   \node[main node] (5) [right of =4]  {};
  \node[main node] (6) [right of =5] {$\pop 0$};
  \node[main node] (7) [right of =6] {$\pop 1$};
  \node[main node] (8) [right of =7]  {$\push 0$};
  \node[main node] (9) [right of =8] {$\push {-1}$};
    \node[main node] (10) [right of =9]  {$\push {-2}$};
  \node[main node] (11) [right of =10] {};
    \node[main node] (12) [right of =11] {$\pop 0$};
  \node[main node] (13) [right of =12] {$\pop 1$};
  \node[main node] (14) [right of =13]  {$\push 1$};
  \node[main node] (15) [right of =14] {$\push 0$};
    \node[main node] (16) [right of =15]  {$\push {-1}$};
      \node[main node] (17) [right of =16]  {$\push {-2}$};
  \node[main node] (18) [right of =17] {$\push -3$};
      \node[main node] (19) [right of =18]  {};
  \node[main node] (20) [right of =19] {$\pop 0$};
  \node[main node] (21) [right of =20] {$\pop 1$};
  \node[main node] (22) [right of =21]  {$\push 0$};
      \node[main node] (23) [right of =22]  {};
  \node[main node] (24) [right of =23] {$\pop 0$};
  \node[main node] (25) [right of =24] {$\pop 1$};
      \node[main node] (26) [right of =25]  {};
  \node[main node] (27) [right of =26] {$\pop 0$};
  \node[main node] (28) [right of =27] {$\pop 1$};
    \node[main node] (29) [right of =28] {$\push 0$};
      \node[main node] (30) [right of =29]  {};
  \node[main node] (31) [right of =30] {$\pop 0$};
  \node[main node] (32) [right of =31] {$\pop 1$};
      \node[main node] (33) [right of =32] {$\push 0$};

  \node[main node] (34) [right of = 33]{};
      \node[main node] (35) [right of =34] {$\pop 0$};
  \node[main node] (36) [right of =35] {$\pop 1$};
  \node[main node] (37) [right of =36]  {$\push 1$};
  \node[main node] (38) [right of =37] {$\push 0$};
  \node[main node] (39) [right of =38] {$\push {-1}$};
  \node[main node] (40) [right of =39]  {$\push {-2}$};
  \node[main node] (41) [right of =40] {$\push {-3}$};
    \node[main node] (42) [right of =41] {$ \, \dots $};

\node[main node] (i0p) at (0, 0) [ rotate = 90, draw=black!30, rectangle] { { $ p$ } };
\node[main node] (1p) at (.6, .2) [ rotate = 90, draw=black!30, rectangle] { { $ \actiona:\epsilon$ } };
\node[main node] (5p) at (3,.4) [ rotate = 90, draw=black!30, rectangle] {{ $\actionb: u \mapsto 1$ }};
\node[main node] (11p) at (6.6, .2) [rotate = 90, draw=black!30, rectangle] {{ $\actiona:\epsilon$ }};
\node[main node] (19p) at (11.4, .4) [ rotate = 90, draw=black!30, rectangle]  {{ $\actionc: u \mapsto 1$ }};
\node[main node] (23p) at (13.8, 1) [ rotate = 90, draw=black!30, rectangle] {{ $\actiond: u_1 \mapsto 0, u_2 \mapsto 1$ }};
\node[main node] (26p) at (15.6,1)[ rotate = 90, draw=black!30, rectangle]{{ $\actiond: u_1 \mapsto 1, u_2 \mapsto 0$ }};
\node[main node] (30p) at (18,1) [ rotate = 90, draw=black!30, rectangle, distance =4cm] {{ $\actiond: u_1 \mapsto 1, u_2 \mapsto 1$ }};
\node[main node] (34p) at (20.4,.2) [ rotate = 90, draw=black!30, rectangle]{{ $\actiona:\epsilon$ }};

\draw[->] (33) -- ++(0,-0.5) --++(1.2,0)--(35);

\draw[->] (29) -- ++(0,-0.5) --++(1.2,0)--(31);
\draw[->] (22) -- ++(0,-0.5) --++(1.2,0)--(24);
\draw[->] (18) -- ++(0,-0.5) --++(1.2,0)--(20);
\draw[->] (17) -- ++(0,-0.6) --++(2.4,0)--(21);
\draw[->] (16) -- ++(0,-0.7) --++(5.4,0)--(25);
\draw[->] (15) -- ++(0,-0.8) --++(7.2,0)--(27);
\draw[->] (14) -- ++(0,-0.9) --++(8.4,0)--(28);
\draw[->] (10) -- ++(0,-0.5) --++(1.2,0)--(12);
\draw[->] (9) -- ++(0,-0.6) --++(2.4,0)--(13);

\draw[->] (8) -- ++(0,-1) --++(14.4,0)--(32);
\draw[->] (4) -- ++(0,-0.5) --++(1.2,0)--(6);
\draw[->] (3) -- ++(0,-0.6) --++(2.4,0)--(7);
\draw[->] (2) -- ++(0,-1.1) --++(20.4,0)--(36);

\draw[-] (41) -- ++(0,-0.5) --++(.5,0);
\draw[-] (40) -- ++(0,-0.6) --++(.9,0);
\draw[-] (39) -- ++(0,-0.7) --++(1.3,0);
\draw[-] (38) -- ++(0,-0.8) --++(1.7,0);
\draw[-] (37) -- ++(0,-0.9) --++(2.1,0);

\draw [-, dashed, red] (.4,2) --(.4, -2);
\draw [-, dashed, red] (2.8,2) --(2.8, -2);
\draw [-, dashed, red] (6.4,2) --(6.4, -2);
\draw [-, dashed, red] (11.2,2) --(11.2, -2);
\draw [-, dashed, red] (13.6,2) --(13.6, -2);
\draw [-, dashed, red] (15.4,2) --(15.4, -2);
\draw [-, dashed, red] (17.8,2) --(17.8, -2);
\draw [-, dashed, red] (20.2,2) --(20.2, -2);

\draw [decorate,decoration={brace,amplitude=10pt},yshift=0pt, red]
(2.8,-1.5) -- (.4,-1.5) node [midway,red,yshift=-10pt] 
{\small \begin{tabular}{c} $B_1 =$\\
$\block(\actiona, \epsilon,0,\emptyset)$ \end{tabular} };

\draw [decorate,decoration={brace,amplitude=10pt},yshift=0pt, red]
(6.4,-1.5) -- (2.8,-1.5) node [midway,red,yshift=-10pt] 
{\small \begin{tabular}{c} $B_2 =$\\$\block(\actionb,\{u \mapsto 2\},2,\{0\})$\end{tabular}};

\draw [decorate,decoration={brace,amplitude=10pt},yshift=0pt, red]
(11.3,-1.5) -- (6.4,-1.5) node [midway,red,yshift=-10pt] 
{\small $B_3$};

\draw [decorate,decoration={brace,amplitude=10pt},yshift=0pt, red]
(13.6,-1.5) -- (11.2,-1.5) node [midway,red,yshift=-10pt] 
{\small $B_4$};

\draw [decorate,decoration={brace,amplitude=10pt},yshift=0pt, red]
(15.4,-1.5) -- (13.6,-1.5) node [midway,red,yshift=-10pt] 
{\small $B_5$};

\draw [decorate,decoration={brace,amplitude=10pt},yshift=0pt, red]
(17.8,-1.5) -- (15.4,-1.5) node [midway,red,yshift=-10pt] 
{\small $B_6$};

\draw [decorate,decoration={brace,amplitude=10pt},yshift=0pt, red]
(20.2,-1.5) -- (17.8,-1.5) node [midway,red,yshift=-10pt] 
{\small $B_7$};

\draw [decorate,decoration={brace,amplitude=10pt},yshift=0pt, red]
(24,-1.5) -- (20.2,-1.5) node [midway,red,yshift=-10pt] 
{\small $B_8$};

\end{tikzpicture}
}
\caption{Nested word encoding of the run in Figure~\ref{fig:run-of-example-1}}
\label{fig:2-run-nw}
\end{figure*}

\textbf{Encoding. } As alluded to in Section~\ref{sec:symbolic-abstraction}, we need to enrich the abstract generating sequences.
We go for a richer encoding where each step is followed by an encoding of the effect of the action on the database. The effect of an action involves
\begin{enumerate*}[label=\alph*)]
\item adding some relational tuples to the current database instance;
\item deleting some relational tuples from the current database instance.
\end{enumerate*}
The above two items can induce
\begin{enumerate*}[label=\arabic*)]
\item adding new elements to the current active domain.
\item  deleting some elements from the current active domain;
\end{enumerate*}

The effects a) and b) are explicitly mentioned in the action $\action$. The number of newly added fresh elements is also explicit in $\action$.
  Hence the induced effect 1) as well as effects a) and b) can be deduced from the action encoding $\frameletter{\action:s}$.

However, the induced effect 2 is not predictable from $\frameletter{\action:s}$. The reason is that, even when an element is involved only  in deletions, it is not clear whether this element can be removed from the current active domain since it may be participating in some other relations which were not tested by the action $\action$. Thanks to recency boundedness, we know that if some element is deleted then it must be from the $\uabound$ most recent elements.  

Another subtle problem is that at every configuration, the active domain need not contain $\uabound$ elements. 
Let $m$ be $\min \{\uabound , |\ADOM{\I}|\}$, which gives the cardinality of the set $\brecentof{\I}{\seqnof}$. The value of $m$ at a configuration is not defined from an action encoding $\frameletter{\action:s}$. Hence our encoding will also guess the value of $m$. Later, we will use $\msonw$ to ensure that our guesses were indeed right.

We will  provide an encoding which will ``guess''  the following: 1) the size of $\brecentof{\I,\seqnof}$ at any configuration and 2) those recent elements which are deleted from the active domain, (or equivalently, it will ``guess'' those recent elements which are surviving in the active domain).

Suppose that $|\brecentof{\I}{\seqnof}| = m$ in the current configuration $\tup{\I,\history, \seqnof} $.  Consider an action $\action$ under an abstract substitution   $s: \getfree{\action} \rightarrow \{ 0, 1, \ldots m-1\}$. Further suppose that the elements with the recency index $J = \{i_1, i_2, \ldots i_\ell\}$ with $J \subseteq \{0, 1, \ldots m-1\}$ are surviving after the action. That means, the elements with recency index in $ \{0, 1, \ldots m-1\} \setminus J$ are deleted from the current database. The action along with its effect is encoded by the following visible word, where $n = |\getnew{\action}|$:
$$ \frameletter{\action:s} \pop 0 \pop{1} \ldots \pop {m-1} \push{i_1} \ldots \push{i_\ell} \push{-1}  \dots \push{-n}$$
 with $m-1 \ge i_1 > \dotsb >  i_\ell \ge 0$. The above word is parametrised by $\action, s, m $ and $J$. We denote it by $\block(\alpha, s, m, J)$.
 
 Intuitively, we delete all the elements from $\recent{\I}$ temporarily, and insert back all the surviving ones (as dictated by $J$). Notice that the order of the indices of the elements from $J$ make sure that  in the later blocks a more-recent element is popped before a less-recent one. Finally, the fresh elements are pushed in.

 Our encoding of a $\uabound$-bounded run is a sequence of such blocks prefixed by $\frameletter{\idb}$:
 $$\frameletter{\idb}\, \block(\action_1, s_1, m_1, J_1)\, \block(\action_2, s_2, m_2, J_2)\, \dotsm $$
 The nesting edges are induced on the word due to the visibility of the alphabet. 
 Our encoding has an interesting feature: the number of unmatched pushes in the prefix upto $\frameletter{\action_j:s_j}$ is $| \ADOM{I_{j}}  |$ where $\I_{j}$ is the database instance at which $\actiona_j$ is executed. The set $\recent{\I_{j}}$ corresponds to the innermost (rightmost)  $| \recent{\I_{j}} |$ unmatched pushes in the prefix. Note that, here an unmatched push in the prefix means it is not matched within the prefix; it may be matched after the prefix.
 
 \begin{example}\label{ex:encoding-of-run}
  The nested-word encoding of the run from Figure~\ref{fig:run-of-example-1} is depicted in Figure~\ref{fig:2-run-nw}. It is 2-recency bounded. The indices 0 and 1 refer to the most recent and second most recent elements. Negative indices refer to freshly added elements.

 Notice that in this example $B_1$ is the only block where $|\recent{\I}| < \uabound =2$. On all successive blocks $|\recent{\I}| = 2$. In block $B_1$, indices 0 and 1 are not used.
 
 In block $B_2$, the substitution uses only the second-most recent element, denoted by $u \mapsto 1$. However, the entire $\recent{\I}$ is popped. Since the second-most recent element is deleted in $B_2$, it is not pushed back, but the most recent element is pushed back (denoted by $\push 0$). Hence for $B_2$, we have $J_2 = \{0\}$. 
 
 Action $\actiona$ of block $B_3$ does not use/modify any element from $\recent{\I}$. However, since $\recent{\I}$ is non-empty, it is popped entirely and pushed back. Notice the inversion in the order  of the sequence of pops and that of pushes. This inversion maintains that less-recent elements are  pushed before the more-recent elements. 
 
 Notice also that the number of pushes on the left of a block which are not matched on the left correspond to the number of elements in the active domain before the execution of the block. For example, the database instance $\I_4$ just before the execution of block $B_5$ has 6 elements, and the $\adom{\I_7}$ has just two elements. 
 
 Notice that the abstract substitution need not be injective (cf. block $B_7$), and need not assign  recent values to  variables in the order of their recency (cf. blocks $B_5$ and $B_6$). 
 
 Notice also that the set $J$ is not determined by the action name nor the  abstract  substitution $s$.
 \end{example}

\begin{figure*}
\includegraphics[width=\textwidth]{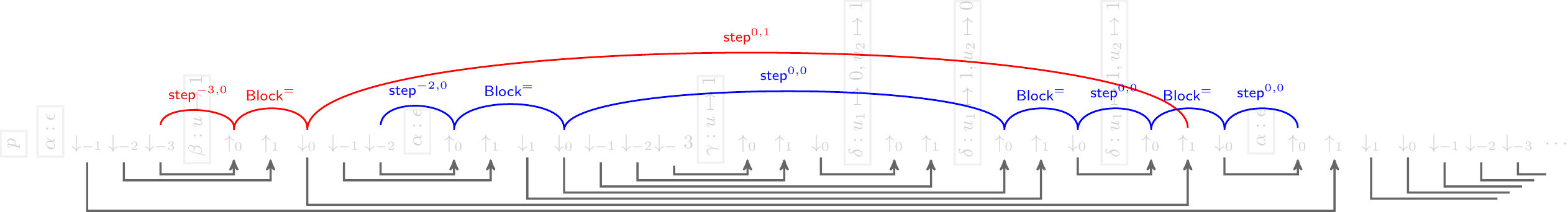}
\caption{ The relation $\Eqij{i}{j}$ tracks the occurrences of the same element in the nested word encoding of a run. In a way it is a transitive closure of the $\step^{i,j}$ relations and the $\sameblock$ relations.  Note that if such a transitive closure path enters a block via the relation $\step^{i,j}$ and exits the block via $\step^{i',j'}$ then $j = i'$.}
\label{fig:2-run-nw-equality-tracing}
\medskip
\scalebox{.95}{\framebox{$
\Eqij{i}{j}(x, y)  \equiv  \\
\forall X_{-\maxy} \forall X_{-\maxy+1} \dots  \forall X_{\uabound-1} 
\left( 
\left(
x \in X_i 
\wedge \forall x_1 \forall x_2.  
\left( \begin{array}{l}
 \bigwedge_{\maxy \le \ell, m \le  \uabound-1 }  (\step^{\ell, m}(x_1, x_2) \wedge {x_1 \in X_\ell}  ) \Rightarrow {x_2 \in X_m}  \\
 \wedge  \bigwedge_{-\maxy \le \ell\le  \uabound-1 }  (\sameblock(x_1, x_2) \wedge {x_1 \in X_\ell}  ) \Rightarrow {x_2 \in X_\ell}  \\   
\end{array} \right)
\right)\\
 \Rightarrow y \in X_j  
 \right)$}}
\caption{Formula $\Eqij{i}{j}$ which states that the element indexed by $i$ in the block of the first argument is same as the element indexed by $j$ in the block of the second argument. This is pictorially depicted in Figure~\ref{fig:2-run-nw-equality-tracing}. }\label{fig:eqij}
\end{figure*}

\subsubsection{Conditions for valid encodings} \label{sec:conditions-for-valid-encodings}

Consider any nested word $W$ over the visible alphabet $\Anw = \Aint \uplus \Apop \uplus \Apush$ of the form  $\frameletter{\idb}\, \block(\action_1, s_1, m_1, J_1)\, \block(\action_2, s_2, m_2, J_2)\, \ldots$.  Let $w \in (\symAlph{\s}{\uabound})^\omega$ be obtained by the $\Aint$ projection of $W$.  Let $W_i$ denote the prefix of $W$ upto $\block_i$, and $w_i$ be the corresponding projection.

For $i \ge 0$, we say that a prefix $W_{i+1}$ is \emph{good} if
$\concret(w_i)$ is defined.  Let $C_i = \tup {I_i, \history_i, \seqnof_i
}$ be the last configuration of $\concret(w_i)$ in this case. Further we require the following:
\begin{enumerate}
\item $m_{i+1} = |\brecentof{\I_i}{\seqnof_i}|$; 
\item  $j \in J_i $ iff,  letting $e$ be the element of recency-index $j$ in $C_i$,  there are a relation $R \in \schema$ and a tuple $t$ of $R$ involving $e$ such that $t$ is present in $\I_i$ but not in instantiated $\getdel{\action_{i+1}}$, or $t$ is present in instantiated $\getadd{\action_{i+1}}$; and 

\item letting $\sigma_{i+1}$ be the instantiation of $s_{i+1}$ at $C_i$, we have $\I_i, \sigma_{i+1} \models \getguard{\action_{i+1}}$. 
\end{enumerate}

We say $W$ is a \emph{valid encoding} of a $\uabound$-recency bounded run of $\s$ if $W_i$ is good for every $i \ge 0$.

Observe that,  if $W_i$ is good then $\concret(w_i)$ is defined. Hence, if a nested word is not a valid encoding, it can be detected at the first index $i$ such that $W_i$ is not good by observing that conditions (1), (2) or (3) is violated. In this case $\concret(w_{i-1})$ is defined since $W_{i-1}$ is good. We will exploit this observation to express valid encodings in \msonw.

In the remainder of this section we will use the above-set indexing convention for the intuitive explanations. That is, $C_i = \tup {I_i, \history_i, \seqnof_i
}$ is the last configuration of $\concret(w_i)$. This means that the previous configuration of $\block_i$ (or the configuration where it is being executed) is $C_{i-1}$.

\begin{remark}\label{rem:pending-push-size}
If $W$ is a valid encoding then, the number of unmatched pushes in the prefix upto $\block_{j+1}$ (excluding) is $| \ADOM{I_{j}}  |$. The set $\brecentof{\I_{j}}{\seqnof_j}$ corresponds to the innermost (rightmost)  $| \brecentof{\I_{j}}{\seqnof_j} |$ unmatched pushes in the prefix. Note that, here an unmatched push in the prefix means it is not matched within the prefix; it may be matched after the prefix.
\end{remark}
We will now provide $\msonw$ formulae stating that these three conditions are satisfied by a nested word over $\Anw$ at all of its blocks. The conjunction of the these formulae will characterise $\brunset{\s}$ (which we denote by $\varphi_{\uabound, \s}^\textsf{valid}$).

\subsection{Expressing valid encodings in \msonw} \label{sec:validity}
We first describe a few $\msonw$ predicates that turn out handy when stating the validity of an encoding in MSO. Such predicates are macros/abbreviation helping towards the readability of the formula describing validity.

\subsubsection{Preliminary formulae}

We write $\Aint(x)$ as a shorthand for $\bigvee_{a \in \Aint} a(x)$. Similarly we define $\Apush(x) \equiv \bigvee_{a \in \Apush} a(x)$ and $\Apop(x) \equiv \bigvee_{a \in \Apop} a(x)$. 

 We write $\sameblock(x,y)$ to indicate that positions $x$ and $y$ belong to the same block. This is a shorthand for 
$$\forall z .\, ((\neg \Aint(z)) \lor (z\le x \wedge z \le y) \vee ( x < z \wedge y < z))$$ 
Notice that a block has exactly one internal letter, which indicates the action and the abstract substitution. The position labelled by such an internal letter is called \emph{head}, and every block has a unique head.
The formula $\sameblock(x,y)$ says that $x$ and $y$ must not be separated by an internal letter (or a head).

We now define a unary predicate with a free variable $x$ for each relation name  $R/a \in \schema$ and choice of $a$ recency indices  $i_1, \dotsc i_a \in  \{0, \ldots ,\uabound-1\}$. The predicate holds at a position  if it is the head of a block and its block deletes a tuple $\tuple{ e_1, \ldots e_a}$ from the relation $R$ where $e_j$ is indexed by $i_j$ in its block, for all $j : 1 \le j \le a$. This predicate is denoted $\del(R(i_1, \ldots i_a))@ x$. 
$$\del(R(i_1, \ldots i_a))@ x \equiv \bigvee_{\frameletter{\action,s} \in \Gamma} \frameletter{\action,s}(x)$$
where $\Gamma = \{  ~\frameletter{\action,s} \mid  \getdel{\action}$ \text{ contains a tuple } $R(u_1 \ldots u_a) $\text{ and } $s(u_j) = i_j$ \text{ for all } $1 \le j \le a\}$.

Similarly we define a unary predicate for adding a tuple to a relation as well. However in this case, the indices may refer to the fresh data values as well. Hence we have unary predicate $\add(R(i_1, \ldots i_a))@ x$ for each 
relation name  $R/a \in \schema$ and choice of $a$ indices  $i_1, \dotsc i_a \in  \{-n, \ldots, 0, \ldots ,\uabound-1 \}$,
where $n := \max_{\actiona\in\act}{|\getnew\actiona|}$.
$$\add(R(i_1, \ldots i_a))@ x \equiv \bigvee_{\frameletter{\action,s} \in \Gamma} \frameletter{\action,s}(x)$$
where $\Gamma = \{  ~\frameletter{\action,s} \mid  \getadd{\action}$ \text{ contains a tuple } $R(\xi_1 \ldots \xi_a) $ and  for all  $1 \le j \le a$  if $\xi_j \in \getfree{\action}$ then  $ s(\xi_j) = i_j $
 and if $\xi_j$ is the $k$th fresh input variable $v_k$ then $ i_j = -k\}$.

\textbf{Equality between indexed elements of different blocks. } Consider the encoding of the run in Figure~\ref{fig:2-run-nw}. Notice that the index $-2$ in the block $B_1$ and index $1$ in the block $B_2$ refer to the same element ($e_2$ in the concrete run of Figure~\ref{fig:run-of-example-1}). Notice also that the element referred to by index $-2$ in Block $B_2$ is the same as the element referred to by index $0$ in  block $B_7$ ($e_5$ in the concrete run of Figure~\ref{fig:run-of-example-1}). 

Given two positions $x$ and $y$ and indices $i$ and $j$, consider following question: I\textit{s the element referred to by index $i$ in the block of $x$ the same as the element referred to by index $j$ in the block of $y$?} In fact, this property can be expressed in \msonw. We define below a binary predicate $\Eqij{i}{j}(x,y)$ for the same.
Indeed we will define such a predicate for every pair $i,j$ with $-\maxy \le i ,j \le \uabound-1$. 

Towards this, first notice that the predicate must hold if there is a $\push i $-labelled position in the block of $x$ that is $\nestrel$-related to $\pop j$-labelled position in the block of $y$. This forms the basic step relation towards defining $\Eqij{i}{j}(x,y)$. 
\begin{align*}
\step^{i,j}(x, y)  \equiv & \exists z_1 \exists z_2 .  \sameblock(z_1, x)  \wedge \sameblock(z_2, y) \\
& \wedge z_1  \nestrel z_2
  \wedge {\push i} (z_1)  \wedge {\pop j} (z_2)  
\end{align*}
Recall that ${\push i }(x)$ means that the position $x$ is labelled by the letter $\push i$. Notice that our definition of $\step^{i,j}(x, y) $ is directional, in the sense that $x$ must necessarily be before $y$ for $\step^{i,j}(x, y) $ to hold.
The transitive closure of the above $\step$ relation gives us the required predicate $\Eqij{i}{j}(x,y)$. Suppose the element indexed $i$ in the block of $x$ is $e$. The element $e$ may appear with different indices at the intermediate steps. 
Hence we need to take a zig-zag transitive closure. Our formula uses $\f \maxy$ second-order position variables. Intuitively the set $X_k$ contains the set of positions such that the element $e$ is indexed by $k$ in its block. Using the universal quantifier, we require that the minimal of such sets which are closed under the zig-zag transitive closure must contain $y$ in the set $X_j$. The formula $\Eqij{i}{j}(x,y)$ is depicted in Figure~\ref{fig:eqij}.

Notice that since $\step$ is directional, so is $\Eqij{i}{j}(x,y)$. I.e, if $\Eqij{i}{j}(x,y)$ then necessarily $x \le y$ or $\sameblock(x,y)$.

\textbf{Recent elements participating in a relation.} Consider a relation $R$ of arity $\arity$. We define a predicate $\relm{R}{x_1, i_1, x_2, i_2, \dots x_\arity, i_\arity}{ y}$ which holds iff the database instance \emph{before} the execution of the block of $y$ has the tuple $\tup {e_1, e_2, \dotsc e_\arity}$ in relation $R$ where, the element $e_j$ is indexed by $i_j$ in the block of $x_j$ for all $j: 1 \le j \le \arity$. 
 This predicate can be expressed in MSO, as given below. 
 \begin{eqnarray*}
\exists x\, x  < y  \wedge \neg \sameblock(x,y) \wedge \\
\bigvee_{-\maxy \le \ell_1, \ldots \ell_\arity  \le \uabound -1 }  \, 
  \add(R(\ell_1, \ldots \ell_\arity) @ x) \wedge 
   \bigwedge_{1 \le j \le \arity} 
 \Eqij{\ell_j}{i_j} (x, x_j) \\ 
\wedge   \forall z \, \neg 
{\color{red}(} x \le z \wedge z < y \wedge \neg \sameblock(z, y) \wedge \\
 \bigvee_{0 \le m_1, \ldots m_\arity \le \uabound-1} 
{\color{blue}(} \del(R(m_1, \ldots m_\arity) )@ z \wedge 
  \bigwedge_{1 \le j \le \arity}
\Eqij{\ell_j}{m_j}(x,z) 
 {\color{blue})}{\color{red})}  
 \end{eqnarray*}
 The formula essentially says that the relation tuple  has been added to the database instance at some point in the past of $y$, and since then it has not been deleted. 
 
Similarly we define  $\reln{R}{x_1, i_1, x_2, i_2, \dots x_n, i_n}{ y}$ 
 which holds iff the database instance \emph{after} the execution of the block of $y$ has a tuple in relation $R$ as before. It   can be expressed as given follows:

\begin{eqnarray*}
\exists x\, x  \le y  \wedge \bigvee_{-\maxy \le \ell_1, \ldots \ell_\arity  \le \uabound -1 }  \, \add(R(\ell_1, \ldots \ell_\arity) )@ x \wedge  \\
 \bigwedge_{1 \le j \le \arity} 
 \Eqij{\ell_j}{i_j} (x, x_j)     
\wedge  \forall z \, \neg 
{\color{red}(} x \le z \wedge z \le y \wedge  \\
 \bigvee_{0 \le m_1, \dotsc , m_\arity \le \uabound-1} 
 {\color{blue}(} \del(R(m_1, \ldots m_\arity) ) @ z  \wedge \bigwedge_{1 \le j \le \arity} 
 \Eqij{\ell_j}{m_j}(x,z) 
  {\color{blue})} {\color{red})}   
 \end{eqnarray*}

\medskip
\subsubsection{Expressing valid encodings in \msonw}

We now show that the conditions given in Section~\ref{sec:conditions-for-valid-encodings} can be expressed in $\msonw$. 

\textbf{0. Well-formedness } We need to check the local consistency of each block appearing in the word, which means 
 $s_i$ must not assign a variable to a recency index higher than or equal to $m_i$, and that $J_i \subseteq \set{ 0, \ldots , m_i }$. Further it must of the form described in Section~\ref{sec:conditions-for-valid-encodings}.  This is a syntactic check inside a block and can be easily stated in \msonw.

\textbf{1. Consistency of $m$. }
We write a formula $\formRecent_m (x)$ to state that, just before executing the block of $x$, the current database $\I$ has at least $m+1$ elements in $\ADOM{\I}$.  Thanks to Remark~\ref{rem:pending-push-size},  this can be expressed in $\msonw$ by saying that there are $m+1$ distinct pushes before the block of $x$ which are not popped until $x$.
\smallskip \noindent
\begin{tabular}{c}
$\formRecent_m (x)  \equiv \exists y \,  \sameblock(x, y) \wedge \Aint(y) \wedge \exists x_1, \ldots x_{m} 
$\\
 $ \bigwedge_{i \neq j} x_i \neq x_j  \wedge  \bigwedge_i \big(\Apush (x_i) \wedge x_i < y \wedge  \forall z  (x_i \nestrel z \Rightarrow y < z)\big) $
\end{tabular}
Now, consistency of $m$ can be stated as:
$$ \forall x \bigwedge_{0 \le i \le \uabound-1} \neg  \formRecent_i (x) \vee \exists y \,( {\pop i }(y) \wedge \sameblock(x,y))$$

\textbf{2. Consistency of J. } Towards this we first need to write a formula $\Live(x,i)$ which holds only if the element with recency index $i$ in the block of $x$ is in $\adom{\I}$ after the execution of the block of $x$. This is expressed similarly to the formula $\Active(u)$ of Example~\ref{ex:formula-active}.
$\Live(x,i) \equiv \bigvee_{R/a \in \schema} \, \exists x_1, \dotsc , x_{\arity} \bigwedge_{ 1 \le j \le \arity} x_i \le x  \wedge \bigvee_{-\maxy \le i_1, i_2, \dotsc , i_{\arity} \le \uabound - 1 }\, \\
 \bigvee_{1 \le j \le \arity} 
\reln{R}{
\dotsc , x_{j-1},i_{j-1}, x, i, x_{j+1},i_{j+1}, \dots 
}{x} $
 where $\maxy = \max_{\action \in act} |\getnew{\action}|$. 
 
Now the consistency of $J$ can be stated by saying that a recency index is pushed  in a block iff it is $\Live$:
$$\forall x \bigwedge_ {0 \le i \le \uabound-1}  
\Live(x,i)  \Leftrightarrow  \big(\exists y \, {\push i }(y) \wedge \sameblock(x, y)\big)    $$

\textbf{3. Consistency of action guards. } We  first present a syntactic translation of an $\folang$ formula  into an $\msonw$ formula. The translation also depends on the current block (a block is represented by the head position of the  block which we denote by a free position variable $x$) as well as the action $\action$ and  the abstract substitution $s$ used in the block. The translation of an $\folang$ formula $\fodb{Q}$ at $x$ wrt.\ $s$ and $\action$ is a $\msonw$ formula denoted $\translate{\fodb{Q}}$. 

In our translation, a  first-order data variable $\fodb{u}$ is represented by the position $x^u$ and an index $i^u$, which is an number between $-\maxy$ and $\uabound -1$ where $\maxy = \max_{\action \in \act} |\getnew{\action}|$. Intuitively, instead of reasoning about an element in the domain, we reason about it symbolically by means of a (past) position where it is live and its recency index at that position. 
Given $\action$, $s$ and $x$, we distinguish between variables belonging to $\getfree{\action}$ and the other variables. For a variable $u \in \getfree{\action}$ we set $x^u = x$ and $i^u = s(u)$. 

The translation is defined inductively as follows: (In the following $x^u =x$ and $i^u = s(u)$ if $u \in \getfree{\action}$, and $\maxy = \max_{\action \in \act} |\getnew{\action}|$ )
\begin{itemize}[leftmargin=13pt]
\item $\translate{\fodb{R(u_1, \ldots u_n)}} \equiv \relm{R}{x^{u_1}, i^{u_1}, \ldots, x^{u_n}, i^{u_n}}{ x}$
\item $\translate{\fodb{u_1 = u_2}} \equiv \Eqij{i^{u_1}}{i^{u_2}}(x^{u_1}, x^{u_2}) $ 
\item $\translate{ \fodb{\exists u\ldotp Q}}  \equiv \exists x^u\ldotp x^u < x \wedge  \bigvee _{ -\maxy \le i^u \le \uabound -1} \translate{\fodb{Q}}$
\item $\translate{\fodb{Q_1 \wedge Q_2 }} \equiv \translate{\fodb{Q_1}} \wedge \translate{\fodb{Q_2}}$
\item $\translate{\fodb{\neg Q}} \equiv \neg \translate{\fodb{Q}}$
\end{itemize}

Now we are ready to express the consistency of action guards with the run. It can be expressed by the following formula:
$ \forall x \, \bigwedge_{\frameletter{\action:s} \in \Aint}  \frameletter{\action,s}(x) \Rightarrow \translate{\fodb{\getguard{\action}}}$.

\smallskip

Let $\varphi_{\uabound, \s}^\textsf{valid}$ be the conjunction of the four conditions listed above. It characterises valid encodings of $\uabound$-bounded runs of a $\sys$ $\s$.

\subsection{Translating {\large \msodb} specifications into {\large \msonw} specifications} \label{sec:mso-translation}

Here we provide a translation of the specifications in $\msodb$ to an equivalent one  in $\msonw$ over valid encodings of runs. This is similar in spirit to the translation described for action guards. As done there, we represent a first-order data variable $\fodb{u}$ by the pair $x^u, i^u$. We extend this representation to incorporate the globally quantified data variables as well. We  distinguish neither between free or bound variables, nor between globally quantified and locally quantifies variables: A data variable $u$ is represented by the pair $x^u, i^u$. 

 A first-order position variable of $\msodb$ will correspond to a block in the nested word encoding. A block is represented by its head. Thus in our translation, a first-order position variable of $\msodb$ corresponds to a first-order position variable of $\msonw$ which varies over heads of blocks.
 A set variable of $\msodb$ corresponds to a set variable in $\msonw$ relativized to head positions. The translation of a $\msodb$ formula $\varphi$ is denoted $\translation{\varphi}$, and is defined inductively as follows: 
\begin{itemize}[leftmargin=15pt]
\item $\translation{\fodb{Q}@x} \equiv \Aint(x) \wedge \bigvee_{\frameletter{\action:s} \in \Aint } \frameletter{\action:s} (x) \Rightarrow \translate{\fodb{Q}}$
\item $\translation{\exists x\ldotp \varphi} \equiv \exists x \ldotp \Aint(x) \wedge \translation{\varphi}$
\item $\translation{\exists X\ldotp \varphi} \equiv \exists X\ldotp (\forall x\ldotp x \in X \Rightarrow \Aint(x)) \wedge \translation{\varphi}$
\item $\translation{\existsg u.  \forun{\varphi}} \equiv  \exists x^u \ldotp \Aint(x^u) \wedge \bigvee _{ -\maxy \le i^u \le \uabound -1}  \translation{\varphi}$ where $\maxy = \max_{\action \in \act} |\getnew{\action}|$
\item $\translation{x < y} \equiv x < y $ \hfill $\bullet$ \enskip $\translation{x \in X} \equiv x \in X$
\item $\translation{{\varphi_1 \wedge \varphi_2 }} \equiv \translation{{\varphi_1}} \wedge \translation{{\varphi_2}}$ \hspace{.8cm} $\bullet$ \enskip  $\translation{{\neg \varphi}} \equiv \neg \translation{{\varphi}}$
\end{itemize}

\subsection{Concluding the reduction}

We reduce  the recency bounded model checking problem to the satisfiability checking of \msonw. 
Given a \sys $\s$, recency bound $\uabound$ and an $\msodb$ specification $\psi$, we construct $\varphi_{\uabound, \s}^\textsf{valid}$ and $\translation{\psi}$ as described in the previous subsections. The $\uabound$ bounded model checking problem reduces to the non-satisfiability checking of $\varphi_{\uabound, \s}^\textsf{valid} \wedge \neg \translation{\psi}$. The satisfiability checking of \msonw is decidable \cite{AlurM09}, and hence by our reduction  \textsc{Recency-bounded-\texttt{MSO/DMS}-MC} is decidable.

The construction of the $\msonw$ formula $\varphi_{\uabound, \s}^\textsf{valid} \wedge \neg \translation{\psi}$ takes time $\mathcal O ((\uabound + |\schema| + |\act|)^{\mathcal{O} (\maxarity + \maxvars)})$ where $|\schema|$ denotes  the number of relations in $\schema$,  $|\act|$ denotes the number of actions of $\s$, $\maxarity$ is the maximum arity of the relations in the schema (i.e., $\maxarity = \max_{R/a \in \schema} a$) and $\maxvars$ is the number of data-variables appearing in the action guards and $\psi$.  

\section{Related Work}
As pointed out in the introduction, \sys{s} belong to the series of
works on the verification of dabase-driven dynamic systems whose
initial state is known
\cite{ICSOC10,ICSOC11,Bagheri2011:Artifacts,BLP:KR:12,BCDD13,SMTD13}. In
\cite{BLP:KR:12}, \emph{artifact-centric multi-agent systems} are
proposed to simultaneously account for business artifacts and for the
specification of agents operating over them. Building on \cite{ICSOC10,ICSOC11},
decidability of verification of FO-CTLK properties with active domain FO
quantification is obtained, under the assumption that the size of the databases maintained by
agents and artifacts never exceeds a pre-defined bound. This notion of
\emph{state-boundedness} is thoroughly studied in \cite{BCDD13}
on top of the framework of \emph{data-centric dynamic systems}
(DCDSs). There, decidability of verification is obtained for a
sophisticated variant of FO $\mu$-calculus with active domain FO quantification, in which the possibility
of quantifying over individual objects across time is limited to those
objects that persist in the active domain of the system. Our logic \msodb differs to those used in
\cite{BLP:KR:12,BCDD13} since it captures linear
properties over runs, as opposed to branching properties over
RTSs. Furthermore, it leverages the full power of MSO to express
sophisticated temporal properties, and is equipped with unrestricted
FO quantification across positions of the run, as well as the
possibility of quantifying over the objects present in the whole run,
as opposed to only those present in the active domain of the current
state. 

Both in \cite{BLP:KR:12} and \cite{BCDD13}, decidability is obtained
by constructing a faithful, finite-state abstraction that preserves
the properties to be verified.
This shows that state-bounded dynamic
systems are an interesting class of \emph{essentially
  finite-state systems} \cite{Abd10}. On the other hand,
state-boundedness is a too restrictive requirement when dealing with
systems such as that of Example~\ref{ex:agency}. In fact, allowing for
unboundedly many tuples to be stored in the database is required to deal with
\emph{history-dependent dynamic systems}, whose behavior is influenced
by the presence of certain patterns in the
  (unbounded) history of the system (cf.~the definition of \emph{gold
    customer} in Example~\ref{ex:agency}). It is also essential to
  capture \emph{last-in first-out dynamic systems}, where the currently
  executed task may be interrupted by a task with a
  higher-priority, and so on, resuming the execution of the original task
only when the (unbounded) chain of higher-priority
  tasks is completed. See, e.g., the pre-emptive offer handling
  adopted in Example~\ref{ex:lifo}.
Notably, as argued in Example~\ref{ex:lifo}, such classes of unbounded systems can all be subject to \emph{exact} \msodb model checking, by choosing a sufficiently large bound for recency.

\section{Conclusion} 
We have proposed an under-approximation of dynamic database-driven systems  that allows unbounded state-space, under which we have shown decidability of model checking against {\msodb}. 
The decidability is obtained by a reduction to the satisfiability checking of MSO over nested words. 
 The complexity of our model checking procedure is non-elementary in the size of the specification and the \sys. A fine-grained analysis of complexity with respect to various input parameters such as arity of the relations, size of schema, number of variables in the queries etc. is left for future work. It is interesting to study whether one can obtain model checking algorithms with elementary complexity by using other specification formalisms, like temporal logics. Expressing valid runs in temporal logic would be important in this case, and it is interesting problem on its own.  Another direction for future work would be to identify other meaningful under-approximation parameters. For example, does bounding the most recently \emph{accessed} elements as opposed to most recently added elements yield decidability?
We also aim at applying under-approximation
techniques in the case where the initial database is not known, and
model checking is studied for every possible initial database,
in the style of \cite{DHPV09,Damaggio2011:Artifact,BoST13}.

\section*{Acknowledgments}
We acknowledge the support of the \emph{Uppsala Programming for
Multicore Architectures Research Center} (UPMARC), the \emph{Programming
Platform for Future Wireless Sensor Networks Project} (PROFUN), and the
EU project \emph{Optique - Scalable End-user Access to Big Data} (FP7-IP-318338).

\clearpage

\bibliographystyle{abbrv}
\bibliography{bibliography.bib}

\appendix
\section{Semantics of {\Large${\folang}$} queries}\label{app:query-semantics}

Given a database instance $\I $ over $\schema$ and $\domain$,  a $\folang$
query $\fodb{Q}$ over $\schema$, and a substitution $\sigma: \freevars{\fodb{Q}} \to \domain$, we write $\I, \sigma \models \fodb{Q}$ if the query $\fodb{Q}$ under the substitution $\sigma$ holds  in  database $\I$. This is defined inductively:
\begin{itemize}
\item $\I, \sigma \models \fodb{\true}$
\item $\I, \sigma \models \fodb{R(u_1, \dotsc , u_{\arity})}   \, \, \text{ if }\begin{array}[t]{l}  \text{ $(e_1, 				\dotsc, e_\arity) \in R^\I$ where}\\
		\text{ $e_i = \sigma(u_i)$ for all $i:1 \le i \le \arity$.}
		\end{array} $
\item $\I, \sigma \models \fodb{u_i = u_j}   \, \,$  if $\begin{array}[t]{l}  \sigma(u_i)= \sigma(u_j ).
\end{array}$  
\item $\I, \sigma \models \fodb{\neg \fodb{Q}}  \, \, $ if $\I, \sigma \not\models \fodb{ \fodb{Q}}  $.
\item $\I, \sigma \models \fodb{\fodb{Q_1} \wedge \fodb{Q_2}} $ if $\I, \sigma \models \fodb{Q_1} $, and $\I, \sigma \models \fodb{Q_2}.$
\item $\I, \sigma \models \fodb{\exists u. \fodb{Q}  } \, \, \text{ if }\begin{array}[t]{l}
\text{there exists }e \in \adom{\I}\\
\text{ such that }  \I, \sigma' \models \fodb{ Q} \text{ where }\\
\text{  $\sigma' $ is obataind from $\sigma$ as follows:}\\
\text{ $ \sigma'$ is defined on $u$ and } \sigma'(u) =e\\
\text{ and $\sigma'(u') = \sigma(u')$ if $u' \neq u$.}
\end{array}  $ 
\end{itemize}

\section{Semantics of {\Large\msodb}}\label{app:mso-semantics}
 A run $\rho$ is an infinite sequence of database instances over $\schema$ and $\domain$: $$\rho = \I_0 , \I_1, \I_2, \I_3 \dots$$
The \emph{global active domain} of the run $\rho$, denoted $\gadom{\rho}$ is the union of all active domains along the run. $\gadom{\rho} = \bigcup_{i \ge 0} \adom{\I_i}$.

An MSO formula $\forun{\phi}$ is interpreted over a  run $\rho$ under a substitution $\sigma$ of  the free variables $\freevars{\forun{\phi}}$. If the formula holds in the run $\rho$ under the substitution $\sigma$,
 we write $\rho, \sigma \models \forun{\phi}$. The semantics is defined inductively:
\begin{itemize}
\item $\rho, \sigma \models \forun{\fodb{Q}@x}   \, \, \text{ if }\begin{array}[t]{l}
\text{$\I_i, \sigma' \models \fodb Q$ where}\\
\text{$i = \sigma(x)$ and $\sigma' = \sigma|_{\freevars{\fodb{Q}}}$}
\end{array}  $ 
\item $\rho, \sigma \models \forun{x < y}  $ if $\sigma(x) < \sigma(y)$
\item $\rho, \sigma \models \forun{x \in X}  $ if $\sigma(x) \in  \sigma(X)$
\item $\rho, \sigma \models \forun{\neg \phi}  $ if $\rho, \sigma \not\models \forun{ \phi}$
\item $\rho, \sigma \models \forun{\forun{\phi_1} \wedge \forun{\phi_2}}  $ if $\rho, \sigma \models \forun{ \phi_1}$, and $\rho, \sigma \models \forun{ \phi_2}$
\item $\rho, \sigma \models {\exists x. \forun{\phi} } \, \, \text{ if }\begin{array}[t]{l}
\text{there exists $i \in \Nat$, such that}\\
 \rho, \sigma[x \mapsto i] \models \forun{ \phi}
\text{ where}\\
 \sigma[x \mapsto i](x) = i\text{, and }\\
 \sigma[x \mapsto i](\xi) = \sigma(\xi)\text{ if }\xi \neq x\\
\end{array}  $ 
\item $\rho, \sigma \models {\exists X. \forun{\phi} } \, \, \text{ if }\begin{array}[t]{l}
\text{there exists $J \subseteq \Nat$, such that }\\
 \rho, \sigma[X \mapsto J] \models \forun{ \phi}
\text{ where}\\
 \sigma[X \mapsto J](X) = J\text{, and }\\
 \sigma[X \mapsto J](\xi) = \sigma(\xi)\text{ if }\xi \neq X\\
\end{array}  $ 
\item $\rho, \sigma \models {\existsg u. \forun{\phi}  } \, \, \text{ if }\begin{array}[t]{l}
\text{there exists $e \in \gadom{\rho}$, such that }\\
 \rho, \sigma[u \mapsto e] \models \forun{ \phi}
\text{ where}\\
 \sigma[u \mapsto e](u) = e\text{, and }\\
 \sigma[u \mapsto e](\xi) = \sigma(\xi)\text{ if }\xi \neq u\\
\end{array}  $ 
\end{itemize}

The semantics of database query ($\I , \sigma \models \fodb{Q}$) is as expected (see Appendix~\ref{app:query-semantics}). The substitution of free variables is always restricted to the active domain of $\I$ in this case. That is,  $\textsf{Image}(\sigma) \subseteq \adom{\I}$ is necessary; just having  $\textsf{Image}(\sigma) \subseteq \gadom{\rho}$ is not sufficient.

Intuitively, $\fodb{Q}@x$
evaluates the $\folang$ query $Q$ over the database instance present
at position $x$ in the run. Formula $x < y$ asserts that position $x$
comes before $y$ along the run. Formula $x \in X$ states that position
$x$ belongs to the set $X$ of positions. Formula $\exists
x. \forun{\phi}$ states that there exists a position $x$ in the run where
$\forun{\phi}$ holds, whereas formula  $\exists X. \forun{\phi}$
models that there exists a set $X$ of positions in the run where
$\forun{\phi}$ holds. Finally, $\existsg u.  \forun{\phi}$ states that
there exists a data value $u$ that is active in some database instance
of the run and that makes $\forun{\phi}$ true. In this light, the quantifier
$\existsg$ ranges over the \emph{global active domain} of the run,
obtained by composing all active domains of the database instances
encountered therein.

\section{Booking Offers Example}
\label{sec:example}

\newcommand{\avstate}{\cval{avail}}
\newcommand{\inbstate}{\cval{booking}}
\newcommand{\clstate}{\cval{closed}}
\newcommand{\drstate}{\cval{drafting}}
\newcommand{\sustate}{\cval{subm}}
\newcommand{\cbstate}{\cval{canceled}}
\newcommand{\fistate}{\cval{finalized}}
\newcommand{\tbistate}{\cval{tbi}}
\newcommand{\acstate}{\cval{accepted}}
\newcommand{\ohstate}{\cval{onhold}}

\newcommand{\cust}{\rel{Cust}}
\newcommand{\agent}{\rel{Ag}}
\newcommand{\city}{\rel{City}}
\newcommand{\comp}{\rel{Rest}}
\newcommand{\offer}{\rel{Offer}}
\newcommand{\booking}{\rel{Booking}}
\newcommand{\pass}{\rel{Pass}} 
\newcommand{\ticket}{\rel{Ticket}}
\newcommand{\ostate}{\rel{OState}}
\newcommand{\bstate}{\rel{BState}}
\newcommand{\hosts}{\rel{Hosts}}
\newcommand{\prop}{\rel{Prop}}

\newcommand{\newflight}{\exact{newO}}
\newcommand{\newbooking}{\exact{newB}}
\newcommand{\closeoffer}{\exact{closeO}}
\newcommand{\cancelbooking}{\exact{cancel}}
\newcommand{\closebooking}{\exact{submit}}
\newcommand{\addperson}{\exact{addP}}
\newcommand{\checkperson}{\exact{checkP}}
\newcommand{\detprice}{\exact{detProp}}
\newcommand{\accept}{\exact{accept}}
\newcommand{\reject}{\exact{reject}}
\newcommand{\confirm}{\exact{confirm}}
\newcommand{\resume}{\exact{resume}}

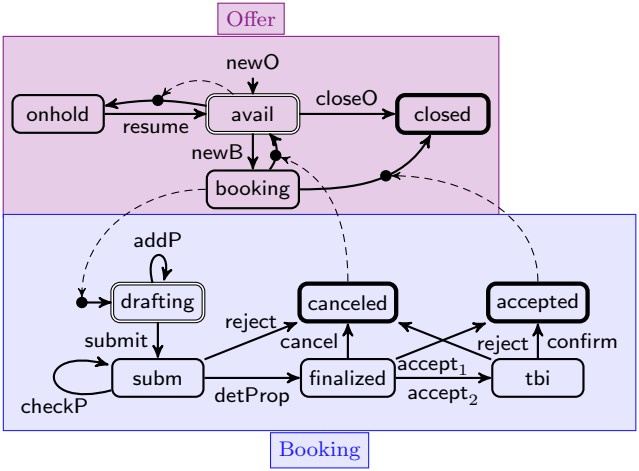
\begin{figure}[t]
\centering
\begin{tikzpicture}[->,>=stealth',auto,x=2.5cm,y=1cm,thick]
\tikzstyle{every node}=[font=\small];
\tikzstyle{state}=[draw,rectangle,rounded corners=3pt,minimum height=.5cm,minimum width=1.2cm];
\tikzstyle{istate}=[double,state,thin];
\tikzstyle{fstate}=[state,ultra thick];
\tikzstyle{offer}=[violet!90,fill=violet!20];
\tikzstyle{booking}=[blue!90,fill=blue!10];
\tikzstyle{elem}=[circle, draw, thick, scale=0.4];

\node (newoffer) at (.5,.7) {$\newflight$};
\node[istate] (av) at (.5,0) {$\avstate$};
\node[state] (inb) at (.5,-1) {$\inbstate$};
\node[fstate] (cl) at (1.5,0) {$\clstate$};
\node[state] (oh) at (-.525,-0) {$\ohstate$};
\node[elem,fill] (toav) at (.62,-.55) {};
\node[elem,fill] (tooh) at (0,.18) {};
\node[elem,fill] (tocl) at (1.2,-.82) {};

\node[elem,fill] (newbooking) at (-.4,-2.5) {};
\node[istate] (dr) at (0,-2.5) {$\drstate$};
\node[fstate] (cb) at (1,-2.5) {$\cbstate$};
\node[state] (su) at (0,-3.5) {$\sustate$};
\node[state] (fi) at (1,-3.5) {$\fistate$};
\node[state] (tbi) at (2,-3.5) {$\tbistate$};
\node[fstate] (ac) at (2,-2.5) {$\acstate$};

 \path[->,thick] 
(newoffer) edge (av)
(av) edge node {$\closeoffer$} (cl)
(av) edge[bend right=0] node[left] {$\newbooking$} (inb)
(inb) edge[bend right=40] (av)
(inb) edge[out=0,in=-120] (cl)
(av) edge[bend right=10] (oh)
(oh) edge node[below] {$\resume$} (av)
(newbooking) edge (dr)
(dr) edge[loop above] node[above] (addp) {$\addperson$} (dr)
(dr) edge node[left] {$\closebooking$} (su)
(su) edge[loop left] node[below,yshift=-.15cm] (checkp) {$\checkperson$} (su)
(su) edge node[below] {$\detprice$} (fi)
(su) edge node[above] {$\reject$} (cb)
(fi) edge node[left] {$\cancelbooking$} (cb)
(fi) edge node[below,xshift=-.1cm,yshift=-.1cm] {$\accept_1$} (ac)
(fi) edge node[below] {$\accept_2$} (tbi)
(tbi) edge node[right,xshift=.3cm,yshift=-.05cm] {$\reject$} (cb)
(tbi) edge node[right] (confirm) {$\confirm$} (ac)
;

\path[->,thin,densely dashed]
(inb) edge[out=180,in=90] (newbooking)
(cb) edge[out=90,in=-20] (toav)
(ac) edge[out=90,in=0] (tocl)
(av) edge[out=135,in=45] (tooh)
;

\begin{scope}[on background layer]
\node[draw,offer,fit=(oh) (cl) (newoffer) (inb)] (offerartifact) {};
\node[above,offer,draw] at (offerartifact.north) {Offer} ;
\node[draw,booking,fit=(checkp) (ac) (confirm) (addp)] (bookingartifact) {};
\node[below,booking,draw] at (bookingartifact.south) {Booking} ;
\end{scope}
\end{tikzpicture}
\caption{Business artifact lifecycles of the running
  example. Solid arrows indicate state transitions
  for an artifact instance. A transition may be explicitly triggered
  by an action applied to the artifact instance, or implicitly because
another artifact instance is entering into a specific state (this
dependency is rendered using dahshed arrows).
\label{fig:artifacts}}
\end{figure}

To show the richness of our framework, we model a data-centric process
used by an agency to advertise restaurant offers and manage corresponding
bookings. Specifically, the process supports B2C interactions where
agents select and publish restaurant offers, and manage booking
requests issued by customers. To describe the process, we adopt the
well-established artifact-centric approach
\cite{Nigam03:artifacts,DBLP:journals/debu/CohnH09,Hull2008:Artifact}. In
particular, the process is centred around the two key (dynamic)
entities of \emph{offer} and
\emph{booking}. Intuitively, each agent can publish a dinner offer
related to some restaurant; if another, more interesting offer is
received by the agent, she puts the previous one on hold, so that it
will be picked up again later on by that or another agent (when it will be among the most
interesting ones). Each offer can result in a corresponding booking by
a customer, or removed by the agent if nobody is interested in
it. Offers are customizable, hence each booking goes through
a preliminary phase in which the customer indicates who she wants to
bring with her to the dinner, then the agent proposes a customized
prize for the offer, and finally the customer decides whether to
accept it or not.

The relational information structure characterizing an
offer contains the following relations:
\begin{itemize}
\item $\offer$ tracks the different offer artifact
  instances, where $\offer(\cval{o},\cval{r},\cval{a})$ indicates that
  offer $\cval{o}$ is for restaurant $\cval{r}$, and is being (or has
  been lastly) managed by agent $\cval{a}$ (other attributes of an
  offer are omitted for simplicity). Two
  read-only unary relations $\comp$ and $\agent$ are used to keep
  track of restaurants that can make offers.
\item $\ostate$ tracks the current state of each offer, where
  $\ostate(\cval{o},\cval{s})$ indicates that offer $\cval{o}$ is
  currently in state $\cval{s}$.
\end{itemize}
As for bookings, the following relations are used:
\begin{itemize}
\item $\booking$ tracks the different booking artifact instances,
  where $\booking(\cval{b},\cval{o},\cval{c})$ indicates that customer
  $\cval{c}$ engaged in booking  $\cval{b}$ for offer $\cval{o}$. A
  read-only unary relation $\cust$ is used to track the registered customers of
  the agency. 
\item $\bstate$ tracks the current state of each booking, similarly to
  the case of $\ostate$.
\item $\hosts$ holds the information about persons participating to a
  dinner: $\hosts(\cval{b},\cval{p})$ indicates that person $\cval{p}$
  is involved in booking $\cval{b}$ - where $\cval{p}$ may refer
  either to a customer or to a non-registered person.
 Like for the other relations, we omit
  additional information related to hosts and only keep their
  identifiers.
\item $\prop$ maintains information about the final offer proposal
  (including price) related to booking: $\prop(\cval{b},\cval{u})$
  indicates that URL $\cval{u}$ points to the final offer proposal for
  booking $\cval{b}$.
\end{itemize}

Figure~\ref{fig:artifacts} characterizes the lifecycle of such business entitites,
i.e., the states in which they can be and the possible state
transitions, triggered by actions. A state transition is triggered for
an artifact instance either by the explicit application of an action,
or indirectly due to a transition triggered for another artifact
instance. We substantiate such actions and implicit effects using \sys
actions working over the schema described before. For the sake of
readability, we use bold to highlight fresh input variables.

A new dinner offer can be inserted into the system by an agent that is
not currently engaged in a booking interaction with a customer. If the
agent is currently idle (i.e., not managing another offer), this has
simply the effect of creating a new offer artifact instance and mark
it as ``available''; if
instead the agent is managing another available offer, such an offer
is put on hold. To deal with these two cases, two \sys actions are
employed. The first case is modeled by action $\newflight_1$ as:
\begin{itemize}[leftmargin=*]
\item $\getguard{\newflight_1} = \comp(r) \land \agent(a) \land \neg \exists o,r'.\offer(o,r',a)$
\item $\getdel{\newflight_1} = \emptyset$
\item $\getadd{\newflight_1} = \set{\offer(\inputvar{y},r,a),\ostate(\inputvar{y},\avstate)}$
\end{itemize}
The second case is instead captured by action $\newflight_2$ as:
\begin{itemize}[leftmargin=*]
\item $\getguard{\newflight_2} = 
\begin{array}[t]{@{}r@{}l@{}}\comp(r) \land \agent(a) & {}\land
  \exists r'.\offer(o,r',a) \\&{}\land \ostate(o,\avstate)
\end{array}$
\item $\getdel{\newflight_2} = \set{\ostate(o,\avstate)}$
\item $\getadd{\newflight_2} = \{
\begin{array}[t]{@{}r@{}l@{}}
\offer(\inputvar{y},r,a),&\ostate(\inputvar{y},\avstate),\\
&\ostate(o,\ohstate) \}
\end{array}
$
\end{itemize}
If an agent is idle and there exist offers that are currently on hold,
the agent may decide to $\resume$ one such offer (consequently
becoming its responsible agent):
\begin{itemize}[leftmargin=*]
\item $\getguard{\resume} = 
\begin{array}[t]{@{}l@{}}
\agent(a) \land \offer(o,r,a') \land \ostate(o,\ohstate) \\
{}\land \neg \exists
  o',r'.\offer(o',r',a)
\end{array}
$
\item $\getdel{\newflight_1} = \set{\offer(o,r,a'),\ostate(o,\ohstate)}$
\item $\getadd{\newflight_1} = \set{\offer(o,r,a),\ostate(o,\avstate)}$
\end{itemize}
An available offer may be explicitly closed by an agent when it
expires or is no longer valid, through the $\closeoffer$ action:
\begin{itemize}[leftmargin=*]
\item $\getguard{\closeoffer} = \exists r,a.\offer(o,c,a) \land \ostate(o,\avstate)$
\item $\getdel{\closeoffer} = \set{ \ostate(o,\avstate)}$
\item $\getadd{\closeoffer} = \set{ \ostate(o,\clstate)}$
\end{itemize}
The last possible evolution of an available action is to be booked by
a customer. The booking process starts by applying the $\newbooking$
action over the available offer and by creating a corresponding
booking artifact instance, and can eventually terminate in two possible ways:
either the booking is canceled, which causes the offer to become
available again, or the booking is finalized and accepted, which
causes the offer to become closed. We go into the details of this
process, starting from the $\newbooking$ action:
\begin{itemize}[leftmargin=*]
\item $\getguard{\newbooking} = 
\begin{array}[t]{@{}r@{}l@{}}
\cust(c) &{} \land \exists r,a.\offer(o,r,a)\\ 
&{}\land \ostate(o,\avstate)
\end{array}
$
\item $\getdel{\newbooking} = \set{ \ostate(o,\avstate)}$
\item $\getadd{\newbooking} = 
\{ 
\begin{array}[t]{@{}r@{}l@{}}
\ostate(o,\inbstate), &
  \booking(\inputvar{y},o,c),\\
& \bstate(\inputvar{y},\drstate)\}
\end{array}
$
\end{itemize}

A drafted booking can be modified by its customer by adding or removing
related persons (for simplicity, we deal with addition only). As for addition, we use two \sys actions $\addperson_1$ and
$\addperson_2$, to respectively account for the case where the host
to be added is a customer or an external person:
\begin{itemize}[leftmargin=*]
\item $\getguard{\addperson_1} = 
\begin{array}[t]{@{}r@{}l@{}}
\exists o,c. \booking(b,o,c) &{} \land \bstate(b,\drstate)\\
&{}\land \cust(h)
\end{array}
$
\item $\getdel{\addperson_1} = \emptyset$
\item $\getadd{\addperson_1} = \set{\hosts(b,h)} 
$
\item $\getguard{\addperson_2} = 
\exists o,c. \booking(b,o,c) \land \bstate(o,\drstate)
$
\item $\getdel{\addperson_2} = \emptyset$
\item $\getadd{\addperson_2} = \set{\hosts(b,\inputvar{y})} 
$
\end{itemize}
Notice that the two actions are
identical, except for the fact that the first creates a new $\hosts$
tuple for a customer, whereas the second does it for a fresh person
identifier. Hence, this style of modeling shows how to capture actions
in which inputs that are not necessarily fresh are used.

Action $\closebooking$ has the simple effect of updating the state of the
considered booking from drafting to submitted. Here the agent
responsible for the offer checks the content of the booking person by
person, finally deciding whether to reject it or to make a customized
price proposal to the customer. Since the host-related information is
finally embedded in the proposal itself (or not needed in case of
rejection), such information is removed from the database. Since
{\sys} do not directly support ``bulk operations'' affecting at once all
tuples that meet a certain criterion, we model it through a loop:
persons are checked and removed through the $\checkperson$ action,
until no more person is hosted by the booking:
\begin{itemize}[leftmargin=*]
\item $\getguard{\checkperson} = 
\begin{array}[t]{@{}r@{}l@{}}
\exists o,c. \booking(b,o,c) &{} \land \bstate(b,\drstate)\\
&{}\land \hosts(b,h)
\end{array}
$
\item $\getdel{\checkperson} = \set{\hosts(b,h)}$
\item $\getadd{\checkperson} = \emptyset$
\end{itemize}
When no more person has to be checked, the agent may decide to $\reject$
the offer. This implicitly makes the offer to which the booking belongs to make
a transition back to the available state:
\begin{itemize}[leftmargin=*]
\item $\getguard{\reject} = 
\begin{array}[t]{@{}r@{}l@{}}
\exists c. \booking(b,o,c) &{} \land \bstate(b,\drstate)\\
&{}\land \neg \exists h.\hosts(b,h)
\end{array}
$
\item $\getdel{\reject} = \set{\bstate(b,\drstate),\ostate(o,\inbstate)}$
\item $\getadd{\reject} =\set{\bstate(b,\cbstate),\ostate(o,\avstate)}$
\end{itemize}
If instead the agents acknowledges the booking, she injects a proposal
URL into the system through the $\detprice$ action:
\begin{itemize}[leftmargin=*]
\item $\getguard{\detprice} = 
\begin{array}[t]{@{}r@{}l@{}}
\exists o,c. \booking(b,o,c) &{} \land \bstate(b,\drstate)\\
&{}\land \neg \exists h.\hosts(b,h)
\end{array}
$
\item $\getdel{\detprice} = \set{\bstate(b,\drstate)}$
\item $\getadd{\detprice} =\set{\bstate(b,\fistate),\prop(b,\inputvar{y})}$
\end{itemize}
Once the booking is finalized, three outcomes may be triggered by the
customer. In the first case, the customer decides to
$\cancelbooking$ the booking (which can be formalized similarly to the
$\reject$ action). In the second and third case, the customer accepts
the proposal. The outcome of the acceptance is conditional, and it
depends on whether the restaurant for which the booking has been made
considers the customer to be a ``gold customer'': if so, the
booking is immediately accepted, and the corresponding offer closed;
if not, then a final validation is required before acceptance. We use $Gold_k(c,r)$
to compactly indicate the query that determines whether $c$ is a
gold customer for $r$ by checking that the customer already completed
in the past at least $k$ bookings related to $r$ and accepted
(where $k$ is a fixed number):
\[
Gold_k(c,r) = 
\begin{array}[t]{@{}l@{}}
\exists o_1,\ldots,o_k,b_1,\ldots,b_k. \\
\bigwedge_{i,j \in \set{1,\ldots,k},i\neq j} \left( o_i \neq o_j \land b_i
  \neq b_j \right) \land{}\\
\bigwedge_{i \in \set{1,\ldots,k}}
\left(
\begin{array}{@{}l@{}}
\booking(b_i,o_i,c) \land{}\\
  \ostate(o_i,\acstate) \land{}\\ \offer(o_i,r)
\end{array}
  \right)
\end{array}
\]
We then use this query to model acceptance using an ``if-then-else'' pattern.
The case where a gold customer accepts a finalized booking is captured
by the $\accept_1$ action:
\begin{itemize}[leftmargin=*]
\item $\getguard{\accept_1} = 
\begin{array}[t]{@{}r@{}l@{}}
\exists c. &\booking(b,o,c) \land \bstate(b,\fistate)\\
&{}\land \exists r.\offer(o,r) \land Gold_k(c,r) 
\end{array}
$
\item $\getdel{\detprice} = \set{\bstate(b,\drstate)}$
\item $\getadd{\detprice} =\set{\bstate(b,\fistate),\prop(b,\inputvar{y})}$
\end{itemize}
The case of a non-gold customer is managed by action $\accept_2$ symmetrically, just checking whether
$\neg Gold_k(c,r)$ holds. The effect of $\accept_2$ is simply to
induce a state transition of the booking from finalized to
``to-be-validated''. The remaining transitions depicted in
Figure~\ref{fig:artifacts} are finally modeled following already
presented actions. 

We conclude this realistic example by stressing that it is unbounded
in many dimensions. On the one hand, unboundedly many offers can be
advertised over time. On the other hand, unboundedly many bookings for
the same offer can be created (and then canceled), and each such
booking could lead to introduce unboundedly many hosts during the
drafting stage of the booking. 

\section{Undecidability of Propositional Reachability}
\label{reach:undecidable:proofs}

We show in the present section that  the \sys propositional reachability is undecidable
as soon as the considered $\sys$ has one of the following:

\begin{itemize}
\item a binary predicate in $\schema$ even though the guards are only union of conjunctive queries $\clang{UCQ}$.
\item two unary predicates in $\schema$ and the guards allow $\clang{FOL}$.
\end{itemize}

To prove these two results, we use a reduction from the control state reachability of a Minsky two counter machine.
To do so, we first recall the control state reachability problem of a two counter machine.
Then, we provide a reduction for each case.

\paragraph{Counter machines}
A \emph{counter machine (\cmac)} $\cm$ is a tuple
$\tup{\cstate,q_0,\creg,\cprog}$, where:
\begin{itemize}
\item  $\cstate$ is a finite set of
  \emph{states}, 
\item $q_0 \in \cstate$ is the \emph{initial state}, 
\item $\creg$ is the number of
\emph{counters} manipulated by the machine (where each counters holds
a non-negative integer value), and 
\item $\cprog \subseteq \cstate \times \set{\inc,\dec,\ifz} \times
  \set{1,\ldots,\creg} \times \cstate$ is a finite set of
  \emph{instructions}. 
\end{itemize}
An instruction of the form $\tup{q_1,i,op,q_2}$ is applicable when the machine is in
state $q_1$, and has the effect of moving it to $q_2$, while operating
on counter $i$ according to operation $op$. Operations are as follows:
\begin{itemize}
\item $\inc$ increases the value of counter $i$ by 1;
\item $\dec$ decreases the value of counter $i$ by 1, and makes the transition
  applicable only if the counter holds a value strictily greater than 0;
\item $\ifz$ does not manipulate the counter, but makes the transition
  applicable only if counter $i$ holds value 0.
\end{itemize}

The execution semantics of $\cm$ is defined in terms of a (possibly
infinite) \emph{configuration graph}.
A configuration of $\cm$ is a pair $\tup{q,V}$, where $q \in \cstate$
and $V: \set{1,\ldots,\creg} \rightarrow \mathbb{N}$ is a function
that maps each
counter to a corresponding natural number.  The configuration graph is
then constructed starting from the \emph{initial configuration}
$\tup{q_0,V_0}$, where $V_0$ assigns $0$ to each counter, and then
inductively applying the transition relation $\ctrans$ defined next.

For every pair $\tup{q,V}$, $\tup{q',V'}$ of
configurations, we write $\cgoto{\tup{q,V}}{\tup{q',V'}}$ if and only if
one of the following conditions holds:
\begin{itemize}
\item $\tup{q,\inc,i,q'} \in \cprog$, $V'(i) = V(i)+1$, and $V'(j) =
  V(j)$ for each $j \in \set{1,\ldots,\creg}\setminus \set{i}$;
\item $\tup{q,\dec,i,q'} \in \cprog$, $V(i) > 0$, $V'(i) = V(i)-1$, and $V'(j) =
  V(j)$ for each $j \in \set{1,\ldots,\creg}\setminus \set{i}$;
\item $\tup{q,\ifz,i,q'} \in \cprog$, $V(i) = 0$, and $V' = V$.
\end{itemize}
We denote by $\ctranscl$ the reflexive transitive closure of $\ctrans$.

The \emph{control state reachability} problem for counter machines is then defined as:
\begin{itemize}
\item \emph{Input}: a \cmac $\cm =
  \tup{\cstate,q_0,\creg,\cprog}$ and a control state $q_f\in\cstate$.
\item \emph{Question}: is it the case that
  $\tup{q_0,V_0}\ctranscl{\tup{q_f,V_f}}$ for some mapping $V_f$?
\end{itemize}
We call \tcreachproblem the \creachproblem problem where
the input \cmac is a \emph{two-counter machine} (\tcmac).
It is well known that \tcreachproblem is undecidable.

\paragraph{Two unary relations with $\folang$ queries}

We provide in the following a reduction from the \tcreachproblem
to the control state reachability of a \sys which schema is composed 
of two unary relations, a number of nullary relations
and which actions use $\folang$ as a query language.

Let $\cm = \tup{\cstate,q_0,q_f,2,\cprog}$ be a \tcmac and let $q_f\in\cstate$.

We define the  data domain $\Delta$,
the relational schema $\schema$ and
the \sys $\s_{\tuple{\cm,q_f}} = \tup{\idb,\act}$ as follows:
\begin{itemize}
\item $\Delta=\Nat$,
\item $\schema = \set{C_1/1,C_2/1} \cup \bigcup_{q \in \cstate}
  \set{S_q/0}$ contains two unary relations to simulate the two counters,
  and a set of nullary relations, one for each control state of $\cm$.
\item $I_0 = \set{S_{q_0}}$.
\item $\act$ is the smallest set satisfying the following conditions:
  for every instruction $\tup{q,i,\mathit{cmd},q'} \in \cprog$,
\begin{itemize}
\item if $\mathit{cmd} = \inc$, $\act$ contains
\begin{align*}
\tup{\emptyset,\set{v},S_q,\set{S_q},\set{C_i(v),S_{q'}}}
\end{align*}
\item if $\mathit{cmd} = \dec$, $\act$ contains
\begin{align*}
\tup{\set{u},\emptyset,S_q
    \land C_i(u),\set{C_i(u),S_q},\set{S_{q'}}}
\end{align*}
\item if $\mathit{cmd} = \ifz$, $\act$ contains
\begin{align*}
\tup{\set{u},\emptyset,S_q
    \land \neg \exists u.C_i(u),\set{S_q},\set{S_{q'}}}
\end{align*}
\end{itemize}
\end{itemize}
All runs of $\s_\cm$ are equivalent regarding how they evolve the
number of tuples in $C_1$ and $C_2$, and they faithfully reproduce the
runs of $\cm$.
We consequently have that $\tcreachproblem(\cm, q_f)$ if and only if the proposition $S_{q_f}$
is reachable by the \sys $\s_{\tuple{\cm,\state_f}}$.

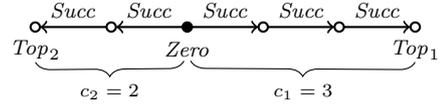
\begin{figure}[t]
\centering
\begin{tikzpicture}
\tikzstyle{every node}=[font=\small]
\tikzstyle{elem}=[circle, draw, thick, scale=0.4];
\node[elem,fill] (zero) at (0,0) {};
\node[elem] (c21) at (-1,0) {};
\node[elem] (c22) at (-2,0) {};
\node[elem] (c11) at (1,0) {};
\node[elem] (c12) at (2,0) {};
\node[elem] (c13) at (3,0) {};

\node at ($(zero) - (0,0.3)$)    {$\mathit{Zero}$};
\node at ($(c13) - (0,0.3)$)    {$\mathit{Top}_1$};
\node at ($(c22) - (0,0.3)$)    {$\mathit{Top}_2$};

\path[->,thick] 
(c21) edge node [above] {$\mathit{Succ}$} (c22)
(zero) edge node [above] {$\mathit{Succ}$} (c21)
(zero) edge node [above] {$\mathit{Succ}$} (c11)
(c11) edge node [above] {$\mathit{Succ}$} (c12)
(c12) edge node [above] {$\mathit{Succ}$} (c13);

\draw[decoration={brace,amplitude=2mm,mirror,raise=4mm},decorate]
(c22.south) -- (zero.south west) node [black,midway,yshift=-.8cm] 
{\footnotesize $c_2 = 2$}; 
\draw[decoration={brace,amplitude=2mm,mirror,raise=4mm},decorate]
(zero.south east) -- (c13.south)
node [black,midway,yshift=-.8cm] 
{\footnotesize $c_1 = 3$}; 
\end{tikzpicture}
\caption{Encoding of 2 counters with a binary relation,
  and two unary relations
\label{fig:2c-three-rel}}
\end{figure}

\paragraph{One binary relation with $\ucqlang$ queries}

We provide in the following a reduction from the \tcreachproblem
to the control state reachability of a \sys which schema is composed 
of one binary relation, three unary relations, a number of nullary relations
and which actions use $\ucqlang$ as a query language.

Let $\cm = \tup{\cstate,q_0,2,\cprog}$ be a \tcmac and let $q_f\in\cstate$.
We define the  data domain $\Delta$,
the relational schema $\schema$ and
the \sys $\s_{\tuple{\cm,q_f}} = \tup{\idb,\act}$ as follows:
\begin{itemize}
\item $\Delta = \Nat$
\item $\schema = \set{\text{Top}_1/1,\text{Top}_2/1, \text{Zero}/1, \text{Succ}/2} \cup \bigcup_{q \in \cstate} \set{S_q/0} \cup \set{S_{init}/0}$
contains four relations ($\text{Top}_1/1,\text{Top}_2/1, \text{Zero}/1, \text{Succ}/2$) encoding the two counters following the intuition in Figure~\ref{fig:2c-three-rel},
a set of nullary relations, one per state of the $\cm$,
plus an initial state $S_{init}/0$.
\item $I_0 = \set{S_{init}}$.
\item $\act$ contains two subsets of actions $\act = \act_{\text{init}} \uplus \act_{\text{cmd}}$.
The set $\act_{\text{init}}$ contains one action that is in charge of starting the simulation:
\begin{align*}
\langle &\emptyset, \set{v}, S_{init}, \set{S_{init}},\\
&\set{ S_{q_0}, \text{Top}_1(v),\text{Top}_2(v), \text{Zero}(v)} \rangle
\end{align*}

The set of commands $\act_{cmd}$ simulate the counter machine transitions and is defined as
the smallest set satisfying the following conditions:
\begin{itemize}
\item for every instruction $\tup{q,i,\inc,q'} \in \cprog$, $\act_{cmd}$ contains
\begin{align*}
\langle & \set{u},\set{v},\\
&S_q \land \text{Top}_i(u),\\
&\set{S_q,\text{Top}_i(u)},\\
&\set{S_{q'}, \text{Succ}(u,v), \text{Top}_i(v)}\rangle;
\end{align*}
\item for every instruction $\tup{q,i,\dec,q'} \in \cprog$, $\act_{cmd}$ contains
\begin{align*}
\langle & \set{u_1,u_2},\emptyset,\\
&S_q \land \text{Succ}(u_1,u_2) \land \text{Top}_i(u_2),\\
&\set{S_q,\text{Succ}(u_1,u_2), \text{Top}_i(u_2)},\\
&\set{S_{q'}, \text{Top}_i(u_1)}\rangle;
\end{align*}
\item for every instruction $\tup{q,i,\ifz,q'} \in \cprog$, $\act_{cmd}$ contains
\begin{align*}
\langle & \set{u},\emptyset,
S_q \land \text{Top}_i(u) \land \text{Zero}(u),
\set{S_q},
\set{S_{q'}}\rangle;
\end{align*}
\end{itemize}

\end{itemize}

The initialization phase sets the pointers of the counters:
$\text{Top}_1$ points to the head of the first counter chain,
$\text{Top}_2$ points to the head of the second counter chain
while 
$\text{Zero}$ points to the tail of both counters chains.
After the initial command has been performed, all three relations $\text{Top}_1$, $\text{Top}_2$ and $\text{Zero}$ contain each one of them exactly one element.
Each run of $\s_{\tup{\cm, q_f}}$ manipulates the extension of $\text{Succ}$ as a
linear order.
At every moment of the run we have that:
\begin{itemize}
\item the distance between the element pointed by $\text{Zero}$ and the element pointed by $\text{Top}_1$ gives the value of the first counter;
\item the distance between the element pointed by $\text{Zero}$ and the element pointed by $\text{Top}_2$ gives the value of the second counter;
\end{itemize}
We consequently have that $\tcreachproblem(\cm, q_f)$ if and only if the proposition $S_{q_f}$
is reachable by the \sys $\s_{\tuple{\cm,\state_f}}$.

\section{Runs which are equivalent modulo permutations}
\label{claim:permutation:proof}

\begin{lemma}
Two runs match on their abstraction, if and only if they are equivalent modulo permutations of the data domain.
\end{lemma}
\noindent
Let
$\hat \rho = \left(\kgotos{\uabound}{\tup{\I_j,\history_j, \seqnof_j}}{\tup{\I_{j+1},\history_{j+1}, \seqnof_{j+1}}}{\action_j:\sigma_j} \right)_{j \ge 0}$
and 
$\hat \rho' = \left(\kgotos{\uabound}{\tup{\I'_j,\history'_j, \seqnof'_j}}{\tup{\I'_{j+1},\history'_{j+1}, \seqnof'_{j+1}}}{\action'_j:\sigma'_j} \right)_{j \ge 0}$
be two extended runs such that
$\abstraction(\hat \rho) = \abstraction(\hat \rho') = (\tuple{\action_j:s_j})_{j\ge 0}$.
That implies that $\action_j = \action'_j$ for every $j\ge0$.

We prove in what follows the existence of a bijection $\lambda: \gadom{\hat \rho} \rightarrow \gadom{\hat \rho'}$
such that, for every $i\ge0$, $\lambda$ is an isomorphism from $\I_i$ onto $\I'_i$.

Notice that the global active domain of a given run $\hat \rho$ amounts to the set that contains all fresh input elements introduced in the database all along the run.
That is $\gadom{\hat \rho} = \cup_{i\ge0} \I_i = \cup_{i\ge0} \setcomp{\sigma_i(v)}{v\in\getnew{\action_i}}$
and $\gadom{\hat \rho} = \cup_{i\ge0} \I'_i = \cup_{i\ge0} \setcomp{\sigma'_i(v)}{v\in\getnew{\action'_i}} = \cup_{i\ge0} \setcomp{\sigma'_i(v)}{v\in\getnew{\action_i}}$ (since $\action'_i = \action_i$).

\textbf{Definition of $\lambda$.}  Let $e\in\Delta$.
We have that $e\in\gadom{\hat \rho}$ if and only if $\exists i\ge 0$ and $\exists v\in\getnew{\action_i}$ such that $\sigma_i(v) = e$. 
Since $\action'_i = \action_i$, $\sigma'_i(v) \in \gadom{\hat \rho'}$ is also well defined. We set $\lambda(e) = \sigma'_i(v)$.
\textbf{Injectivity.} Let $e_1,e_2\in\gadom{\hat\rho}$ such that $e_1 \ne e_2$.
That means that there are $i_1, i_2 \ge 0$, $v_1 \in\getnew{\action_{i_1}}$ and $v_2 \in\getnew{\action_{i_2}}$
such that $\sigma_{i_1}(v_1) = e_1$ and $\sigma_{i_2}(v_2) = e_2$.
We have that $\lambda(e_1) = \sigma'_{i_1}(v_1)$ and $\lambda(e_1) = \sigma'_{i_2}(v_2)$.
Notice that, since $e_1 \ne e_2$, we can not have $i_1 = i_2$ and $v_{1} = v_{2}$ at the same time.
That, the fact that substitutions have to be injective when applied to the fresh variables
and together with the freshness condition for the newly input values,
imply that $\lambda(e_1) \ne \lambda(e_2)$. Thus, $\lambda$ is injective.
\textbf{Surjectivity.}
Let $e'\in\gadom{\hat\rho'}$. That implies the existence of $i\ge0$ and $v\in\getnew{\action'_i}$
such that $\sigma'_{i}(v) = e'$.
Since $\action_i = \action'_i$, $\sigma_{i}(v)$ is also well defined and we set $e = \sigma_{i}(v)$.
We have that $\lambda(e) = e'$.
Thus, $\lambda$ is surjective.

Thus, $\lambda$ is a bijection. In a similar fashion, we can define  yet another bijection from 
$\setcomp{\seqnof_i(e)}{e\in\gadom{\hat\rho}}$ onto $\setcomp{\seqnof'_i(e)}{e\in\gadom{\hat\rho'}}$
 that we denote by $\beta$ and such that $\beta(\seqnof_i(e)) = \seqnof'_i(\lambda(e))$.
Moreover, we can show that  $\beta$ is monotonic.

Now that we have defined $\lambda$ (and $\beta$),
we shall prove by induction on $i\in\Nat$ that $\lambda$ is an isomorphism from $\I_i$ onto $\I'_i$.

For the base case ($i = 0$), we trivially have that $\I_0 = \I'_0$ and $\I_0 \cap \gadom{\hat \rho} = \emptyset = \I_0 \cap \gadom{\hat \rho'}$.
Assume now that for some $i\in\Nat$ we have that $\lambda$ is an isomorphism from $\I_i$ onto $\I'_i$.
Let's prove that $\lambda$ is also an isomorphism from $\I_{i+1}$ onto $\I'_{i+1}$.
Let $R(\sigma_i(u_1),\ldots,\sigma_i(u_n))$
(with $R\in\schema$ and $u_1,\ldots,u_n\in\getfree{\action_i}\uplus\getnew{\action_i}$)
be one of the facts that will be added to $\I_i$ in order to obtain $\I_{i+1}$.
Since $\action_{i} = \action'_{i}$,
the fact $R(\sigma'_i(u_1),\ldots,\sigma'_i(u_n))$ is also added to $\I'_i$ in order to obtain $\I'_{i+1}$.
If $u_j\in\getnew{\action_i}$ then we have that $\sigma'_i (u_j) = \lambda(\sigma_i(u_j))$.
In the case where $u_j\in\getfree{\action_i}$, since we have that 
i) the symbolic substitutions $s_i$ and $s'_i$ are equal, which free variables parts
are respectively defined by the sequence numbering functions $\seqnof_i$ and $\seqnof'_i$,
and that
ii) $\beta(\seqnof_i(e)) = \seqnof'_i(\lambda(e))$ for every $e\in\adom{\I_i}$,
we deduce that $\sigma'_i(u_j) = \lambda(\sigma_i(u_j))$.
Thus, the added fact amounts to $R(\lambda(\sigma_i(u_1)),\ldots,\lambda(\sigma_i(u_n)))$.
The case of deleted facts is simpler, and the inverse reasoning (from $\I'_i$ to $\I_{i}$) is identical.
Thus $\lambda$ is also an isomorphism from $\I_{i+1}$ onto $\I'_{i+1}$.

\section{Generality of the Model}

We discuss the generality of the model. In particular, we consider
variants of {\sys}s that supports:
\begin{itemize}
\item constant values;
\item (SQL-like) \emph{standard variable substitution} for
  guards and input variables, consequently tackling the repetition of
  matching values;
\item Weakening the freshness requirement for input
  values, to support the possibility of matching input
  variables with already existing values.
\end{itemize}
In summary, we show that all such variants can be reduced back to the
standard model presented in Section~\ref{sec:dms-definition}, while preserving the
original system behavior.

\subsection{Constant Removal}\label{constant:less}

Let $\s = \tup{\idb^{\s},\act^{\s}}$ be a \sys defined over the data domain $\Delta$,
the constant data domain $\Delta_0$ and the schema $\schema^{\s}$.

In the following, we show first how to construct from $\s$ a constant-free \textit{compacted} \sys $\s' = \tup{\idb^{\s'},\act^{\s'}}$
defined over the data domain $\Delta'=\Delta\setminus\Delta_0$ and the schema $\schema^{\s'}$.
Then, we show that the \syss $\s$ and $\s'$ are bisimilar,
i.e. their respective configuration graphs $\cgraph_\s$ and $\cgraph_{\s'}$ are isomorphic.

\paragraph{Constructing the constant-free \sys}
In this part we show how to construct the \textit{compacted} constant-free \sys $\s'$.
To do so, we start by rewriting the set of relations from $\schema^{\s}$ defined over $\Delta$
into a set of \textit{compacted} relations $\schema^{\s'}$
defined over the set of non-constant elements $\Delta'=\Delta\setminus\Delta_0$.
Later on, we show how to express facts and database instances from $\dbset{\schema}{\Delta}$
into facts and database instance from $\dbset{\schema'}{\Delta'}$.
Finally, we show how to rewrite the actions from $\act^{\s}$ so that their guards and $\del$ and $\add$ facts can be expressed
in terms of relations from $\schema'$ and elements from $\Delta'$.

\textbf{Compacting relations.}
Let $\rell/\arity\in\schema$ be a relation of non null arity $\arity \ge 1$ and let
$\sigma: \set{1,\ldots,\arity} \rightarrow \Delta_0 \cup \set{-}$ be a mapping.
Here, the symbol $-$ represents a place holder.
Intuitively, $\sigma$ is used to associate to every argument $e_i$ of a fact $\rell(e_1,\ldots,e_\arity)$
either the argument $e_i$ itself, if $\sigma(i) = -$,
or the constant $\sigma(i)\in\Delta_0$ otherwise.
The indices for which the arguments are kept identical are given by the set
$\setcomp{i:1\le i \le \arity }{ \sigma(e_i) = -}$.
Let $\arityb_\sigma$ be the size of that set, i.e. $\arityb_\sigma = |\setcomp{i:1\le i \le \arity }{ \sigma(e_i) = -}|$,
and, assuming that $\arityb_\sigma > 0$,
let $\pos_\sigma$ be the unique bijection from $\setcomp{i:1\le i \le \arity }{ \sigma(e_i) = -}$ onto $\set{i:1\le i \le \arityb_\sigma}$
such that $\pos_\sigma(i) < \pos_\sigma(j)$ for every $i < j$.
We associate to $\rell/\arity$ and to $\sigma$ a new relation of arity ${\arityb_\sigma}$, denoted by $\rell^\sigma/\arityb_\sigma$ and defined as follows.
Let $\mathsf{e'} = \tup{e'_1,\ldots,e'_{\arityb_\sigma}}\in({\Delta'}\cup\datavars)^{\arityb_\sigma}$ be a tuple of non constant values and variables.
The fact $\rell^{\sigma}(e'_1,\ldots,e'_{\arityb_\sigma})$ of the new relation $\rell^\sigma/\arityb_\sigma$
on the tuple $\mathsf{e'} = \tuple{e'_1,\ldots,e'_{\arityb_\sigma}}$
is defined as the fact $\rell(e_1,\ldots,e_\arity)$ of $\rell/\arity$ on the tuple
$\mathsf{e} = \tup{e_1,\ldots,e_\arity}$ by: for every $i:1\le i \le \arity$,
$e_i := e'_{\pos_\sigma(i)}$ if $\sigma(i) = -$ and $e_i := \sigma(i)$ otherwise.
In other words, the arguments of $\rell^\sigma/\arityb_\sigma$ are mapped to the arguments of $\rell/\sigma$
that are specified by $\sigma$ as being place holders,
while the rest of the arguments of $\rell/\arity$ are constant values from $\Delta_0$ given by $\sigma$.
For instance, if we consider a ternary relation $\rell/3$, the set of constants $\Delta_0 = \set{c_1, c_2}$
and the mapping $\sigma: \set{1,2,3} \rightarrow \Delta_0 \cup \set{-}$
defined by $\sigma(1) := -$, $\sigma(2) := c_2$ and $\sigma(3) := -$,
then the corresponding compacted relation $\rell^\sigma$ is the binary relation
defined by $\rell^{\sigma}(e_1,e_2) := \rell(e_1,c_2,e_2)$ for any $(e_1,e_2)\in({\Delta'}\cup\datavars)^2$.

We associate to $\rell/\arity$ the set of \textit{compacted relations}
$\compactrel{\rell/\arity} := \setcomp{\rell^{\sigma}}{\sigma: \set{1,2,\ldots,\arity} \rightarrow \Delta_0 \cup \set{-}}$.
In the case where $\arity = 0$ (i.e. $\rell/\arity$ is a nullary relation), the set of compacted relations derived from $\rell$ is defined as $\rell$ itself,
i.e. $\compactrel{\rell/\arity} := \set{\rell/\arity}$.
Finally, we define $\schema^{\s'}$ to be the union of the sets of compacted relations:
$\schema^{\s'} := \cup_{\rell/\arity \in \schema^\s} \compactrel{\rell/\arity}$.
%

\textbf{Rewriting facts.}
The purpose here is to describe how to rewrite facts defined using relations from $\schema^{\s}$ over values from $\Delta$
into facts defined using relations from $\schema^{\s'}$ over values from $\Delta' = \Delta \setminus \Delta_0$,
and vice et versa.

Let $\fakt= \rell(e_1,\ldots,e_\arity)$ be a fact defined using a relation $\rell/\arity$ from $\schema^{\s}$
and a tuple $\mathsf{e} = \tup{e_1,\ldots,e_\arity}\in\pair{\Delta\cup\datavars}^\arity$ of values and variables.
We associate to $\fakt$ the mapping
$\sigma_\fakt:\set{1,\ldots,\arity} \rightarrow \Delta_0 \cup \set{-}$
defined by $\sigma_{\fakt}(i) = e_i$ if $e_i$ happens to be a constant value, i.e. if $e_i\in\Delta_0$, and
$\sigma_{\fakt}(i) = -$ if not (i.e. if $e_i$ is a variable or a value from ${\Delta'}$).
We define the \textit{compacted fact} of $\fakt$
to be the fact $\rell^{\sigma_{\fakt}}(e'_1,\ldots,e'_{\arityb})$,
where $\tup{e'_1,\ldots,e'_{\arityb}} = \tup{e_i|\sigma_\fakt(e_i)\notin\Delta_0}$,
and we denote it by $\compactfact{\fakt}$.

Furthermore, the reverse transformation is well defined and we call it \textit{fact expansion}.
More precisely, let $\rell^{\sigma}(e'_1,\ldots,e'_\arityb)$ be a compact fact such that
$\rell/\arity\in\schema$, $\sigma\in\pair{\Delta_0 \cup \set{-}}^{\set{1,\ldots,\arity}}$,
$\arityb = |\setcomp{i:1\le i \le \arity}{\sigma(i) = -}|$
and $\textsf{e'} = \tup{e'_1, \ldots, e'_{\arityb}}\in{\Delta'\cup\datavars}^\arityb$.
We use  $\pos_\sigma$ to denote the unique bijection from $\setcomp{i:1\le i \le \arity }{ \sigma(e_i) = -}$
onto $\set{i:1\le i \le \arityb}$ such that $\pos_\sigma(i) < \pos_\sigma(j)$ for every $i < j$.
The expanded fact is denoted by $\expandfact{r^{\sigma_{\fakt}}(e'_1,\ldots,e'_{\arityb})} := \rell(e_1,\ldots,e_\arity)$
and is defined by
$e_i = \sigma(i)$ if $\sigma(i)\in\Delta_0$
and $e_i = e'_{\pos_\sigma(i)}$ if $\sigma(i) = -$.

Finally note that in both cases we have that
$\tup{e'_1,\ldots,e'_{\arityb}} = \tup{e_1, \ldots, e_\arity}\setminus\pair{\Delta_0\cup\datavars}$.

\textbf{Compacting database instances.}
Let $\I\in\dbset{\schema^{\s}}{\domain}$ be a database instance.
Let $\fakt = \rell(e_1,\ldots,e_\arity)$ be a fact from $\I$.
As mentioned in the previous paragraph,
the compacted fact $\compactfact{\fakt} = r^{\sigma_{\fakt}}(e'_1,\ldots,e'_{\arityb_\sigma})$
is defined using the relation $\rell^{\sigma_{\fakt}}$ from $\schema'$
and the tuple $\tup{e'_1,\ldots,e'_{\arityb_\sigma}} = \tup{e_i|\sigma_\fakt(e_i)\in{\Delta'}}$
of elements from ${\Delta'}$.
Therefore, we can associate to $\I$ the database instance $\I'\in\dbset{\schema^{\s'}}{\Delta'}$
built using relations from the schema $\schema^{\s'}$ with elements from the set of non constant values $\Delta'$
and defined by $\setcomp{\compactfact{\fakt}}{\fakt\in\I}$.
We call this database instance the \textit{compacted} database instance and we denote it by $\compactdb{\I}$.
In particular, we define $\idb^{\s'} := \compactdb{\idb^{\s}}$.
Notice that $\ADOM{\compactdb{\I}} = \ADOM{\I} \setminus \Delta_0$.

The reverse operation, that we call \textit{database expansion}, is defined by
$\expanddb{\I'} :=  \setcomp{\expandfact{\fakt}}{\fakt\in\I'}$ for every database instance $\I'\in\dbset{\schema'}{\Delta'}$.
Notice that the operation is well defined and that
$\ADOM{\expanddb{\I'}} = \ADOM{\I'} \cup \Delta_0$ for every $\I'\in\dbset{\schema'}{\Delta'}$.

\textbf{Rewriting actions}.
The first purpose of this paragraph is to show how to rewrite actions from $\act^{\s}$
so that:
\begin{itemize}
\item bound variables used in the action guard are quantified over non-constant values from $\Delta$;
\item every free variable is either fixed to be a constant from $\Delta_0$
or is constrained by the guard to not be evaluated to a constant;
\item facts from the guard, the deletion and the addition parts of the action are written using compacted relations;
\end{itemize}

The second purpose of this paragraph is to show that the \sys $\s'$ is well defined.

Let $\actiona$ be an action from $\act^{\s}$.
In the following,
we use $\getbound{\actiona}$ to denote the set of bound variables that appear in the guard of action $\actiona$,
we use $u\notin\Delta_0$ to denote the variable constraint defined by $\bigwedge_{c\in\Delta_0} \lnot(u = c)$
and we use $\actiona_{\set{u_1/e_1,\ldots,u_m/e_m}}$ to denote the action obtained from $\actiona$ after
replacing every variable $u_i$ with the expression $e_i$.

\begin{figure*}
\begin{gather*}
\left(
\bigcup_{\actiona\in\act^{\s} \land |\getfree{\actiona}|>0}
\setcomp{\compactact{\fixof{\boundof{\actiona},\cons}}}
{\cons \in {\Delta_0 \cup \set{-}}^{\getfree{\actiona}}}
\right)
\\\bigcup
\left(
\bigcup_{\actiona\in\act^{\s} \land |\getfree{\actiona}| = 0}
\set{\compactact{\boundof\actiona}}
\right)
\end{gather*}
\caption{New set of actions $\act^{\s'}$}
\label{set:actions:compact}
\end{figure*}

The set of constants $\Delta_0$ being a finite set,
we have that:
\begin{itemize}
\item every $\folang$ expression of the form $\forall u.  Q(u)$ is equivalent to
$\left(\forall u. {u\notin\Delta_0 \Rightarrow Q(u)}\right) \land \bigwedge_{c\in\Delta_0} Q(c)$; and
\item every $\folang$ expression of the form $\exists u.  Q(u)$ is equivalent to
$\left(\exists u. {u\notin\Delta_0 \land Q(u)}\right) \lor \bigvee_{c\in\Delta_0} Q(c)$;
\end{itemize}
Following this schema, the guard of action $\actiona$ is rewritten for every bound variable $u\in\boundvars{\actiona}$.
The obtained action $\boundof{\actiona}$ is equivalent to $\actiona$ and
is such that every bound variable $u\in\boundvars{\actiona'}$ satisfies $u\notin\Delta_0$.
Moreover, the length of the rewritten guard $\getguard{\boundof{\actiona}}$ is at most $|\Delta_0|^{|\boundvars{\actiona}|}$ times
the size of the original guard $\getguard{\actiona}$.
Finally, notice that $\getfree{\actiona} = \getfree{\boundof\actiona}$.

Assume that $\getfree{\actiona} \ne \emptyset$,
and let $\cons: \getfree{\actiona} \rightarrow \Delta_0 \cup \set{-}$
be a mapping that associates to each free variable of $\actiona$
either a constant from $\Delta_0$ or the placeholder symbol $-$.
Using this mapping, we transform action $\boundof\actiona$ into a new action denoted by $\fixof{\boundof\actiona,\cons}$.
The new action is obtained after performing the following transformations for every free variable $u$ of $\actiona$.
If $\cons$ maps $u$ to a constant (i.e. $\cons(u)\in\Delta_0$),
then every occurrence of $u$ in $\boundof\actiona$ is replaced with its constant valuation. 
Otherwise, if $\cons$ maps $u$ to the placeholder $-$,
then $u$ is constrained to not be a constant.
Formally, we have that
$\getguard{\fixof{\boundof{\actiona},\cons}} := \getguard{\subof{\boundof{\actiona},\sigma}} \land
\bigwedge_{u\in\setcomp{\getfree{\boundof{\actiona}}}{\cons(u) = -}} u\notin\Delta_0$,
$\getdel{\fixof{\boundof{\actiona},\cons}} := \getdel{\subof{\boundof{\actiona},\sigma}}$ and
$\getadd{\fixof{\boundof{\actiona},\cons}} := \getadd{\subof{\boundof{\actiona},\sigma}}$,
where
$\subof{\boundof{\actiona},\sigma} := a_{\setcomp{u/\cons(u)}{u\in\getfree{\boundof{\actiona}} \land \cons(u)\in\Delta_0}}$.
Finally, we perform the transformation that consists in replacing every fact
$\fakt = \rell(e_1,\ldots,e_\arity)$ that appear in $\fixof{\boundof{\actiona},\cons}$ by its compacted version $\compactfact{\fakt}$.
The obtained action $\actiona'$ is denoted by $\compactact{\fixof{\boundof{\actiona},\cons}}$.
Notice that:
\begin{itemize}
\item $\getfree{\actiona'} = \tuple{u\in\getfree\actiona|\cons(u)=-}$,
\item $\getnew{\actiona'} = \getnew\actiona$
\item If there exist $\sigma:\getfree{\actiona}\uplus\getnew{\actiona}\rightarrow\Delta$
and a database instance $\I\in\dbset{\schema^{\s}}{\domain}$
such that
$\I,\sigma \models \getguard{\actiona}$
and
$\sigma(u) = \cons(u)$ for every $u\in\tuple{u\in\getfree{\action}|\cons(u)\in\Delta_0}$,
then we have that
$\compactdb{\I},\sigma|_{\getfree{\actiona'}\uplus\getnew{\actiona}} \models \getguard{\actiona'}$
\end{itemize}

If $\getfree{\actiona} = \emptyset$ then $\getfree{\boundof{\actiona}} = \emptyset$
and the set of compacted actions derived from $\actiona$ is defined as the singleton $\set{\compactact{\boundof{\actiona}}}$
which contains the action that we obtain from $\boundof\actiona$ after replacing every fact $\fakt = \rell(e_1,\ldots,e_\arity)$
that appear in $\actiona$ by its compacted version $\compactfact{\fakt}$.
In this case we have that:
\begin{itemize}
\item $\getfree{\actiona'} = \getfree\actiona = \emptyset$,
\item $\getnew{\actiona'} = \getnew\actiona$,
\item If there exist $\sigma:\getnew{\actiona}\rightarrow\Delta$
and a database instance $\I\in\dbset{\schema^{\s}}{\domain}$
such that
$\I,\sigma \models \getguard{\actiona}$,
then we have that
$\compactdb{\I},\sigma \models \getguard{\actiona'}$.
\end{itemize}

We define $\act^{\s'}$ to be the union of actions obtained from the actions in
$\act^{\s}$ by compacting them after fixing subsets of their free variables to constant values
(see formula in Figure~\ref{set:actions:compact}).

Note that any relational fact $\rell(e_1,\ldots,e_m)$ involved in the definition of any action $\actiona$ from $\act^{\s'}$
is defined only by using a relation $\rell$ from $\schema^{\s'}$
over a tuple $e = \tup{e_1,\ldots,e_m}$ of elements from ${\Delta'}$ and variables from $\datavars$.

Moreover, all free variables used in those actions are constrained to not belong to the set of constant values.
Therefore, facts that could possibly be added to a database instance after the execution of an action $\actiona$ from  $\act^{\s'}$
are facts defined only by using a relation $\rell$ from $\schema^{\s'}$ over a tuple $e = \tup{e_1,\ldots,e_m}$
of values from ${\Delta'}$, fresh values or non-constant valuations of free variables.

Thus, $\s' = \tup{\idb^{\s'},\act^{\s'}}$ is a well defined \sys
over the set of non constant values $\Delta'$ and the compacted schema $\schema^{\s'}$.

\paragraph{Bi-similarity between the two configuration graphs}
We show in this second part of the proof that the two configuration graphs $\cgraph_\s$ and $\cgraph_\s'$,
respectively generated by the \sys $\s = \tup{\idb^\s,\act^\s}$ over $\Delta$ and $\schema^{\s}$
and by the \sys $\s' = \tup{\idb^{\s'},\act^{\s'}}$ over $\Delta'$ and $\schema^{\s'}$,
where $\schema^{\s'}$, $\idb^{\s'}$ and $\act^{\s'}$ are defined as above,
are isomorphic.
To do so, we proceed by:
\begin{itemize}
\item defining a bijection between the set of configurations of $\s$
and the set of configurations of $\s'$;
\item showing that $\cgraph_\s'$ simulates $\cgraph_\s$; and
\item showing that $\cgraph_\s$ simulates $\cgraph_\s'$;
\end{itemize}

\textbf{Bijection.}
The respective sets of configurations of the \syss $\s$ and $\s'$
are given by $\dbset{\schema}{\Delta} \times 2^{\Delta}$
and 
$\dbset{\schema'}{\Delta'} \times 2^{\Delta'}$.
Let $\isoconf:
\dbset{\schema}{\Delta} \times 2^{\Delta}
\rightarrow
\dbset{\schema'}{\Delta'} \times 2^{\Delta'}$
be the mapping defined by
$\isoconfof{\tup{\I, \history}} := \tup{\compactdb{\I}, \history\setminus\Delta_0}$.
Note that since $H\setminus\Delta_0\subseteq\Delta'$
and $\compactdb{\I}\in\dbset{\schema'}{\Delta'}$,
$\isoconf$ is well defined.
We show that $\isoconf$ is a bijection.

\textit{Injectivity.} Let $\tup{\I_1, \history_1}$ and $\tup{\I_2, \history_2}$ be two configurations
from $\dbset{\schema}{\Delta} \times 2^{\Delta}$.
Assume that $\isoconfof{\tup{\I_1, \history_1}} = \isoconfof{\tup{\I_2, \history_2}}$.
This is equivalent to $a)$ $\compactdb{\I_1} = \compactdb{\I_2}$
and $b)$ $\history_1\setminus\Delta_0 = \history_2\setminus\Delta_0$.
We already know that $\Delta_0\subseteq\history_1$ and that $\Delta_0\subseteq\history_2$.
Together with $b)$, this implies that $\history_1 = \history_2$.
Moreover, $a)$ implies that $\setcomp{\compactfact{\fakt}}{\fakt\in\I_1} = \setcomp{\compactfact{\fakt}}{\fakt\in\I_2}$.
Compacting facts is a bijective operation, which reverse consists in expanding them.
We therefore deduce that $\I_1 = \I_2$.
Thus, $\isoconf$ is injective.

\textit{Surjectivity.} Let $\tup{\I', \history'}$ be a configuration from $\dbset{\schema'}{\Delta'} \times 2^{\Delta'}$.
We set $\I := \setcomp{\expandfact{\fakt}}{\fakt\in\I'}$ and $\history = \history'\cup\Delta_0$
and we have that $\isoconfof{\tup{\I,\history}} = \tup{\I',\history'}$.
Thus, $\isoconf$ is surjective.

We deduce that $\isoconf$ is a bijection.

\textbf{The graph $\cgraph_\s'$ simulates $\cgraph_\s$}.
Let $\confc_1 = \tup{\I_1,\history_1},\confc_2 = \tup{\I_2,\history_2}$
be two configurations defined over $\s$,
i.e. $\confc_1,\confc_2\in\dbset{\schema}{\Delta}\times2^\Delta$,
let $\actiona\in\act^{s}$ be an action, $\vec{u}_{\actiona} = \getfree{\actiona}$, $\vec{v}_{\actiona} = \getnew{\actiona}$, 
and let $\sigma \in \Delta^{\vec{u}_{\actiona}\uplus\vec{v}_{\actiona}}$ be a substitution such that
$\gotos{\tup{\I_1,\history_1}}{\tup{\I_2,\history_2}}{\actiona:\sigma}$.
Let $\confc'_1 = \tup{\I'_1,\history'_1}, \confc'_2 = \tup{\I'_2,\history'_2}\in\dbset{\schema'}{\Delta'}\times2^{\Delta'}$
be the respective images of $\confc_1$ and $\confc_2$, i.e.
$\tup{\I'_1,\history'_1} := \tup{\compactdb{\I_1}, \history_1\setminus\Delta_0}$
and
$\tup{\I'_2,\history'_2} := \tup{\compactdb{\I_2}, \history_2\setminus\Delta_0}$.
We shall prove that there exists an action $\actiona'\in\act^{s'}$ and a substitution $\sigma'\in\Delta'^{\getfree{{\actiona'}}\uplus\getnew{{\actiona'}}}$
such that $\gotos{\tup{\I'_1,\history'_1}}{\tup{\I'_2,\history'_2}}{\actiona':\sigma'}$.

Assume that $|\getfree{\actiona}| > 0$.
We can then define the mapping $\cons:\getfree{\actiona}\rightarrow\Delta_0\cup\set{-}$ given by
$\cons(u) = \sigma(u)$ if $\sigma(u)$ is a constant value, i.e. if $\sigma(u)\in\Delta_0$,
and $\cons(u) = -$ otherwise.
Let $\actiona' = \compactact{\fixof{\boundof\actiona,\cons}}$,
$\vec{u}_{\actiona'} = \getfree{\actiona'}$ and
$\vec{v}_{\actiona'} = \getnew{\actiona'}$.
Note that
$\vec{u}_{\actiona} = \getfree{\actiona} = \getfree{\boundof{\actiona}}$ and
$\vec{v}_{\actiona} = \getnew{\actiona} = \getnew{\boundof{\actiona}}$.
Since $\vec{u}_{\actiona'} = \tup{u\in\vec{u}_{\actiona}|\cons(u)=-} = \tup{u\in\vec{u}_{\actiona}|\sigma(u)\notin\Delta_0}$
and since $\vec{v}_{\actiona'} = \vec{v}_{\actiona}$
are such that $\sigma(v)\notin\Delta_0$ for any $v\in\vec{v}_{\actiona}$,
we can define the mapping $\sigma': \vec{u}_{\actiona'}\uplus\vec{v'}_{\actiona} \rightarrow \Delta'$
as the restriction of $\sigma$ to $\vec{u}_{{\actiona'}}\uplus\vec{v}_{{\actiona'}}$ on $\Delta' = \Delta\setminus\Delta_0$,
i.e. $\sigma' := \sigma|_{\vec{u}_{\actiona'}\uplus\vec{v}_{\actiona'}}$.
We can check that
$\gotos{\tup{\I'_1,\history'_1}}{\tup{\I'_2,\history'_2}}{a':\sigma'}$.
In fact:
\begin{itemize}
\item for every $u_i\in\vec{u}_{\actiona'}$, $\sigma'(u_i)=\sigma(u_i)\in\ADOM{\I_1}$. Since $\sigma'(u_i)\notin\Delta_0$,
we deduce that $\sigma'(u_i)\in\ADOM{\I_1}\setminus\Delta_0 = \ADOM{\compactdb{\I_1}} = \ADOM{\I'_1}$;
\item for every variable $v_i\in\vec{v}_{\actiona'}$, $\sigma'(v_i) = \sigma(v_i)\notin\history_1$. Since $\history'_1 = \history_1 \setminus \Delta_0$,
we deduce that $\sigma(v_i)\notin\history_1$;
\item $\sigma'|_{\vec{v}_{\actiona'}}$ being the partial mapping of an injective function is also injective;
\item $\history'_2 = \history_2\setminus\Delta_0 = (\history_1 \cup \ADOM{\I_2})\setminus\Delta_0
=(\history_1\setminus\Delta_0) \cup (\ADOM{\I_2}\setminus\Delta_0)
=\history'_1 \cup \ADOM{\I'_2}$;
\item $\I'_1,\sigma'|_{\vec{u}_{\actiona'}}  \models \getguard{\actiona'}$ holds because $\actiona'$ consists in re-writing action $\actiona$
by using relations from $\schema'$ instead of $\schema$,
after replacing variables from $\tup{u_i\in\vec{u}_{\actiona}|\cons(u_i)\in\Delta_0}$ by their constant $\sigma$-valuation
$\sigma(u)=\cons(u)\in\Delta_0$;
\item We have that
\begin{align*}
\I'_2 =& \compactdb{\I_2} \\
=& \compactdb{
	( \I_1 - \substDB{\sigma}{\getdel{\actiona}}}\\
&	+ \substDB{\sigma}{\getadd{\actiona}}
	) \\
=& (
	\compactdb{\I_1}- \\
	&\compactdb{\substDB{\sigma}{\getdel{\actiona}}}) \\
& + \compactdb{\substDB{\sigma}{\getadd{\actiona}}} \\
=& (\I'_1- \substDB{\sigma'}{\getdel{\actiona'}}) + \substDB{\sigma'}{\getadd{\actiona'}}
\end{align*}
\end{itemize}

Otherwise, we have that $|\getfree\actiona| = 0$.
In that case, if $\actiona' = \compactact{\boundof\actiona}$,
then we have that $\getfree\actiona' = \getfree\actiona = \emptyset$
and $\getnew\actiona' = \getnew\actiona$
and $\sigma'$ is defined as $\sigma$ it self.

\textbf{The graph $\cgraph_\s$ simulates $\cgraph_\s'$}.
Suppose now that
$\confc'_1 = \tup{\I'_1,\history'_1},\confc'_2 = \tup{\I'_2,\history'_2}\in\dbset{\schema'}{\Delta'}\times2^{\Delta'}$
are two configurations of $\s'$,
that $\actiona'\in\act^{s'}$ is an action such that $\vec{u}_{\actiona'} = \getfree{\actiona'}$ and $\vec{v}_{\actiona'} = \getnew{\actiona'}$, and let
$\sigma' \in \Delta'^{\vec{u}_{\actiona'}\uplus\vec{v}_{\actiona'}}$ be a substitution such that
$\gotos{\tup{\I'_1,\history'_1}}{\tup{\I'_2,\history'_2}}{a':\sigma'}$.

We set $\I_1 := \expanddb{\I'_1}$,
$\I_2 := \expanddb{\I'_2}$,
$\history_1 = \history'_1 \cup \Delta_0$ and
$\history_2 = \history'_2 \cup \Delta_0$.

Since $\actiona'\in\act^{\s'}$, there are two cases.

First case is that there exists an action $\actiona\in\act^{\s}$ such that $\actiona' = \compactact{\boundof\actiona}$.
In that case $u_{\actiona'} = \getfree{\actiona'} = \getfree{\actiona} = \emptyset$.
Moreover, since $\actiona$ and $\actiona'$ share the same set of fresh variables,
the mapping $\sigma:\getfree{\actiona}\uplus\getnew{\actiona}\rightarrow\Delta$ defined by
$\sigma(v) = \sigma'(v)$ for every $v\in\getnew{\actiona}$
is well defined and is such that:
\begin{itemize}
\item for every $u_i\in\getfree{\actiona} =\emptyset $, $\sigma(u_i)\in\ADOM{\I_1}$ (vacuous truth);
\item for every $v_i\in\getnew{\actiona}$, $\sigma(v_i) = \sigma'(v_i)\notin\history'_1$.
Since $\history'_1 = \history_1\setminus\Delta_0$
and $\sigma'(v_i)\notin\Delta_0$ (because $\Delta'=\Delta\setminus\Delta_0$),
we deduce that $\sigma(v_i) \notin\history_1$;
\item $\sigma|_{\getnew{\actiona}} = \sigma = \sigma' = \sigma'|_{\getfree{\actiona'}}$ is injective;
\item we have that:
\begin{align*}
\history_2 =& \history'_2 \cup \Delta_0\\
=& \history'_1 \cup \ADOM{\I'_1} \cup \Delta_0\\
=& \pair{\history_1\setminus\Delta_0} \cup \pair{\ADOM{\I_1}\setminus\Delta_0} \cup \Delta_0\\
=& \history_1 \cup \ADOM{\I_1}
\end{align*}
\item $\I_1,\sigma|_{\vec{u}}  \models \getguard{\actiona}$ holds because $\actiona$ consists in re-writing action $\actiona'$
by expanding relations from $\schema'$ to relations from $\schema$;
\item we have that
\begin{align*}
\I_2 =& \setcomp{\expandfact{\fakt}}{\fakt\in\I'_2} \\
=& \{\expandfact{\fakt}|
\fakt\in\left(\I'_1 - \substDB{\sigma'}{\getdel{\actiona'}}\right)\\
&+ \substDB{\sigma'}{\getadd{\actiona'}}\}\\
=&
\left(\setcomp{\expandfact{\fakt}}{\fakt\in\I'_1}\right.
-\\
&\left.\setcomp{\expandfact{\fakt}}{\fakt\in\substDB{\sigma'}{\getdel{\actiona'}}}\right)
+\\
&\setcomp{\expandfact{\fakt}}{\fakt\in\substDB{\sigma'}{\getadd{\actiona'}}}\\
=&
(\I_1 - \substDB{\sigma}{\getadd{\actiona}}) + \substDB{\sigma}{\getadd{\actiona}}
\end{align*}
\end{itemize}
Thus, $\gotos{\tup{\I_1,\history_1}}{\tup{\I_2,\history_2}}{\actiona:\sigma}$.

In the second case, there exists an action $\actiona\in\act^{\s}$ and a mapping
$\cons:\getfree{\actiona} \rightarrow \Delta_0 \cup \set{-}$
such that $\actiona' = \compactact{\fixof{\boundof\actiona,\cons}}$.
The very same reasoning goes for this case, with the particularity that
$\sigma: \getfree{\actiona}\uplus\getnew{\actiona}\rightarrow\Delta$
is defined as follows:
$\sigma(u) = \sigma'(u)$ for every $u\in\setcomp{\getfree{\actiona}}{\cons(u) = -} = {\getfree{\actiona'}}$,
$\sigma(u) = \cons(u)$ for every $u\in\setcomp{\getfree{\actiona}}{\cons(u)\in\Delta_0}$ and 
$\sigma(v) = \sigma'(v)$ for every $v\in\getnew{\actiona} = \getnew{\actiona'}$;
and we have that
$\gotos{\tup{\I_1,\history_1}}{\tup{\I_2,\history_2}}{\actiona:\sigma}$.

\begin{example}
Let $\s  = \tup{
\idb^{\s},
\set{\actiona, \actionb}
}$
be a \sys  defined over a data domain $\Delta$ containing the set of constants $\set{c_1, c_2}$
and over the schema $\set{\rel{R}/2, \rel{Q}/1}$,
with
$\idb^{\s} = \set{\rel{R}(c_1, c_2), \rel{Q}(c_1)}$,
$\actiona = \tup{\set{u}, \emptyset, \rel{R}(u,u), \set{\rel{R}(u,u)}, \set{\rel{Q}(u)}}$ and
$\actionb = \tup{\emptyset,\set{v},\textsf{True}, \emptyset, \set{\rel{R}(v,v)}}$.

Then the constant-free \sys $\s'$ is defined as the tuple $\tup{\idb^{\s'},\act^{\s'}}$
over the constant-free domain $\Delta' = \Delta\setminus\set{c_1,c_2}$ and over the
the schema $\schema^{\s'}$ by:
\begin{itemize}
\item $\schema^{\s'} = \schema^{\s'}_1  \cup \schema^{\s'}_2$
with
$\schema^{\s'}_1 = \{
\rel{R}^{\sigma_{1.1}}/0 = \rel{R}(c_1, c_1),
\rel{R}^{\sigma_{1.2}}/0 = \rel{R}(c_1, c_2),
\rel{R}^{\sigma_{1.3}}/1 = \rel{R}(c_1, -),
\rel{R}^{\sigma_{1.4}}/0 = \rel{R}(c_2, c_1),
\rel{R}^{\sigma_{1.5}}/0 = \rel{R}(c_2, c_2),
\rel{R}^{\sigma_{1.6}}/1 = \rel{R}(c_2, -),
\rel{R}^{\sigma_{1.7}}/1 = \rel{R}(-, c_1),
\rel{R}^{\sigma_{1.8}}/1 = \rel{R}(-, c_2),
\rel{R}^{\sigma_{1.9}}/2 = \rel{R}(-, -)
\}$, $\schema^{\s'}_2 = \{
\rel{Q}^{\sigma_{2.1}}/0 = \rel{Q}(c_1),
\rel{Q}^{\sigma_{2.2}}/0 = \rel{Q}(c_2),
\rel{Q}^{\sigma_{2.3}}/0 = \rel{Q}(-)
\}
$,
$\setcomp{\sigma_{1.i}}{1 \le i \le 9} =  {\Delta_0 \cup \set{-}}^{\set{1,2}}$
and $\setcomp{\sigma_{2.i}}{1 \le i \le 3} =  {\Delta_0 \cup \set{-}}^{\set{1}}$.
For instance, using the mapping $\sigma_{1.6}: \set{1,2,3} \rightarrow \Delta_0 \cup \set{-}$
defined by $\sigma_{1.6}(1) = c_2$ and $\sigma_{1.6}(2) = -$,
we get the compacted unary relation $\rel{R}^{\sigma_{1.6}}/1$
defined by $\rel{R}^\sigma(e) = \rel{R}(c_2, e)$ for every $e\in\Delta'$.
\item $\idb^{\s'} = \set{\rel{R}^{\sigma_{1.2}}, \rel{Q}^{\sigma_{2.1}}}$.
Both $\rel{R}^{\sigma_{1.2}}$ and $\rel{Q}^{\sigma_{2.1}}$ are nullary relations 
corresponding respectively to the facts $\rel{R}(c_1, c_2)$ and $\rel{Q}(c_1)$.
\item Since $\getfree{\actiona} = \set{u}$ and $\getfree{\actionb} = \emptyset$, the set of actions is then given by:
\begin{align*}
\act^{\s'} 	=& \setcomp{\compactact{\fixof{\actiona,\cons}}}{\cons: {\Delta_0 \cup \set{-}}^{\set{u}}} \\
		& \cup \set{\compactact{\actionb}}\\
		=& \set{\compactact{\fixof{\actiona,\cons_1}},\\
		&\compactact{\fixof{\actiona,\cons_2}},\\
		&\compactact{\fixof{\actiona,\cons_3}}}\\
		& \cup \set{\compactact{\actionb}}
\end{align*}
Where $\cons_1$, $\cons_2$ and $\cons_3$ are defined by 
$\cons_1(u) = c_1$, $\cons_2(u) = c_2$ and $\cons_3(u) = -$. Thus:
\begin{align*}
\act^{\s'} 	=\left\{\right.  &\tup{\emptyset,\emptyset,\rel{R}^{\sigma_{1.1}}, \set{\rel{R}^{\sigma_{1.1}}}, \set{\rel{Q}^{\sigma_{2.1}}}}\\
	&\tup{\emptyset,\emptyset,\rel{R}^{\sigma_{1.5}}, \set{\rel{R}^{\sigma_{1.5}}}, \set{\rel{Q}^{\sigma_{2.2}}}},\\
	&\tup{\set{u},\emptyset,\rel{R}^{\sigma_{1.9}}(u,u), \set{\rel{R}^{\sigma_{1.9}}(u,u)}, \set{\rel{Q}^{\sigma_{2.3}}(u)}},\\
	&\left. \tup{\emptyset,\set{v},\textsf{True}, \emptyset, \set{\rel{R}^{\sigma_{1.9}}(v,v)}} \right\} \\
	=\left\{\right.& \tup{\emptyset,\emptyset,\rel{R}(c_1,c_1), \set{\rel{R}(c_1,c_1)}, \set{\rel{Q}(c_2)}},\\
	& \tup{\emptyset,\emptyset,\rel{R}(c_2,c_2), \set{\rel{R}(c_2,c_2)}, \set{\rel{Q}(c_2)}},\\
	& \tup{\set{u},\emptyset,\rel{R}(u,u), \set{\rel{R}(u,u)}, \set{\rel{Q}(u)}}),\\
	& \left. \tup{\emptyset,\set{v},\textsf{True}, \emptyset, \set{\rel{R}(v,v)}} \,\, \right\}
\end{align*}
\end{itemize}
\end{example}

\subsection{DMSs with Possibly Overlapping Inputs}
\label{gen:injectivity:proof}

According to the \sys execution semantics, the application of an
action is done by substituting the guard answer variables and
by \emph{injectively} substituting the fresh input variables with corresponding values.
We show here that this is not a limitation of the approach,
which can in fact seamlessly account for standard variable substitution, possibly mapping multiple
fresh variables with the same value.

The algorithm in Figure~\ref{alg:non-injective} shows precisely how to build a set of injective actions,
i.e. actions where fresh variables are mapped to different values,
from a set of non-injective actions,
i.e. actions where fresh variables can be mapped to the same value.
A demonstration of the algorithm is given in Example ~\ref{example:injectivity}.

\newcommand{\compllist}{\textsc{complement-list}}
\renewcommand{\gets}{:=}
\newcommand{\actni}{\act_{\mathit{std}}}
\newcommand{\newvar}[1]{#1_{\mathit{n}}}
\algrenewcommand\algorithmicindent{.5em}

\begin{figure}[!t]
\begin{algorithmic}[1]
\small
\Procedure{standard-substitution}{$\act$}
\\\textbf{input:} Set $\act$ of actions, \textbf{output:} Set
$\actni$ of actions
\State $\actni \gets \emptyset$
\ForAll{$\tup{\guard(\vec{u}),\del(\vec{u}),\add(\vec{u},\vec{v})}
  \in \act$}
\ForAll{$\vec{p} = \tuple{\vec{s}_1,\ldots,  \vec{s}_{|p|}  }$ partition of $\vec{v}$}
\State $\add'(\vec{u},\tup{v^{'}_i|1\le i\le |p|}) \gets$\par
	\hskip\algorithmicindent
	\hskip\algorithmicindent
	\hskip\algorithmicindent
	\hskip\algorithmicindent
	\hskip\algorithmicindent
	\hskip\algorithmicindent
	\hskip\algorithmicindent
	\hskip\algorithmicindent
$\add(\vec{u},\vec{v})\textbf{[}v/{v^{'}_i} \textbf{ if } v \in \vec{s}_i| 1\le i \le |p|\textbf{]}$
\State \begin{varwidth}[t]{\linewidth}
	$\actni \gets \actni$ \par
	\hskip\algorithmicindent
	\hskip\algorithmicindent
	\hskip\algorithmicindent
	\hskip\algorithmicindent
	\hskip\algorithmicindent
	\hskip\algorithmicindent
	\hskip\algorithmicindent
	\hskip\algorithmicindent
	\hskip\algorithmicindent
	\hskip\algorithmicindent
	$\cup \set{\tup{\guard(\vec{u}),\del(\vec{u}),\add'(\vec{u},\tup{v^{'}_i|1\le i\le |p|})}}$
	\end{varwidth}
\EndFor
\EndFor
\EndProcedure
\end{algorithmic}
\caption{Procedure for turning a set of actions into another set of
  actions that simulates standard variable substitutions. Notation
  $E(\vec{u})[\vec{u}/\vec{z}]$, where $E$ is a formula or a set of
  facts, indicates $E$ where variables $\vec{u}$ are consistently
  replaced with corresponding variables/constants $\vec{z}$.
  For every partition set element $\vec{s}_i$, all variables belonging to a $\vec{s}_i$
  are replaced with the new fresh variable $v^{'}_i$.
  }
\label{alg:non-injective}
\end{figure}

\begin{example}
Given action
\[
\begin{array}{@{}l@{}l@{}}
\actiona = \langle &\set{u_1,u_2}, \set{v_1,v_2,v_3}, R(u_1,u_2), \\
&\set{Q(u_2)},\set{R(u_2,v_1),R(u_2,v_2),R(u_1,v_3)}\rangle,
\text{ we get:}
\end{array}
\]
\[
\left\{
\begin{array}{l@{}l}
\actiona_1 =& \langle \set{u_1,u_2}, \set{v'_1,v'_2,v'_3}, R(u_1,u_2),\set{Q(u_2)},\\
&\set{R(u_2,v^{'}_1),R(u_2,v^{'}_2),R(u_1,v^{'}_3)}\rangle, \\
\actiona_2 =& \langle \set{u_1,u_2}, \set{v'_1,v'_2}, R(u_1,u_2),\set{Q(u_2)},\\
&\set{R(u_2,v^{'}_1),R(u_2,v^{'}_2),R(u_1,v^{'}_2)}\rangle, \\
\actiona_3 =& \langle \set{u_1,u_2}, \set{v'_1,v'_2}, R(u_1,u_2),\set{Q(u_2)},\\
&\set{R(u_2,v^{'}_2),R(u_2,v^{'}_2),R(u_1,v^{'}_1)}\rangle, \\
\actiona_4 =& \langle \set{u_1,u_2}, \set{v'_1,v'_2}, R(u_1,u_2),\set{Q(u_2)},\\
&\set{R(u_2,v^{'}_2),R(u_2,v^{'}_1),R(u_1,v^{'}_2)}\rangle, \\
\actiona_5 =& \langle \set{u_1,u_2}, \set{v'_1}, R(u_1,u_2),\set{Q(u_2)},\\
&\set{R(u_2,v^{'}_1),R(u_2,v^{'}_1),R(u_1,v^{'}_1)}\rangle
\end{array}
\right\}
\]
to interpret the original action using standard variable substitutions
for fresh input variables $v_1, v_2, v_3$.
In action $\actiona_2$ for instance, $v'_1$ replaces the subset $\tuple{v_1}$, while $v'_2$ replaces the subset $\tuple{v_2, v_3}$.
This replacement corresponds to the partition $\vec{p} = \tuple{ \vec{s}_1 = \tuple{v_1}, \vec{s}_2 =  \tuple{v_2,v_3} }$.
\label{example:injectivity}
\end{example}

\subsection{Weakening Freshness}
\label{gen:freshness:proof}

One may argue that inputs provided to a \sys may not
necessarily be fresh.
In fact, there may be cases in which a \sys
action is meant to establish new relations among already existing
values, but still interacting with the external world to decide
which. We call \emph{arbitrary-input \sys} a \sys that does not
necessarily require the input variables to be assigned to fresh values.
We provide in what follows a proof and an example illustrating this remark.

\begin{proof}
Let  $\s = \tup{\idb,\act}$ be an arbitrary-input \sys defined over the data domain $\Delta$ and the schema $\schema$.
We produce a corresponding standard \sys $\s_{fresh} =  \tup{\idb,\act'}$,
defined over the same data domain and over an extended schema $\schema'$,
such that:
\begin{itemize} 
\item the schema of the $\schema' = \schema\cup\set{\rel{Hist}/1}$ is the extension of the schema of $\s$
by adding the unary relation $\rel{Hist}$, which role is to store all the values seen in during the run of the \sys,
\item the set of actions $\act'$ of the standard \sys $\s'$
is defined as the smallest set satisfying the following property:
for every arbitrary-input action $\tuple{\vec{u}, \vec{i}, \guard\left(\vec{u}\right),\del\left(\vec{u}\right),\add\left(\vec{u},\vec{i}\right)}\in\act$
with abitrary-input variables $\vec{i}$,
and for every possible binary partition $\vec{h}\uplus\vec{f}$ of the set of input variables $\vec{i}$,
$\act'$ contains the standard \sys action $\tup{\vec{u}\uplus\vec{h},\vec{f},\guard'(\vec{u},\vec{h}),\del(\vec{u}),\add'(\vec{u},\vec{h}, \vec{f})}$, where
\begin{itemize}
\item
$\guard'(\vec{u},\vec{h}) = \guard(\vec{u})
	\land
	\bigwedge_{h\in\vec{h}} \rel{Hist}(h)$ and
\item
$\add'(\vec{u},\vec{h}, \vec{v})=\add(\vec{u}, \vec{h}\uplus\vec{f}) \cup \setcomp{\rel{Hist}(f)}{f\in\vec{f}}$
\end{itemize}
\end{itemize}

Thus, every action
with arbitrary-input variables $\vec{i}$ 
is translated into $2^{|\vec{i}|}$ actions, each one
handling the case in which a subset of the uniform input variables are
mapped to fresh values, while the remaining ones are bound to values
present in the history of the run.
It is easy to see that the configuration graph obtained from $\s$ by
removing the requirement that input variables must match with fresh
values, and the standard configuration graph of $\s'$, indeed coincide.
\end{proof}

\begin{example}
Given schema $\schema = \set{R/2,Q/1}$, we replace the arbitrary input action
\[
\tup{\set{u_1,u_2},\set{i_1,i_2},R(u_1,u_2),\set{Q(u_2)},\set{R(u_2,i_1),R(u_2,i_2)}}
\]
with the standard (fresh) input set of actions
\[
\left\{
\begin{array}{@{}l@{}l@{}}
\actiona_1= \langle \set{u_1,u_2},\set{f_1,f_2},R(u_1,u_2),\\
\,\,\,\,\,\,\,\,\set{Q(u_2)},\set{R(u_2,f_1),R(u_2,f_2), \rel{Hist}(f_1), \rel{Hist}(v_2)}\rangle\\
\actiona_2= \langle\set{u_1,u_2,h_1},\set{f},R(u_1,u_2) \land \rel{Hist}(h),\\
\,\,\,\,\,\,\,\,\set{Q(u_2)}, \set{R(u_2,h), R(u_2,f), \rel{Hist}(f)}\rangle\\
\actiona_3= \langle \set{u_1,u_2,h_1,h_2},\emptyset, R(u_1,u_2) \land \rel{Hist}(h_1) \land \rel{Hist}(h_1),\\
\,\,\,\,\,\,\,\,\set{Q(u_2)}, \set{R(u_2,h_1), R(u_2,h_2)}\rangle\\
\end{array}
\right\}
\]
\end{example}

\newcommand{\lock}[1]{\rel{Lock}_{#1}}
\newcommand{\deleting}[1]{\rel{DelPhase}_{#1}}
\newcommand{\adding}[1]{\rel{AddPhase}_{#1}}

\newcommand{\inputrel}[1]{\rel{FreshInput}_{#1}}
\newcommand{\parrel}[1]{\rel{ParMatch}_{#1}}

\newcommand{\nolock}{\Phi_{\mathit{NoLock}}}
\newcommand{\parcomputed}[1]{\Phi_{\beta}^{\mathit{AllSub}}}

\newcommand{\initact}[1]{\exact{Init}_{#1}}
\newcommand{\computeparact}[1]{\exact{CompAns}_{#1}}
\newcommand{\enupdate}[1]{\exact{EnableU}_{#1}}
\newcommand{\dodelact}[1]{\exact{ApplyDel}_{#1}}
\newcommand{\doaddact}[1]{\exact{ApplyAdd}_{#1}}
\newcommand{\enaddact}[1]{\exact{DelToAdd}_{#1}}
\newcommand{\finact}[1]{\exact{Finalize}_{#1}}

\newcommand{\delmarker}{\cval{0}}
\newcommand{\addmarker}{\cval{1}}

\newcommand{\crord}{\exact{NewO}}

\subsection{Simulating Bulk Operations}
\label{app:bulk}
We show how \emph{bulk actions} can be 
simulated by standard \syss. Recall that \syss adopt a
\textit{retrieve-one-answer-per-step} semantics, i.e., a \sys action
$\tup{\vec{u}, \vec{v},
  \guard,\del,\add}$ is applied by nondeterministically grounding its
action parameters $\vec{u}$ with \emph{one} answer to $\guard$. Bulk
actions, instead, require the adoption of  a
\textit{retrieve-all-answers-per-step} semantics, where the update
specified by $\del$ and
$\add$ would be enforced by simultaneously considering \emph{all}
answers to $\guard$.
\begin{example}
\label{ex:bulk}
Consider a \sys handling the replenishment of a warehouse. The \sys
operates over the following relations:
\begin{itemize}
\item $\rel{TBO}/1$, used to store those products that need to be ordered
  so as to replenish the warehouse.
\item $\rel{InOrder}/2$, used to keep track of orders and their
  contained items;  $\rel{InOrder}(\cval{p},\cval{o})$ indicates that
  product $\cval{p}$ belongs to order $\cval{o}$.
\end{itemize}
In this context, we want to model an action that handles the creation
of a new replenishment order, on the one hand ensuring that the newly
created order contains all products present in the $\rel{TBO}$
relation, and on the other hand emptying such a relation, so as to
avoid that the same products are ordered twice. Such an action
involves a bulk operation, since it must guarantee that \emph{for
  every} product to-be-ordered, that product is removed from the
$\rel{TBO}$ relation made part of the created order. Assuming a
\textit{retrieve-all-answers-per-step} semantics, this can be directly
captured by the \sys bulk action $\crord$, defined as follows:
\begin{itemize}
\item $\getguard{\crord} = \rel{TBO}(p)$;
\item $\getdel{\crord} = \set{\rel{TBO}(p)$};
\item $\getadd{\crord} = \set{\rel{InOrder}(p,o)$}.
\end{itemize}
Here, $p$ is  a (universally quantified) matching with all products present in $\rel{TBO}$, while $o$ is a
standard, fresh-input variable used to inject a fresh order
identifier into the system.
\end{example}
Let $\beta = \tup{\vec{u}, \vec{v},
  \guard,\del,\add}$ be a bulk action, i.e., an action where the
action parameters
$\vec{u}$ are implicitly universally quantified. 
We show how the bulk update induced by $\beta$ can be simulated  in a standard \sys
though a complex sequence of actions and the introduction of accessory
relations. 

The following accessory relations are used:
\emph{(i)} A proposition $\lock{\beta}$, used to ``lock'' the sequence of
  actions simulating $\beta$, guaranteeing that it is
not interrupted by other actions.
\emph{(ii)} A relation $\inputrel{\beta}$ with arity $|\vec{v}|$, used to
  store (in a single tuple) the selected substitution for the fresh-input variables of
  $\beta$, enabling the consistent usage of such a substitution when
  reconstrucing the bulk update of $\beta$.
\emph{(iii)} A relation $\parrel{\beta}$ with arity $|\vec{u}|+1$, used to
  incrementally store all answers to $\getguard{\beta}$, then exhaustively
  considering them when reconstructing the bulk update induced by
  $\beta$. The last argument of $\parrel{\beta}$ is used to ``flag''
  tuples that have been already considered for the corresponding
  deletion of tuples within the bulk update (more details are given
  below).
\emph{(iv)} Two propositions $\deleting{\beta}$ and $\adding{\beta}$,
  identifying those portions of the sequence of actions
  respectively dealing with the bulk deletion/addition   of $\beta$.

Whenever $\beta$ is eligible for execution, those accessory relations
are all empty. At the completion of
the sequence of actions simulating $\beta$, such accessory relations
will be empty again.
To ensure the non-interruptibility of the sequence of actions used
to simulate bulk actions, all actions of the \sys of interest must be
modified so as to incorporate the negation of all lock propositions
(like $\lock{\beta}$ above), denoted in the following $\nolock$.

The simulation of $\beta$ is done by structuring the sequence in three
phases. The first consists of the application of a single
initialization action
$\initact{\beta}$, executable when $\getguard{\beta}$ admits at
least one answer. $\initact{\beta}$ sets the lock, and 
stores the selected substitution for the fresh-input variables. Specifically, $\initact{\beta}$
has fresh-input variables $\vec{v}$, and is defined as:
\begin{itemize}
\item $\getguard{\initact{\beta}} = (\exists \vec{u}.\getguard{\beta}(\vec{u})) \land \nolock$;
\item $\getdel{\initact{\beta}} = \emptyset$;
\item $\getadd{\initact{\beta}} = \set{\lock{\beta},\inputrel{\beta}(\vec{v})}$.
\end{itemize}
The second phase deals with the computation of all answers to
$\getguard{\beta}$, storing them into the corresponding accessory relation. Such
a phase is identified by the presence of the lock for $\beta$, and by
the absence of the flags marking the bulk deletion and addition of
tuples. The computation of the answers to $\getguard{\beta}$ is handled by
iteratively executing action $\computeparact{\beta}$. This action is
executable when there is at least one answer to $\getguard{\beta}$ that has not
yet been transferred. If this is the case, it nondeterministically
picks one such answers, and transfers it into the accessory
relation. Specifically, $\computeparact{\beta}$ has exactly $\vec{u}$
as action parameters, no fresh-input variable, and is defined as follows:
\begin{itemize}
\item $
\getguard{\computeparact{\beta}} = \begin{array}[t]{@{}l@{}}
\lock{\beta} \land \neg
  \deleting{\beta} \land \neg \adding{\beta}\\
 {}\land \getguard{\beta}(\vec{u})
  \land \neg \parrel{\beta}(\vec{u}) ;
\end{array}
$
\item $\getdel{\computeparact{\beta}} = \emptyset$;
\item $\getadd{\computeparact{\beta}} = \set{\parrel{\beta}(\vec{u},\delmarker)}$.
\end{itemize}
When inserting an answer tuple into the accessory
relation, the last, additional argument of $\parrel{\beta}$ is set to
$\delmarker$. This witnesses that such an answer tuple has still to be
considered when reconstructing the deletions induced by $\beta$.

Action $\computeparact{\beta}$ becomes non-executable when the FO
sentence $\parcomputed{\beta} = \forall \vec{u}. \getguard{\beta}(\vec{u})
\rightarrow \parrel{\beta}(\vec{u})$ holds in the current database. We
consequently insert a dedicated action $\enupdate{\beta}$, marking the end of reiterated application of
$\computeparact{\beta}$. This action is executable when all answer tuples
have been transferred to the accessory relation, and has the effect of
indicating that it is now time to apply the bulk update induced by
$\beta$. The execution semantics of \syss dictates that additions have
priority over deletions. For this reason, the bulk update first
requires to consider all deletions, and then all additions. Hence,
$\enupdate{\beta}$ raises flag $\deleting{\beta}$. Specifically,
$\enupdate{\beta}$ has no action parameters nor fresh-input variables,
and is defined as follows:
\begin{itemize}
\item $
\begin{array}[t]{@{}l@{}}
\getguard{\enupdate{\beta}} =\\
\qquad  \lock{\beta} \land \neg
  \deleting{\beta} \land \neg \adding{\beta}
  \land \parcomputed{\beta};
\end{array}$
\item $\getdel{\enupdate{\beta}} = \emptyset$;
\item $\getadd{\enupdate{\beta}} = \set{\deleting{\beta}}$.
\end{itemize}

The introduction of flag $\deleting{\beta}$ marks the beginning of the
third phase, which deals with the actual bulk update. As mentioned
before, this phase is split into two sub-phases: a first sub-phase
dealing with deletions, a second sub-phase dealing with
additions. Both sub-phases consists of the iterative application of
one dedicated action, which deals with the tuples to be deleted/added due to a
specific answer tuple to $\getguard{\beta}$. As for deletion, the iteratively
executed action is $\dodelact{\beta}$.
 This action
nondeterministically picks an answer tuple for $\getguard{\beta}$ that still has
to be considered for deletion. This is done by extracting a tuple from
$\parrel{\beta}$, checking that the last argument of such a tuple
corresponds to $\delmarker$. In addition, since the deletion may
depend on the fresh-input of $\beta$ as well, $\dodelact{\beta}$ also needs to
extract the single tuple present in the input accessory relation
$\inputrel{\beta}$. Specifically, $\dodelact{\beta}$ has 
$\vec{u}$ as action parameters, no fresh-input
variables, and is defined as follows:
\begin{itemize}
\item $
\getguard{\dodelact{\beta}} =
\deleting{\beta}
  \land \parrel{\beta}(\vec{u},\delmarker)$;
\item $\getdel{\dodelact{\beta}} = \getdel{\beta} \cup \set{\parrel{\beta}(\vec{u},\delmarker)}$;
\item $\getadd{\dodelact{\beta}} = \set{\parrel{\beta}(\vec{u},\addmarker)}$.
\end{itemize}
Notice that the update over $\parrel{\beta}$ changing the last
argument of the tuple $\vec{u}$ is used to track  that the
selected tuple has been already processed for deletion.
$\dodelact{\beta}$ cannot be applied anymore when all
tuples in $\parrel{\beta}$ are marked with $\addmarker$. This
situation indicates that no more deletions have to be considered, and
that bulk addition must now be handled. Such a transition is captured by the
dedicated action $\enaddact{\beta}$, which does not have action
parameters nor fresh-input variables, and is defined as follows:
\begin{itemize}
\item $
\begin{array}[t]{@{}l@{}}
\getguard{\enaddact{\beta}} =\\
\qquad \deleting{\beta} \land \forall
  \vec{u},m.\parrel{\beta}(\vec{v},m) \rightarrow m = \addmarker;
\end{array}$
\item $\getdel{\enaddact{\beta}} = \set{\deleting{\beta}}$;
\item $\getadd{\enaddact{\beta}} = \set{\adding{\beta}}$.
\end{itemize}

The second sub-phase simulating the bulk update is captured by the
iterative application of action $\doaddact{\beta}$, which closely
resembles  $\dodelact{\beta}$, with three differences:
\emph{(i)}
it requires to consider not only the answers to
  $\getguard{\beta}$ (stored in $\parrel{\beta}$), but also the
  selected matching for the input variables (stored in
  $\inputrel{\beta}$);
\emph{(ii)}
it handles the insertion of tuples, hence refers to $\getadd{\beta}$;
\emph{(iii)}
  it removes a tuple from
$\parrel{\beta}$ to mark that it has been considered for addition. Specifically,  $\doaddact{\beta}$ has 
$\vec{u}$ and $\vec{v}$ as action parameters, no fresh-input
variables, and is defined as:
\begin{itemize}
\item $
\begin{array}[t]{@{}l@{}}
\getguard{\doaddact{\beta}} =\\
\qquad 
\adding{\beta}
  \land \parrel{\beta}(\vec{u},\addmarker) \land
  \inputrel{\beta}(\vec{v});
\end{array}$
\item $\getdel{\dodelact{\beta}} = \set{\parrel{\beta}(\vec{u},\addmarker)}$;
\item $\getadd{\dodelact{\beta}} = \getadd{\beta}$.
\end{itemize}
It is easy to see that this last sub-phase ends where there is no more
tuple in $\parrel{\beta}$. This marks the end of the addition loop,
and triggers the execution of the last action of the sequence, namely
$\finact{\beta}$. This last action mirrors $\initact{\beta}$, and has
in fact a twofold effect: releasing the lock(s), and emptying the
content of $\inputrel{\beta}$ by removing its single tuple (recall, in
fact, that $\parrel{\beta}$ is already empty). Specifically,
$\finact{\beta}$ has $\vec{v}$ as action parameters (since it needs to
match those against the  $\inputrel{\beta}$ relation), has no fresh-input variable, and
is defined as:
\begin{itemize}
\item $
\begin{array}[t]{@{}l@{}}
\getguard{\finact{\beta}} = \\
\qquad \inputrel{\beta}(\vec{v}) \land \neg \exists
  \vec{u},m. \parrel{\beta}(\vec{u},m);
\end{array}
$
\item $\getdel{\finact{\beta}} = \set{\inputrel{\beta}(\vec{v})}$;
\item $\getadd{\finact{\beta}} = \emptyset$.
\end{itemize}

\begin{example}
By applying the general mechanism to simulate bulk actions with
standard ones on the bulk action of Example~\ref{ex:bulk}, we get:
\begin{itemize}
\item The init action $\initact{\crord}$ with fresh-input variable
  $o$, where:
\begin{itemize}
\item  $\getguard{\initact{\crord}} = (\exists p.\rel{TBO}(p)) \land \nolock$;
\item $\getdel{\initact{\crord}} = \emptyset$;
\item $\getadd{\initact{\crord}} = \set{\lock{\crord},\inputrel{\crord}(o)}$.
\end{itemize}
\item The guard answer computation action $\computeparact{\crord}$
  with action parameter $p$, where
\begin{itemize}
\item  $
\begin{array}[t]{@{}l@{}}
\getguard{\computeparact{\crord}} =\\
\qquad \lock{\crord} \land \neg
  \deleting{\crord} \land \neg \adding{\crord}\\
\qquad {}\land \rel{TBO}(p) \land \neg \parrel{\crord}(p);
\end{array}
$
\item $\getdel{\computeparact{\crord}} = \emptyset$;
\item $\getadd{\computeparact{\crord}} = \set{\rel{\parrel{\crord}}(p,\delmarker)}$.
\end{itemize}
\item The update enablement action $\enupdate{\crord}$, where
\begin{itemize}
\item  $\begin{array}[t]{@{}l@{}}
\getguard{\enupdate{\crord}} =\\
\qquad  \lock{\beta} \land \neg
  \deleting{\beta} 
\land \neg \adding{\beta}
\\  \qquad {}\land \forall p. \rel{TBO}(p) \rightarrow \parrel{\crord}(p);
\end{array}$
\item $\getdel{\enupdate{\crord}} = \emptyset$;
\item $\getadd{\enupdate{\crord}} = \set{\deleting{\crord}}$.
\end{itemize}
\item The bulk deletion action $\dodelact{\crord}$ with action parameter
  $p$, where:
\begin{itemize}
\item  
$\begin{array}[t]{@{}l@{}}
\getguard{\dodelact{\crord}} =\\
\qquad \deleting{\crord} \land \parrel{\crord}(p,\delmarker);
 \end{array}
 $
\item $\getdel{\dodelact{\crord}} = \set{\rel{TBO}(p),\parrel{\crord}(p,\delmarker)}$;
\item $\getadd{\dodelact{\crord}} = \set{\parrel{\crord}(p,\addmarker)}$.
\end{itemize}
\item The action $\enaddact{\crord}$, marking the transition between the bulk deletion and
  the bulk addition:
\begin{itemize}
\item  $
\begin{array}[t]{@{}l@{}}
\getguard{\enaddact{\crord}} = \\
\deleting{\crord} \land \forall p,m.\parrel{\crord}(p,m) \rightarrow m
  = \addmarker;
\end{array}
$
\item $\getdel{\enaddact{\crord}} = \set{\deleting{\crord}}$;
\item $\getadd{\enaddact{\crord}} = \set{\adding{\crord}}$.
\end{itemize}
\item The bulk addition action $\doaddact{\crord}$ with action parameter
 s $p$ and $o$, where:
\begin{itemize}
\item  $
\begin{array}[t]{@{}l@{}}
\getguard{\doaddact{\crord}} = \\
\ \adding{\crord} \land \parrel{\crord}(p,\addmarker) \land \inputrel{\crord}(o);
\end{array}
$
\item $\getdel{\dodelact{\crord}} = \set{\parrel{\crord}(p,\addmarker)}$;
\item $\getadd{\dodelact{\crord}} = \set{\rel{InOrder}(p,o)}$.
\end{itemize}
\item The finalization action $\finact{\crord}$ with action parameter
  $o$, where:
\begin{itemize}
\item  $
\begin{array}[t]{@{}l@{}}
\getguard{\finact{\crord}} = \\
\qquad \inputrel{\crord}(o) \land \neg \exists p,m.\parrel{\crord}(p,m);
\end{array}
$
\item $\getdel{\finact{\crord}} = \set{\inputrel{\crord}(o) }$;
\item $\getadd{\finact{\crord}} = \emptyset$.
\end{itemize}
\end{itemize}
\end{example}


\end{document}